\documentclass[usenatbib,useAMS]{mnras}
\usepackage[T2A]{fontenc} 
\usepackage{tablefootnote}
\usepackage{multirow}
\usepackage{xcolor}
\usepackage{graphicx}	
\usepackage{amsmath}	
\usepackage{amssymb}	
\usepackage{multicol}   
\usepackage{bm}		
\usepackage[normalem]{ulem} 
\usepackage{caption} 
\usepackage{url}
\usepackage{longtable}
\usepackage{pdfpages}
\usepackage{subcaption}
\usepackage{hyperref}
\usepackage{pdflscape}

\title[High-redshift radio galaxies]{Radio properties of high-redshift galaxies at $z \geq 1$}

\author[M.~Khabibullina et al.]{ 
M.~Khabibullina,$^{1}$\thanks{E-mail: rita@sao.ru} 
A.~Mikhailov,$^{1}$
Yu.~Sotnikova,$^{1,2}$
T.~Mufakharov,$^{1,2}$
M.~Mingaliev,$^{1,2,3}$
\newauthor
A.~Kudryashova,$^{1}$
N.~Bursov,$^{1}$
V.~Stolyarov,$^{1,4}$
R.~Udovitskij$^{1}$
\\
$^{1}$Special Astrophysical Observatoryy, Russian Academy of Sciences, Nizhny Arkhyz, 369167, Russia\\
$^{2}$Kazan (Volga Region) Federal University, Kazan 420008, Russia\\
$^{3}$Institute of Applied Astronomy, Russian Academy of Sciences, St. Petersburg 191187, Russia\\
$^{4}$Astrophysics Group, Cavendish Laboratory, University of Cambridge, Cambridge, CB3 0HE, UK}

\date{Received July 6, 2023; accepted September 11, 2023}

\begin{document}
\label{firstpage}
\pagerange{\pageref{firstpage}--\pageref{lastpage}}
\maketitle

\begin{abstract}
Study of high-redshift radio galaxies (HzRGs) can shed light on the active galactic nuclei (AGNs) evolution in massive elliptical galaxies. The vast majority of observed high-redshift AGNs are quasars, and there are very few radio galaxies at redshifts $z>3$. We present the radio properties of 173~sources optically identified with radio galaxies at $z\geqslant1$ with flux densities
$S_{1.4}\geqslant20$~mJy.  Literature data were collected for compilation of broadband radio spectra, estimation of radio variability, radio luminosity, and radio loudness. Almost 60\% of the galaxies have steep or ultra-steep radio spectra; 22\% have flat, inverted, upturn, and complex spectral shapes, and 18\% have peaked spectra (PS). The majority of the PS sources in the sample (20/31) are megahertz-peaked spectrum sources candidates, i.e. possibly very young and compact radio galaxies. The median values of the variability indices
at 11 and 5~GHz are $V_{S_{11}}=0.14$ and $V_{S_{5}}=0.13$, which generally indicates a weak or moderate character of the long-term variability of the studied galaxies. The typical radio luminosity and radio loudness are $L_{5}=10^{43}$--\,$10^{44}$~erg\,s$^{-1}$ and $\log R=3$\,--\,$4$ respectively. We have found less prominent features of the bright compact radio cores for our sample compared
to high-redshift quasars at $z\geq3$. The variety of the obtained radio properties shows the different conditions for the formation of radio emission sources in galaxies.
\end{abstract}

\begin{keywords}
                galaxies: active --       
                galaxies: high-redshift --
                quasars: general --
                radio continuum: galaxies
\end{keywords}

\maketitle

\section{INTRODUCTION}

Distant galaxies with redshifts $z > 1$ and radio
luminosities $L_{\rm 500MHz}$ exceeding 10$^{34}$\,erg\,s$^{-1}$ are classified as powerful radio galaxies \citep{2008AARv..15...67M} with large $z$ -- HzRG. These are very rare objects with a density of approximately \mbox{10\,$^{-8}$ Mpc$^{-3}$} for the redshift range ($2\leq z < 5$). Their spectral energy distribution includes radiation from the dust and stellar components as well as radiation of the active nucleus. Studies of the first two components show that HzRGs are among the most massive galaxies in the Universe
\citep{2007ApJS..171..353S, 2009MNRAS.395.1099B, 2010ApJ...725...36D} and are thought to be the progenitors of the massive ellipticals we observe today \citep{1964ApJ...140...35M}. 

These facts allow the radio galaxies to be used as tools for the study of the visible and dark matter distributions, dynamics of the Universe, and the history of its structure formation. The parameters determined from observed data include the age of the galaxy \citep{1999BSAO...48...41V, 2001BSAO...52....5V, 2000A&AT...19..663V}, which is limited by the age of the Universe and cannot exceed the latter, since galaxies need certain time to form after the birth of the Universe. A striking feature of radio galaxies is that we observe them virtually from their formation epoch, i.e., their radio sources are so powerful that modern radio surveys have catalogued almost all objects of this type \citep{Verkhodanov_and_Parijskij}. Such objects can serve as good probes for the study of the formation of galaxy clusters. Therefore, constructing samples of radio galaxies from various redshift ranges is one of the most important tasks in observational cosmology.

A correlation that has frequently been used to search for HzRGs,
but is still under debate, is the spectral index to redshift relation (``$\alpha$\,--\,$z$''), where radio sources with steeper spectra appear to be preferentially associated with higher redshift sources \citep{1979A&AS...35..153T, 2000A&AS..143..303D, 2008AARv..15...67M}. One of the mechanisms that has been suggested to explain this relation is that this is a result of the increased inverse Compton (IC) losses at higher redshifts due to the higher energy density of the cosmic microwave background (CMB) (e.g., \citealt{2014MNRAS.438.2694G} or \citealt{2006MNRAS.371..852K}). Recently, by modelling a large number of sources over a wide range of redshifts using the Broadband Radio Astronomy ToolS\footnote{\url{http://www.askanastronomer.co.uk/brats/}} \citep{2013MNRAS.435.3353H, 2015MNRAS.454.3403H},
\cite{2018MNRAS.480.2726M} found that IC losses could indeed be the primary driver behind the ``$\alpha$\,--\,$z$'' relation.
There are two other scenarios that could result in this relation, one connected with observational biases and the other related to the environment of the galaxy. A denser medium at high redshift would keep the radio lobes more confined and would result in higher surface brightness for older radio emission coming from the low-energy electron populations. If at higher redshifts radio galaxies preferentially lie in denser environments, these effects could lead to the observed relation. Another effect from the increased ambient density is related to the acceleration of electrons in or near the hotspot. 

Although as yet there is no general common consensus on the nature of the ``$\alpha$--$z$'' correlation, the very fact that this empirical dependence exists is important for research. An example of an object selected in this way is the radio galaxy that is the redshift record-holder with $z=5.19$ and its spectral index\footnote{Determined in assumption of the power law relation between the flux density $S_{\nu}$ at a frequency $\nu$ and the spectral index $\alpha$: $S_{\nu}\sim \nu^{\alpha}$.} $\alpha = -1.63$ at \mbox{0.365\,--\,1.4}~GHz range \citep{1999ApJ...518L..61V}. Another example is the object RC 0311$+$0507 investigated in the ``Big Trio'' Program \citep{2006AstL...32..433K}. Its redshift $z=4.514$, it has the record radio luminosity for radio galaxies with $z > 4$, and its spectral index $\alpha = -1.31$ at \mbox{0.365\,--\,4.85}~GHz range.

The peaked spectrum shape in the extragalactic radio sources that are characterized by steep optically thin spectra and a spectral turnover in the GHz domain is considered an indicator of young radio sources due to their small size. These AGNs are subdivided into compact steep-spectrum (CSS), gigahertz peaked-spectrum (GPS), and high-frequency peaked (HFP) sources. The CSS sources have typical turnover frequencies in the observer's frame of reference smaller than 500~MHz and linear sizes (LS) 
of 1--20~kpc \citep{1998PASP..110..493O}. The GPS and HFP sources have LS~$< 1$~kpc with \mbox{$0.5 \leq \nu_{\rm peak,obs}\leq 5$}~GHz 
and $\nu_{\rm peak,obs} > 5$~GHz respectively \citep{2000A&A...363..887D}. Additionally the anticorrelation between the rest frame turnover frequency and the projected linear size for the GPS and CSS sources supports assumptions about their evolutionary connection \citep{1990A&A...231..333F,1997AJ....113..148O}. Compact PS sources with an observed turnover frequency below 1~GHz have also been referred to as megahertz peaked-spectrum (MPS) sources \citep{2015MNRAS.450.1477C,2016MNRAS.459.2455C,2016MNRAS.463.3260C,2004NewAR..48.1157F}. It is believed that they are a combination of relatively nearby CSS/GPS sources and more distant high-frequency peakers with the peak frequency shifted to low frequencies due to the cosmological redshift. 

Investigation of the nature of PS sources can help to understand 
the features of AGN evolution. Some studies have shown that 
the peaked spectrum of high-redshift quasars is formed mainly due to 
the dominant contribution of radio emission from the bright compact core \citep{1998A&AS..131..435S,2021MNRAS.508.2798S}. Many known extremely distant AGNs, at $z>5$, are blazars and have a peaked radio spectrum (for example, \cite{2004ApJ...610L...9R,2010A&A...524A..83F,2016AN....337..101L,2017MNRAS.467.2039C,2020A&A...643L..12S,2021MNRAS.503.4662M}). Such a spectral shape prevails at higher redshifts, where bright jet-beamed blazars are easily discovered. However, the steepening of the spectral index with increasing redshift has not yet been revealed \citep{2012MNRAS.420.2644K,2014A&A...569A..52S,2014MNRAS.443.2590S}. One of the reasons is the poor statistics for distant radio sources. At the same time, it should not be ruled out that AGNs can evolve in different ways and such a correlation is natural.

In order to study the nature of HzRGs and their evolution, it is important to obtain a large sample of such objects. Nowadays databases and identification techniques allow automating the search and identification of distant radio galaxies. The catalog of HzRGs can be used not only to perform just purely cosmological studies but also 
to carry out statistical analysis of identification lists and corresponding populations at different wavelengths \citep{2000BSAO...50..115V,2003ARep...47..119V, 2004Ap.....47..505B,2003BSAO...55...90T}, to search for and analyze the properties of radio galaxies \citep{1999BSAO...48...41V,1994ARep...38..307V, 1999ARep...43..417V, 1999BSAO...48....5P, 2008pc2..conf..247V, 2014AstL...40..606S, 2014ARep...58..506S}, and to
simulate radio astronomical surveys at different telescopes 
\citep{1994AstL...20..671G, 2008AstBu..63...95K, 2008AstBu..63...56M}.

Such a catalog of radio galaxies with  redshifts $z\geq0.3$ 
has been previously constructed by the authors \cite{2009AstBu..64..123K, 2009AstBu..64..276K, 2009AstBu..64..340K}. The catalog contains 2442 sources with  redshifts, photometric magnitudes, 
and flux densities in the radio band; the sizes of the radio sources, 
the coordinates, and the radio spectral indices were calculated from the results of cross-identification with the radio catalogs from the CATS\footnote{Astronomical CATalogs Support System, (\url{http://cats.sao.ru})}  database \citep{1997BaltA...6..275V, 2005BSAO...58..118V}.

In this paper we collect a sample of 173~bright radio galaxies, with spectral flux density more than 20~mJy at 1.4~GHz (hereafter we indicate the frequency in lowercase index), with spectroscopic redshifts $z\geqslant1$, based on the data from the NED\footnote{NASA/IPAC Extragalactic Database (\url{http://ned.ipac.caltech.edu})}, SDSS\footnote{\url{http://www.sdss.org}} \citep{2020ApJS..250....8L}, CATS, and VizieR Information System\footnote{\url{https://vizier.u-strasbg.fr/viz-bin/VizieR}} \citep{2000A&AS..143...23O}. We present their radio properties, radio spectra classification, radio variability parameters at a long time scale up to 30 years, radio loudness and radio luminosity estimates. We propose new MPS, GPS, and HFP candidates and estimate their fraction among bright radio galaxies at high redshifts. Several infrared properties of AGNs have been analyzed.

\section{THE SAMPLE}

\begin{table*} \setlength{\tabcolsep}{3pt}
\centering
\caption{\label{tab:catalogs}The basic radio catalogues used in the study} \medskip
\centering
\begin{tabular}{c|c|c|l}
\hline
{Catalog/Instrument$^*$} & {Epoch} & Frequency, GHz&  \multicolumn{1}{c}{Reference} \\
\hline
TXS & 1974\,--\,1983   &  0.365       &  \protect\cite{1996AJ....111.1945D}\\
GBIMO & 1979\,--\,1996   &  2.5, 8.2    &  \protect\cite{2001ApJS..136..265L}\\
RCSP & 1980\,--\,1981   &  3.9     &    \protect\cite{1996BSAO...42....5B} \\
GB6 & 1986\,--\,1987   & 4.8        & \protect\cite{1996ApJS..103..427G} \\
PKS90 & 1989 & 8.4   &  \protect\cite{1991MNRAS.251..330W}\\
WENSS & 1991         &  0.325, 0.609  &  \protect{\cite{1997A&AS..124..259R}} \\
FIRST & 1993\,--\,1994 &  1.4    &  \protect\cite{1994ASPC...61..165B}\\
NVSS & 1993\,--\,1996 & 1.4       & \protect\cite{1998AJ....115.1693C} \\
JVAS & 1995\,--\,1997 & 8.4   &  \protect\cite{1998AAS...193.4004W}\\
PMNMi & 1995\,--\,1996 & 0.96, 2.3, 3.9, 7.7, 11.2  & \protect\cite{1998BSAO...46...28M} \\
SRCKi & 1995\,--\,1996 & 0.96, 2.3, 3.9, 7.7, 11.2, 21.7   & \protect\cite{2002BSAO...54....5K} \\
KOV97 & 1997 & 0.96, 2.3, 3.9, 7.7, 11.2, 21.7  &  \protect{\cite{1999A&AS..139..545K}}\\
VLSS & 1998 & 0.074          &  \protect\cite{2007AJ....134.1245C}\\
NCPMi & 1999  & 2.3, 3.9, 7.7, 11.2, 21.7  & \protect{\cite{2001A&A...370...78M}} \\
GPSra & 2006\,--\,2010 & 1.1, 2.3, 4.8, 7.7, 11.2, 21.7  & \protect{\cite{2012A&A...544A..25M}} \\
ATCA & 2007       &  1.4       &  \protect\cite{2012MNRAS.427.1830W}\\
CGR15 & 2008\,--\,2009 &  15       &  \protect\cite{2011ApJS..194...29R}\\
BlMin & 2005\,--\,2014 & 1.1, 2.3, 4.8, 7.7, 11.2, 21.7 & \protect\cite{2017AN....338..700M} \\
GPSSt & 2006\,--\,2017 & 1.1, 2.3, 4.8, 7.7/8.2, 11.2, 21.7  &  \protect\cite{2019AstBu..74..348S} \\
TGSS & 2010\,--\,2012 & 0.15       & \protect{\cite{2017A&A...598A..78I}} \\
GLEAM & 2013\,--\,2015 & 0.072\,--\,0.231 & \protect\cite{2017MNRAS.464.1146H} \\
VLASS & 2016\,--\,2019 & 2\,--\,4        & \protect\cite{2020PASP..132c5001L} \\
RATAN-600 & 2017\,--\,2020 & 1.2, 2.3, 4.7, 8.2, 11.2, 22.3 & \protect\cite{{2021MNRAS.508.2798S}} \\
\hline
\multicolumn{4}{l}{\footnotesize$^*$ Names are given in accordance with the CATS identifiers}\\
\end{tabular}
\end{table*}

To build the primary list, we used the NED database, where we selected objects with the following parameters: redshifts $z\geq1$, flux densities at 1.4~GHz $S_{1.4}\geq20$ mJy, classified as a galaxy. The initial list contained 9\,000 objects. This sample was polluted by objects with differing properties. So the next stage consisted in cleaning the initial sample from wrong sources. To this end, we selected the following objects to be removed from the initial sample:
\begin{list}{}{
\setlength\leftmargin{6mm} \setlength\topsep{0mm}
\setlength\parsep{0mm} \setlength\itemsep{1mm} }
    \item[1)] 
 objects without spectroscopic redshifts; \item[2)]  objects which had no the morphological type ``radio galaxy''; \item[3)]  objects without NVSS observations. 
 \end{list}

We performed an extensive scan of the literature including 
the optical, infrared, and radio surveys represented in 
the NED database, VizieR Information System, and CATS. To work with the NED database as well as with the VizieR collection of catalogues, the {\tt Python} module {\tt{astroquery}} was used \citep{2019AJ....157...98G}. It allows batch searches in a variety of databases using a common API\footnote{\url{https://astroquery.readthedocs.io}}. In addition to the radio catalogues, cross-matching with WISE \citep{2010AJ....140.1868W} was performed (see details in Section~\ref{sec:WISE}). The total number of the catalogs used
is more than 170, the vast majority of the measurements are provided by 21 radio catalogues that are summarized in Table~\ref{tab:catalogs}, where we include the epochs, frequencies of observations, and corresponding literature references. The data cover a time period of several dozens of years, and the main data are represented by the NRAO VLA Sky Survey (NVSS, \citealt{1998AJ....115.1693C}), 
Faint Images of the Radio Sky at Twenty-cm Survey (FIRST, \citealt{1994ASPC...61..165B}), Westerbork Northern Sky Survey (WENSS, \citealt{1997A&AS..124..259R}), 
Green Bank 6-cm Survey (GB6, \citealt{1996ApJS..103..427G}), 
Australia Telescope 20~GHz Survey (ATCA20, \citealt{2010MNRAS.402.2403M}), GaLactic and Extragalactic All-sky Murchison Widefield Array (GLEAM) survey at 72\,--\,231~MHz \mbox{(2013\,--\,2014}; \citealt{2017MNRAS.464.1146H}), Giant Metrewave Radio Telescope Sky Survey (TGSS) at 150~MHz (2015; \citealt{2017A&A...598A..78I}), VLA measurements \citep{2007ApJS..171...61H}, and others. 
The time period begins in 1974 with the Texas Survey of Radio Sources, covering $-35\,.\!\!^\circ5 < \delta < 71\,.\!\!^\circ5$ at 365~MHz (TXS, \citealt{1996AJ....111.1945D,1996ApJS..103..427G}).
A significant contribution in the measurements was provided by the Multiyear Monitoring Program of Compact Radio Sources at 2.5 and 8.2~GHz \citep{2001ApJS..136..265L} and by the low frequency  GLEAM{\footnote{GaLactic and Extragalactic All-sky Murchison Widefield Array}} survey at \mbox{72\,--\,231}~MHz \citep{2017MNRAS.464.1146H}. The multi-frequency quasi-simultaneous observations with the RATAN-600 radio telescope are presented by several catalogs at six frequencies from the papers of
\cite{1996BSAO...42....5B,1998BSAO...46...28M,2002BSAO...54....5K,1999A&AS..139..545K,2001A&A...370...78M,2017AN....338..700M,2012A&A...544A..25M,2019AstBu..74..348S,2021MNRAS.508.2798S}.

The sample includes additional objects found in 
the
literature individually. They are:
\begin{itemize}
\item  NVSS\,J142047+120547 \citep{2021AN....342.1092G}, 
\item NVSS\,J160608+312447 \citep{2008ApJS..175...97H}, 
\item RC\,J0311+0507 (\citealt{2014MNRAS.439.2314P}, NVSS\,J031147+050802), 
\item MRC\,0251$-$273 (\citealt{2008AARv..15...67M}, NVSS\,J025316$-$270913), 
\item MRC\,2025$-$218 (\citealt{2008AARv..15...67M}, NVSS\,J202759$-$214057), and 
\item MRC\,2104$-$242 (\citealt{2008AARv..15...67M}, NVSS\,J210658$-$240504). 
\end{itemize}

According to the early data \citep{2015MNRAS.446.2483S}, NVSS\,J142047+120547 was classified as a beamed blazar source,
but \cite{2021AN....342.1092G} concluded from observations that it is most likely a young radio galaxy. For NVSS\,J160608+312447 there is no information in NED about the method for redshift determination; but as can be seen from \cite{2008ApJS..175...97H}, the redshift was determined spectroscopically in the CGRaBS (Candidate Gamma-Ray
Blazar Survey) survey, so we also added it to our sample.

As a result, the sample contains 173~objects which are classified very diversely: 105 (61\% of the sample) QSOs and QSO candidates, five BL\,Lacs, three~BZQs, 59 radio galaxies
and one object of mixed type (Table~\ref{table1_appendix}, a short fragment is presented. The full version can be found in VizieR\footnote{\url{https://cdsarc.cds.unistra.fr/viz-bin/cat/J/other/AstBu/78.443}}). For future analysis we consider two subsamples, one containing objects classified as galaxies with quasars (114 sources, named ``Q''), and the rest 59 objects designated as ``G''. 
The flux densities for the objects range from 0.02 to 9.8~Jy at 1.4~GHz with the median value of 0.12~Jy.

\begin{table*} \setlength{\tabcolsep}{2pt}
\centering
\caption{\label{table1_appendix}  The parameters of the galaxies in the sample: NVSS source name (J2000 format), redshift z, flux density at 1.4~GHz $S_{1.4}$ in Jy from the NVSS database, optical object type, and the literature where the optical classification was taken: [1]---\protect\cite{2018A&A...613A..51P}; [2]---\citet{2019ApJS..242....4D}; [3]---\citet{2015A&A...583A..75S}; [4]---\citet{2007ApJS..171..353S}; [5]---\citet{1999AJ....117.1122S}; [6]---\citet{2019MNRAS.483.1354S}. The redshifts are taken from the NED database, SDSS database, and articles. A short fragment is presented.} 
\begin{tabular}{l|c|c|c|c}
\hline

\multicolumn{1}{c|}{NVSS name} &  $z$  & $S_{1.4}$, Jy & Type & Reference \\ \hline
 \multicolumn{1}{c|}{1} & 2 & 3 & 4 & 5 \\
\hline

003005$+$295706  &  5.199  &  $0.017\pm0.001$  &  G, QSO  &  SDSSdr13\protect\footnote{\url{http://www.sdss.org/dr13/data_access/bulk/}}   \\
003205$-$041417  &  3.161  &  $0.042\pm0.002$  &  G, QSO  &  [1]   \\ 
003818$+$122731  &  1.395  &  $1.034\pm0.031$  &  G, BL Lac  &   [2]   \\ 
010152$-$283119  &  1.694  &  $0.661\pm0.020$  &  G, QSO  &  [3]   \\ 
011322$+$133505  &  2.661  &  $0.033\pm0.001$  &  G, QSO  &  SDSSdr13   \\ 
011651$-$205206  &  1.415  &  $4.091\pm0.123$  &  G  &  [4]   \\ 
011747$+$011407  &  3.696  &  $0.028\pm0.001$  &  G, QSO  &  [1]   \\ 
012229$+$192339  &  1.595  &  $0.449\pm0.014$  &  G  &  [5]   \\ 
012529$+$005407  &  1.711  &  $0.070\pm0.002$  &  G, QSO  &   [1]   \\ 
013028$-$261000  &  2.347  &  $1.464\pm0.049$  &  G  &  [6]   \\ 
\hline
\end{tabular}
\end{table*}

\begin{figure}
\includegraphics[width=0.47\textwidth]{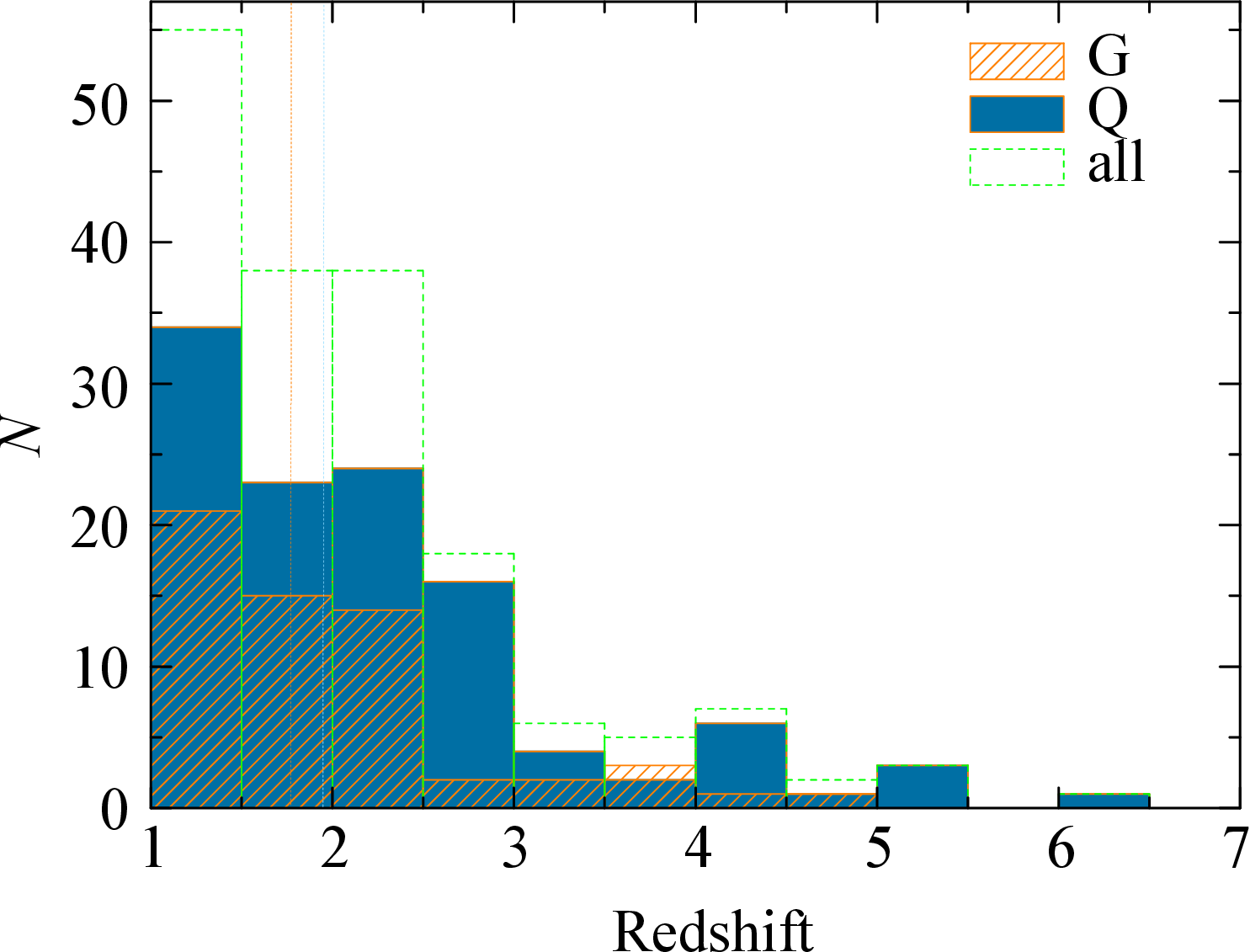}
\caption{\small Redshift distribution for the sample. The median values are marked by the orange and blue lines for the G and Q types, respectively.}
\label{distr_z}
\end{figure}
\begin{figure}
\includegraphics[width=0.44\textwidth]{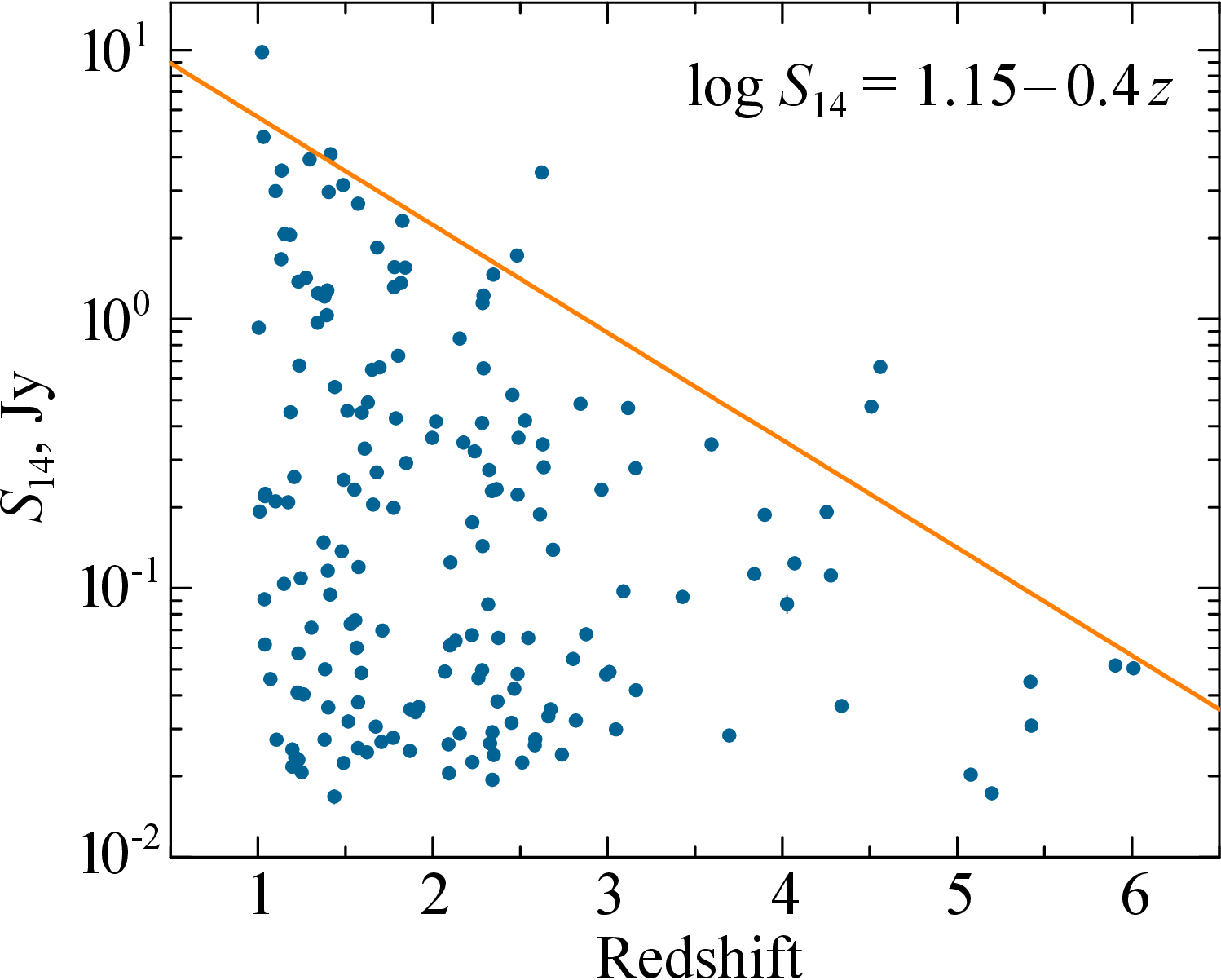} \caption{\small Flux density in Jy at 1.4~GHz versus redshift. The regression fit is drawn for the maximum fluxes inside bins with a redshift step of $\Delta z = 0.5$.}
\label{fig:flux-z}
\end{figure}

The redshifts of the galaxies extend from 1 up to 6.01 with 
median values of 1.78 for the G-type sources and 1.96 for the Q type (Fig.~\ref{distr_z}). The object with the highest redshift in our sample is J1225+4140, that is classified both as a galaxy and alternatively as a quasar, so we consider it as a Q-type source in our analysis. Its photometric redshift $z=0.59\pm0.05$, which strongly contradicts the spectroscopic redshift estimated in SDSS\,DR13\footnote{\url{http://www.sdss.org/dr13/data_access/bulk/}} as $z=6.008\pm0.001$. Its radio spectrum is well approximated with a straight line in the range from 150~MHz to 3~GHz and has a steep slope with $\alpha=-0.7$. We recently (in 2022\,--\,2023) estimated its flux density with RATAN-600 at frequencies higher than 3~GHz for the first time and obtained $19\pm3$ mJy at 4.7~GHz and $21\pm5$~mJy at 11.2~GHz (preliminary results to be published). These flux density values are in good agreement with its steep spectrum shape at 0.15\,--\,3~GHz.

In the ``flux\,--\,redshift'' diagram at 1.4~GHz (Fig.~\ref{fig:flux-z}), an upper flux boundary, which gives the maximum luminosities of the observed radio galaxies at different redshifts, is presented. As it was supposed in \cite{2009AstBu..64..123K}, it could be due to the dynamics of cosmological expansion. Following the above mentioned authors, we fit a linear regression to the maximum values of the distribution in bins of $\Delta z = 0.5$. The regression relation is given by the formula $\log S_{1.4} = p + rz$, where $S_{1.4}$ is the flux density at 1.4~GHz, $p = 1.15$ is a constant intercept, and $r = -0.4$ is the slope of the line. The resulting regression is in good agreement with the regression obtained in \cite{2009AstBu..64..123K}. Here we point out the especially powerful radio galaxies whose fluxes 
are greater by more than half the value of the regression boundary: NVSS\,J043236+413829 (3C\,119, $z=1.023$) \citep{1964ApJ...140..969K}, NVSS\,J074533+101112 (PKS\,0742+10, $z=2.624$) \citep{2007A&A...463...97L}, NVSS\,J160608+312447 (WISEA\,J160608.53+312446.5, $z=4.56$, \citealt{2008MNRAS.390..819G}), 
and NVSS\,J031147+050802 (4C\,+04.11, $z=4.510$) \citep{1967MmRAS..71...49G}.

\begin{figure}
\includegraphics[width=0.45\textwidth]{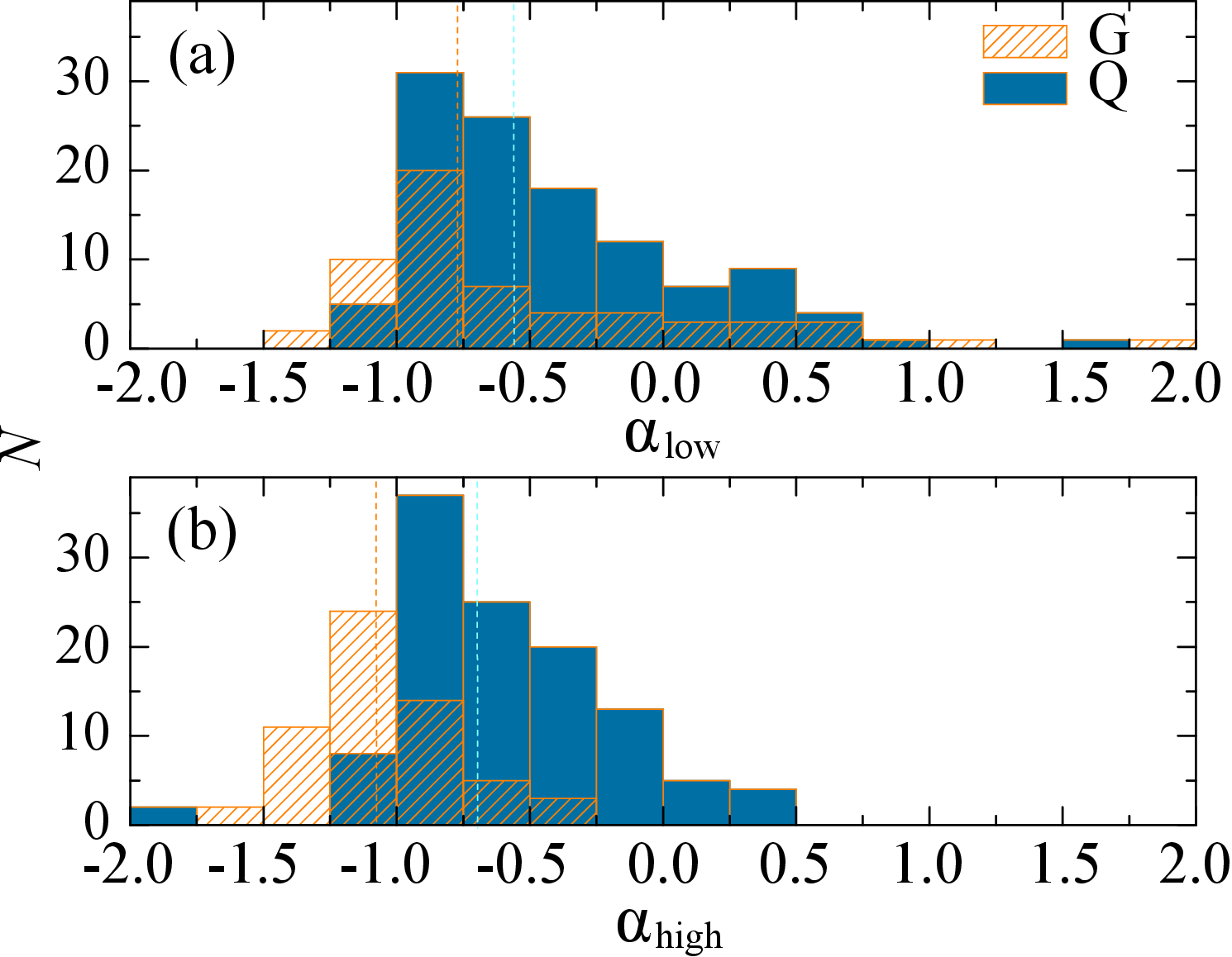}
\caption{\small The low-frequency (panel~a) and high-frequency (panel~b) spectral index distributions for the G and Q source types. The median values are marked by the vertical dashed lines.}
\label{fig:alpha}
\end{figure}

\begin{table} \setlength{\tabcolsep}{2pt}
\caption{\label{tab:type}Spectral types in the sample}
\begin{tabular}{l|c|c|c}
\hline
Type & Criteria  &  $N$ & {\%} \\
\hline
Peaked  & $\alpha_{\rm low}>0$, $\alpha_{\rm high}<0$ & 31 & 17.9  \\ 
Flat    & $-0.5 \leq \alpha \leq 0$  &  28  & 16.2 \\ 
Inverted  & $\alpha>0$ &  3  & 1.7 \\ 
Upturn  & $\alpha_{\rm low}<0$, $\alpha_{\rm high}>0$ &  6  & 3.5 \\ 
Steep  & $-1.1 < \alpha < -0.5$ & 81 & 46.8 \\ 
Ultra-steep  & $\alpha \leq -1.1$ & 23  & 13.3 \\ 
Complex  & Two or more maxima/minima & 1  &  0.6 \\ 
\hline
\end{tabular}
\end{table}

\section{RADIO PROPERTIES}

\subsection{Radio Spectra and Classification}
\label{sec:spectra}

We constructed broadband radio spectra of the galaxies (Fig.~\ref{fig:A1}, Appendix) based on the results of cross identification in the CATS database (within a $30''\times30''$ identification box), NED, and VizieR Information System. To exclude false field radio sources in the specified box, we used a technique of data analysis similar to that described by \cite{2000BSAO...49...53V, 2009AstBu..64...72V}. The essence of the method is to use a joint analysis of the data in the coordinate and spectral spaces to separate out the probable identifications of specific radio sources at various radio frequencies. To this end, we use the \textit{spg} program from the radio continuum data reduction package FADPS implemented at the RATAN-600 radio telescope \citep{1997..conf..1V}.  

We performed a classification of spectral types
based on the common criteria (e.g., \citealt{2006MNRAS.371..898S,2008MNRAS.386.1729T}) listed in Table~\ref{tab:type}. The sample contains sources with a variety of spectral shapes. For the peaked and upturn shapes, the spectral indices $\alpha_{\rm low}$ and $\alpha_{\rm high}$ were calculated for the frequencies lower than or higher than the point where the spectral slope changes its sign from positive to negative or vice versa. For part of the spectra, the low-frequency spectral index matches with 
the high-frequency one.

\begin{figure*}
\includegraphics[width=0.495\textwidth]{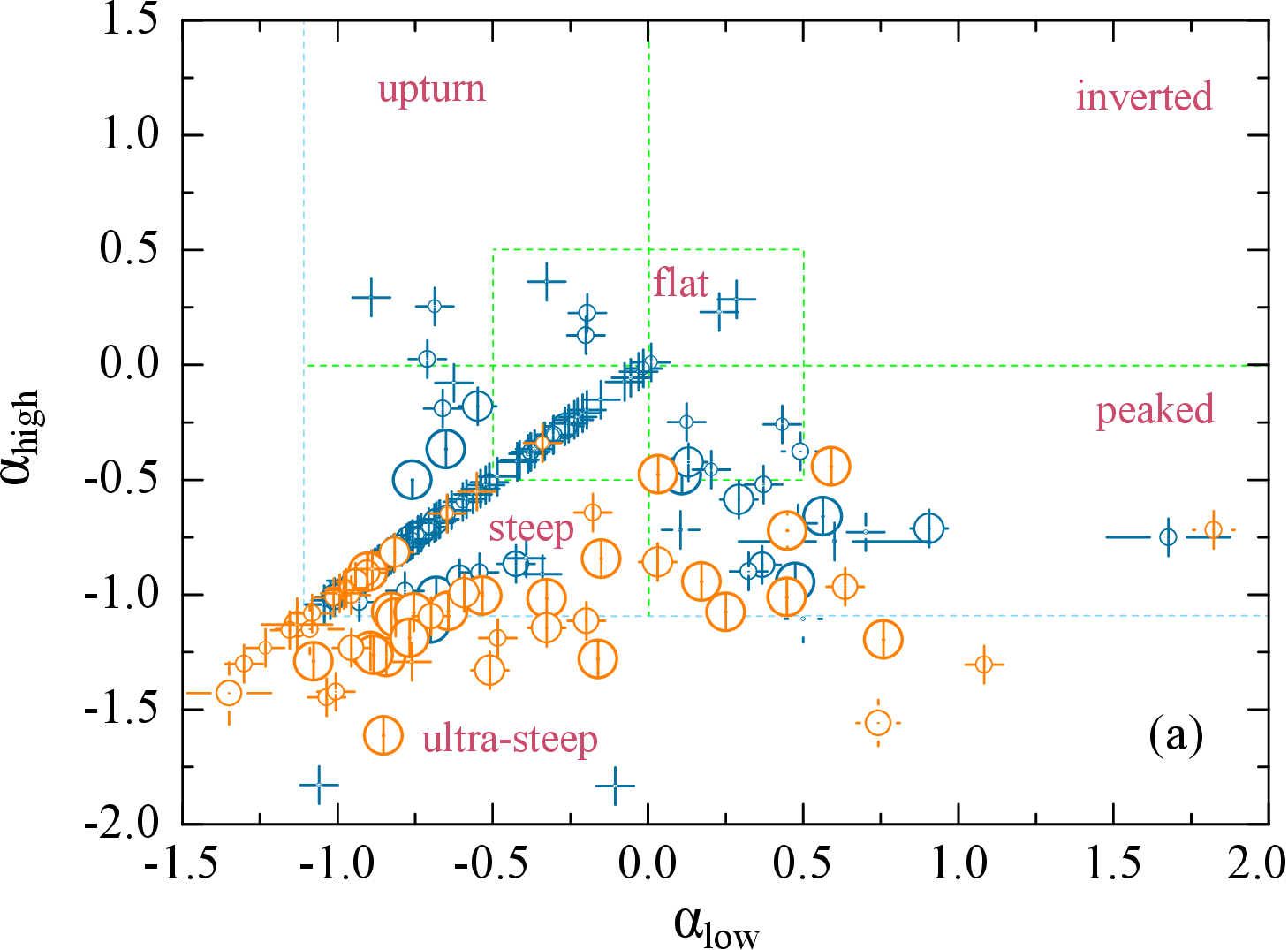}
\includegraphics[width=0.45\textwidth]{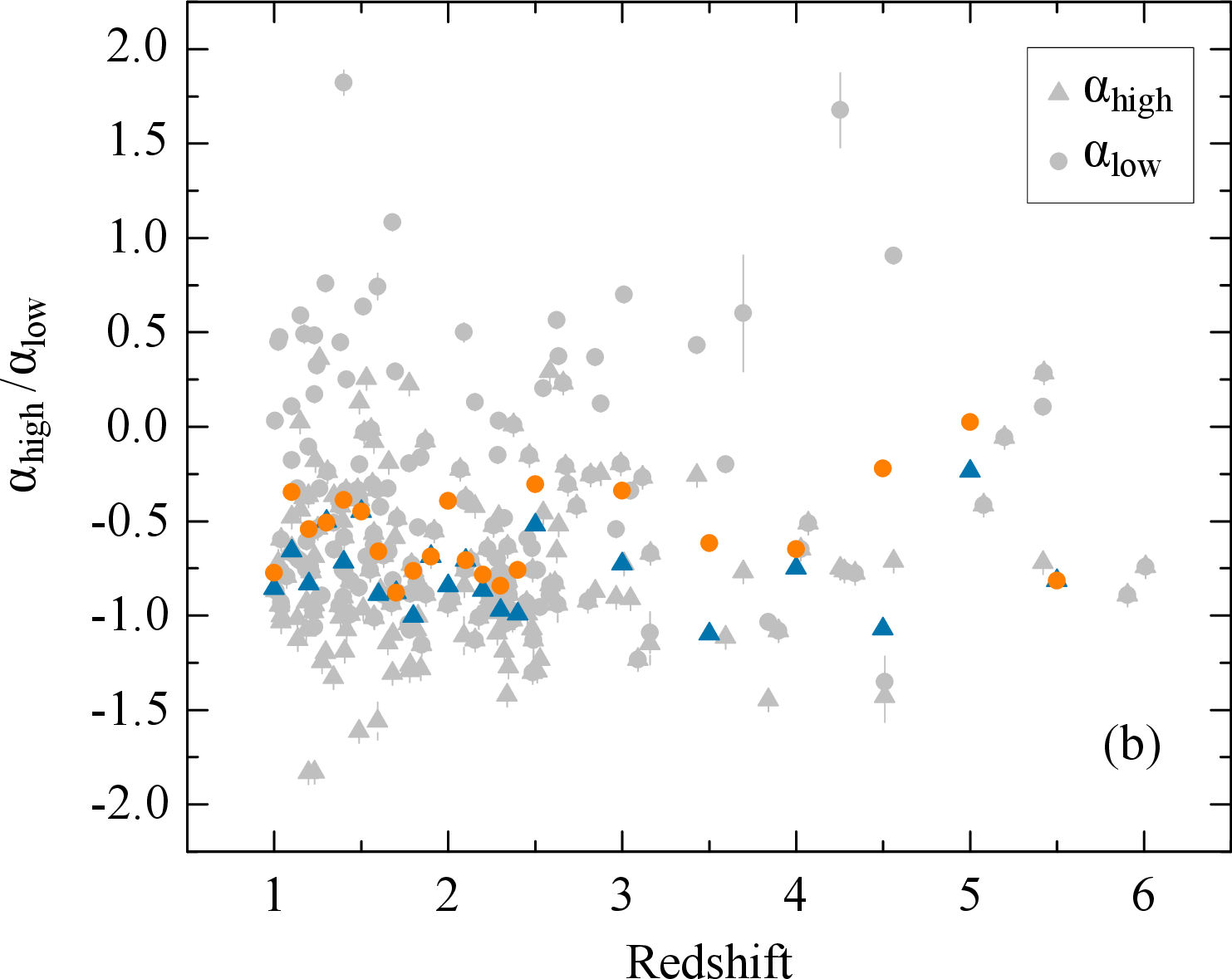}
\caption{\small Panel~(a): the ``colour\,--\,colour'' diagram for the spectral indices in our sample. The square shows the flat spectrum area, the green-coloured line indicates the area of ultra-steep spectrum sources. The point size is proportional to the NVSS flux density at 1.4~GHz. We show galaxies  in orange and quasars in blue.
Panel~(b): the $\alpha$ versus $z$ plot for low-frequency and high-frequency spectral indices. The colored symbols represent the median spectral indices for redshift bins of 0.1 (for $z$ from 1.0 to 2.5) and 0.5 (for $z$ from 2.5 to 5.5).}
\label{alpha_z}
\end{figure*}

Figure~\ref{fig:alpha} shows the distribution of the low-frequency and high-frequency spectral indices for the two types of the sources. There is a clear difference in the distributions of the high-frequency spectral indices,
where quasars demonstrate more flat spectral shapes, confirming their optical classification. The median values are $\alpha_{\rm low}=-0.63$ and $\alpha_{\rm high}=-0.83$ for the whole sample. For the subsample with galaxies (G) the median spectral indices are \mbox{$\alpha_{\rm low}=-0.77$}, $\alpha_{\rm high}=-1.08$, and for the Q subsample \mbox{$\alpha_{\rm low}=-0.56$}, $\alpha_{\rm high}=-0.69$ (Table~\ref{tab:alpha}). The two-colour diagram of the spectral indices in the top panel of Fig.~\ref{alpha_z} presents the distribution of the spectral shapes of the sources. The point sizes correspond to the NVSS flux density level at 1.4~GHz.

\begin{figure}
\includegraphics[width=0.45\textwidth]{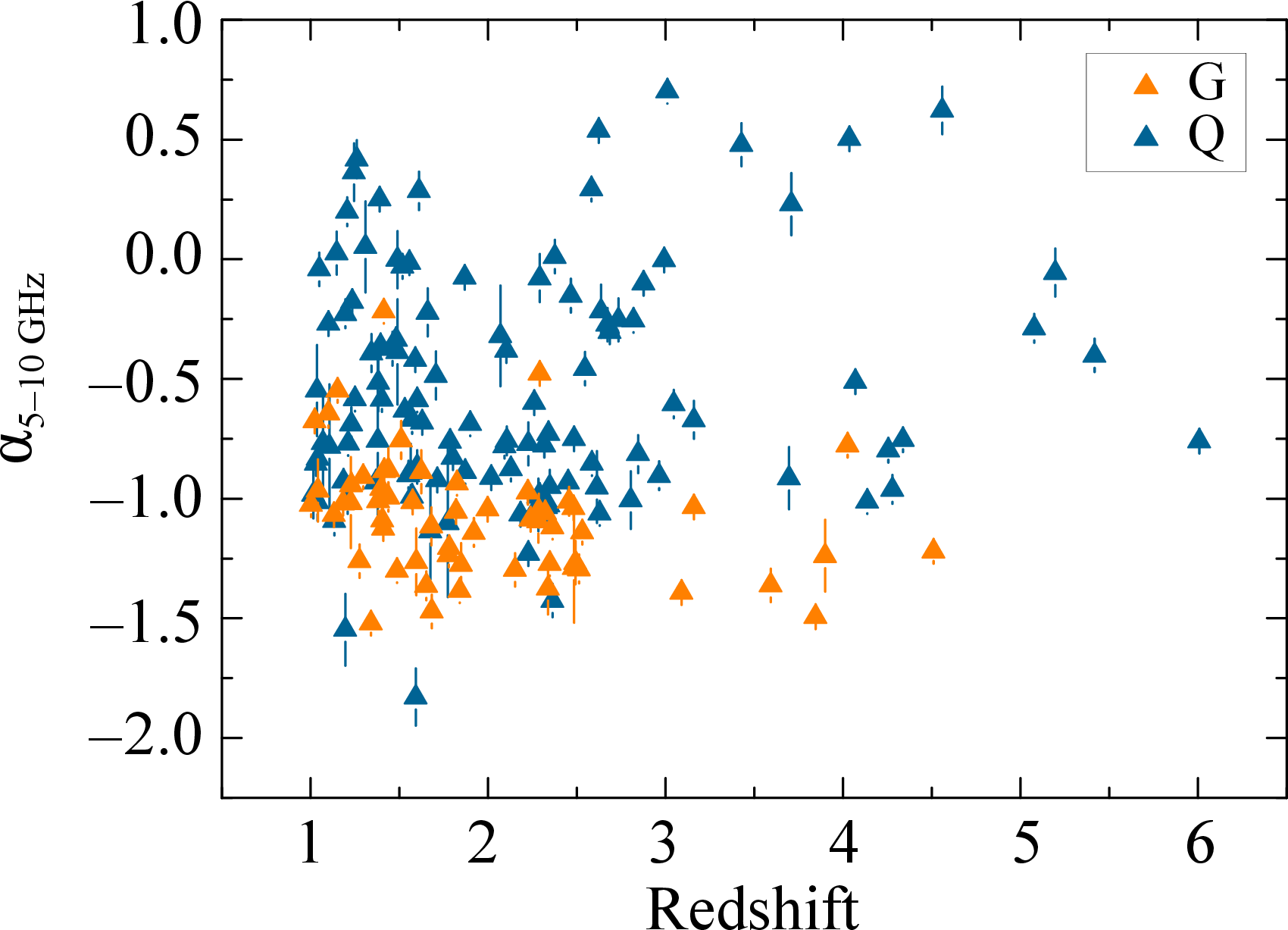}
\caption{\small The $\alpha$ versus $z$ plot for radio spectra in the sources rest frame for G and Q types separately.}
\label{fig:rest}
\end{figure}
\begin{figure*}
\includegraphics[width=.78\textwidth]{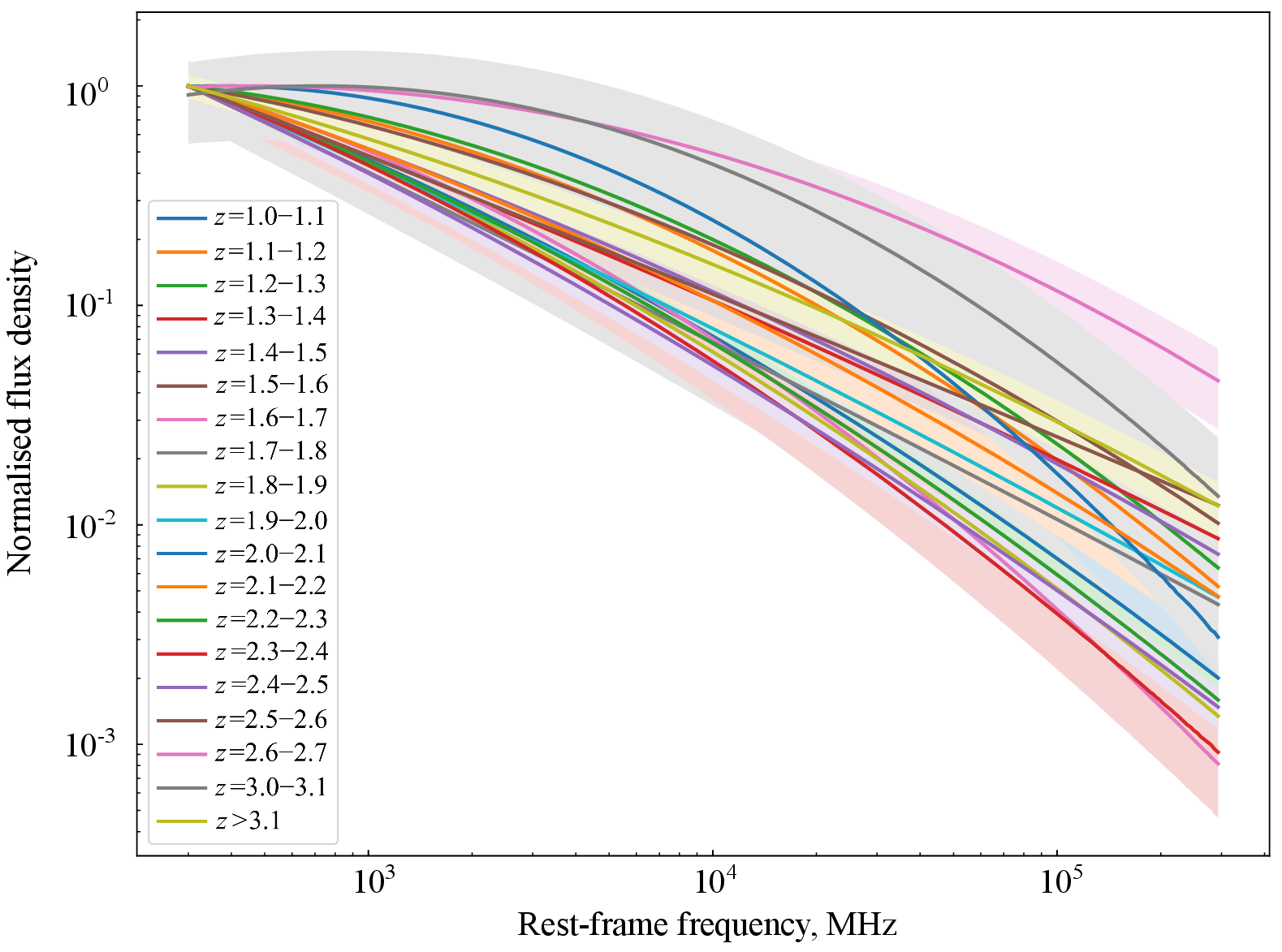}
\caption{\small Average spectra of the sources normalised by the peak flux density. The colored lines relate to the redshift bins \mbox{$z = 1.0$\,--\,$3.1$}, and 
the
corresponding strips represent 
the
$3\,\sigma$ uncertainties.}
\label{fig:aver_spectra}
\end{figure*}

\begin{table}
\caption{\label{tab:alpha} The median and mean 
values of the spectral indices $\alpha_{\rm low}$ and $\alpha_{\rm high}$ for 
the G and Q source types. The standard deviations from the mean are given in parentheses.  The last column shows the Pearson correlation coefficients for the
``$\alpha$\,--\,$z$'' relation}
\begin{tabular}{l|c|c|c|c}
\hline
Type   &$N$ & median & mean & $r$ ($p$-$value$) \\
\hline
\multicolumn{4}{c}{$\alpha_{\rm low}$} & $\alpha_{\rm low}-z$ \\
\hline
G      &  59 & $-$0.77 & $-$0.52 (0.66) & $-$0.41 (0.001) \\
Q  & 114 & $-$0.56 & $-$0.41 (0.50) &~ 0.22~ (0.02) \\
\hline
\multicolumn{4}{c}{$\alpha_{\rm high}$} & $\alpha_{\rm high}-z$ \\
\hline
G      & 59  & $-$1.08 & $-$1.05 (0.26) &  $-$0.25 (0.06) \\
Q  & 114 & $-$0.69 & $-$0.59 (0.39) & ~~0.02 (0.58) \\
\hline
\end{tabular}
\end{table}

The ``$\alpha$\,--\,$z$'' diagram (Fig.~\ref{alpha_z}b) shows the low-frequency (circles) and high-frequency (triangles) spectral indices relative to the redshifts in a wide range from 1 to 6. The blue and orange symbols represent the median spectral indices for redshift bins of 0.1 (for $z$ from 1.0 to 2.5) and for bins of 0.5 (for $z$ from 2.5 to 6). The distributions of $\alpha_{\rm high}$ at redshifts
from 1 to 3 and from 3 to 6 are significantly different (at the 0.05 level according to the Kolmogorov\,--\,Smirnov test), which is due to the deficiency of galaxies at $z>3$. We did not find correlation between the redshift and the low-frequency and high-frequency indices; however, $\alpha_{\rm low}$ and $z$ does correlate when we examine 
the subsample of galaxies: $r=-0.41$ ($p$-$value=0.001$). Other values for the ``$\alpha$\,--\,$z$'' correlations are shown in Table~\ref{tab:alpha}.

We checked the ``$\alpha_{5-10}$\,--\,$z$'' relation for the radio spectra recalculated into the rest frame using the ($1 + z$) correction (Fig.~\ref{fig:rest}). The spectral index was estimated between frequencies \mbox{5\,--\,10}~GHz. We do not find a correlation for the whole sample, as well as for 
the subsample of Q sources. But for the G-type sources, a significant anticorrelation was found (Pearson $r=-0.32$, $p$-$value<0.01$).

\subsection{The Average Spectrum}

To study the population of distant galaxies,
we used the continuum radio spectra averaged over the objects located within tight redshift intervals $\Delta z = 0.1$. The spectrum of each radio source was preliminarily recalculated into the rest frame using the $(1+z)$ correction. The detailed description of the average spectrum method is given in \citealt{2018AstBu..73..393V, 2021MNRAS.508.2798S}. As a result, we have calculated eighteen average spectra in the range \mbox{$z = 1.0$\,--\,$3.1$} and the average radio spectrum for 21 galaxies at $z>3.1$ (Fig.~\ref{fig:aver_spectra}). The number of galaxies in each bin varies from 3 to 12, thus we can not conclude about cosmological evolution of the averaged spectra. We only note some of their characteristic features. Firstly, all 
the averaged spectra have a convex shape 
with flux density decrease with increasing frequency. The absolute value of the spectral index increases with increasing frequency
(see Table~\ref{tab:aver_ind})\footnote{The average spectra were approximated by parabolas, and then we calculated the spectral index as 
a derivative at a given frequency.}.
However, the spectral curvature is small, and only 4 spectra have peaks at frequencies above 100~MHz. Secondly, the spectral indices at 5~GHz and 11 GHz vary approximately from $-0.6$ to $-1.0$, which means the optically thin emission. Thus, we do not detect prominent features of the spectral component related to a bright compact radio core.

\begin{table}
\caption{\label{tab:aver_ind} \small The peak frequency
and the
spectral indices at 5~GHz and 11~GHz 
for the average spectra of the galaxies
from this work (top part) and 
for the average spectra of quasars 
from \protect\cite{2021MNRAS.508.2798S} (bottom part)}
\centering
\begin{tabular}{c|c|c|c}
\hline
 $z$ & $\nu_{peak}$, GHz  &  $\alpha_{5}$ & $\alpha_{11}$ \\
\hline
1.0\,--\,1.1  & -- & $-0.84$ & $-0.91$  \\ 
1.1\,--\,1.2  & -- & $-0.66$ & $-0.79$  \\
1.2\,--\,1.3  & 0.10 & $-0.63$ & $-0.76$  \\
1.3\,--\,1.4  & -- & $-0.67$ & $-0.69$  \\
1.4\,--\,1.5  & -- & $-0.68$ & $-0.73$  \\ 
1.5\,--\,1.6  & -- & $-0.60$ & $-0.68$  \\
1.6\,--\,1.7  & -- & $-0.93$ & $-1.06$  \\
1.7\,--\,1.8  & -- & $-0.78$ & $-0.79$  \\
1.8\,--\,1.9  & -- & $-0.90$ & $-0.98$  \\ 
1.9\,--\,2.0  & -- & $-0.76$ & $-0.78$  \\
2.0\,--\,2.1  & 0.37 & $-0.68$ & $-0.88$  \\
2.1\,--\,2.2  & -- & $-0.72$ & $-0.79$  \\
2.2\,--\,2.3  & -- & $-0.88$ & $-0.95$  \\ 
2.3\,--\,2.4  & -- & $-0.95$ & $-1.03$  \\
2.4\,--\,2.5  & -- & $-0.91$ & $-0.96$  \\
2.5\,--\,2.6  & -- & $-0.63$ & $-0.64$  \\
2.6\,--\,2.7  & 0.44 & $-0.36$ & $-0.47$  \\
3.0\,--\,3.1  & 0.73 & $-0.46$ & $-0.65$  \\
$\geq3.1$  & -- & $-0.60$ & $-0.65$  \\
\hline
3.0\,--\,3.1  & 0.52 & $-0.21$ & $-0.28$  \\ 
3.1\,--\,3.2  & 1.50 & $-0.19$ & $-0.32$  \\
3.2\,--\,3.3  & 0.65 & $-0.26$ & $-0.35$  \\
3.3\,--\,3.4  & -- & $-0.36$ & $-0.37$  \\
3.4\,--\,3.5  & 1.61 & $-0.23$ & $-0.39$  \\ 
3.5\,--\,3.6  & 5.35 & $+0.03$ & $-0.30$  \\
3.6\,--\,3.7  & 2.43 & $-0.12$ & $-0.26$  \\
3.7\,--\,3.8  & 4.78 & $-0.01$ & $-0.14$  \\
3.0\,--\,3.8  & 1.60 & $-0.18$ & $-0.30$  \\
\hline
\end{tabular}
\end{table}

\subsection{Radio Luminosity}

We calculated the radio luminosity at 5 GHz according to the common formula:
 \begin{equation}
L_{\nu} = 4 \pi D_{L}^2 \nu S_{\nu} (1+z)^{-\alpha -1}, 
\end{equation}
where $\nu$ is the frequency, $S_{\nu}$ is the measured flux density, 
$z$ is the redshift, $\alpha$ is the spectral index, and $D_{L}$ is the luminosity distance. We used the $\Lambda$CDM cosmology with \mbox{$H_0 = 67.74$~km\,s$^{-1}$\,Mpc$^{-1}$}, $\Omega_m=0.3089$, and \mbox{$\Omega_\Lambda=0.6911$} \citep{2016A&A...594A..13P} 
to estimate
the luminosity distance.

\begin{figure}
\includegraphics[width=0.49\textwidth]{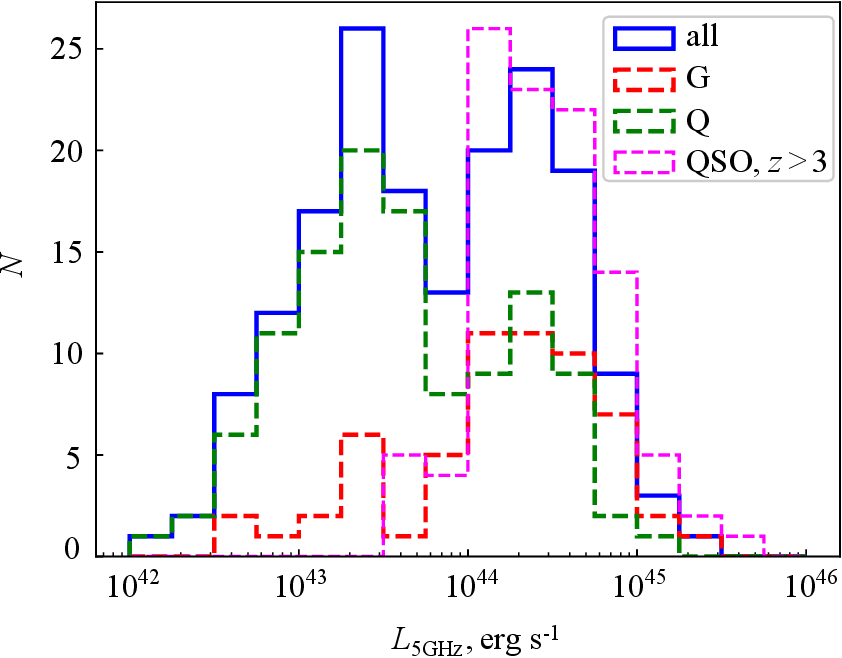}
\caption{\small Distributions of the radio luminosity at 5 GHz. The radio luminosity distribution of distant quasars at $z>3$ is marked in magenta \citep{2021MNRAS.508.2798S}.}
\label{fig:Lum_gist}
\end{figure}

Figure~\ref{fig:Lum_gist} shows the luminosity distribution 
with a
median value of about
$ 6.63\times 10^{43}$~erg~s$^{-1}$ and 
a mean of about 
\mbox{$ 1.84 \times 10^{44}$~erg\,s$^{-1}$}. The source NVSS\,J031147+050802 at $z=4.51$ has the highest luminosity \mbox{$L_{5\text{GHz}} \approx 2.17 \times 10^{45}$~erg\,s$^{-1}$}. We note 
the
bimodal character of the luminosity distribution: it is a superposition of two Gaussian distributions. This fact can be related 
to
the sample heterogeneity. We constructed the radio luminosity distributions for 
the
two subsamples: G and Q which have different distributions (Fig.~\ref{fig:Lum_gist}).

The resulting bimodality of the total distribution is due to the bimodality of the Q group and approximately equal contribution of the G group. This group (G) has a peak in the distribution at the same luminosities as the second peak in the Q group
(see also Table~\ref{lum_loud}). 
The bimodal luminosity distribution of the Q group can mean that some objects from the Q group do not have prominent quasar properties and are misclassified. 

\begin{table*}
\setlength{\tabcolsep}{3.5pt}
\caption{\label{lum_loud} \small Descriptive statistics 
of
the radio luminosity and 
radio loudness 
in
different subsamples}
\begin{tabular}{c|c|c|c|c|c|c|c|c|c}
\hline
   \multirow{2}{*}{Subsample}  & \multicolumn{3}{c|}{$\log L_{5}$,  erg s$^{-1}$} & \multicolumn{3}{c|}{$\log R_{1}$} & \multicolumn{3}{c}{$\log R_{2}$} \\ \cline{2-10}
     & mean & sd & median & mean & sd & median & mean & sd & median \\     
\hline
all (173) & 43.81 & 0.69 & 43.82 & 4.06 & 0.92 & 3.92 & 3.48 & 0.81 & 3.41 \\
G (59) & 44.19 & 0.62 & 44.26 & 4.78 & 0.91 & 5.00 & 4.09 & 0.70 & 4.23 \\
Q (114) & 43.62 & 0.63 & 43.51 & 3.70 & 0.69 & 3.63 & 3.19 & 0.70 & 3.18 \\
QSO, $z>3$ (102) & 44.45 & 0.38 & 44.37 & 3.56 & 0.69 & 3.49 & -- & -- & -- \\
\hline
\end{tabular}
\end{table*}

\begin{figure*}
\includegraphics[width=0.49\textwidth]{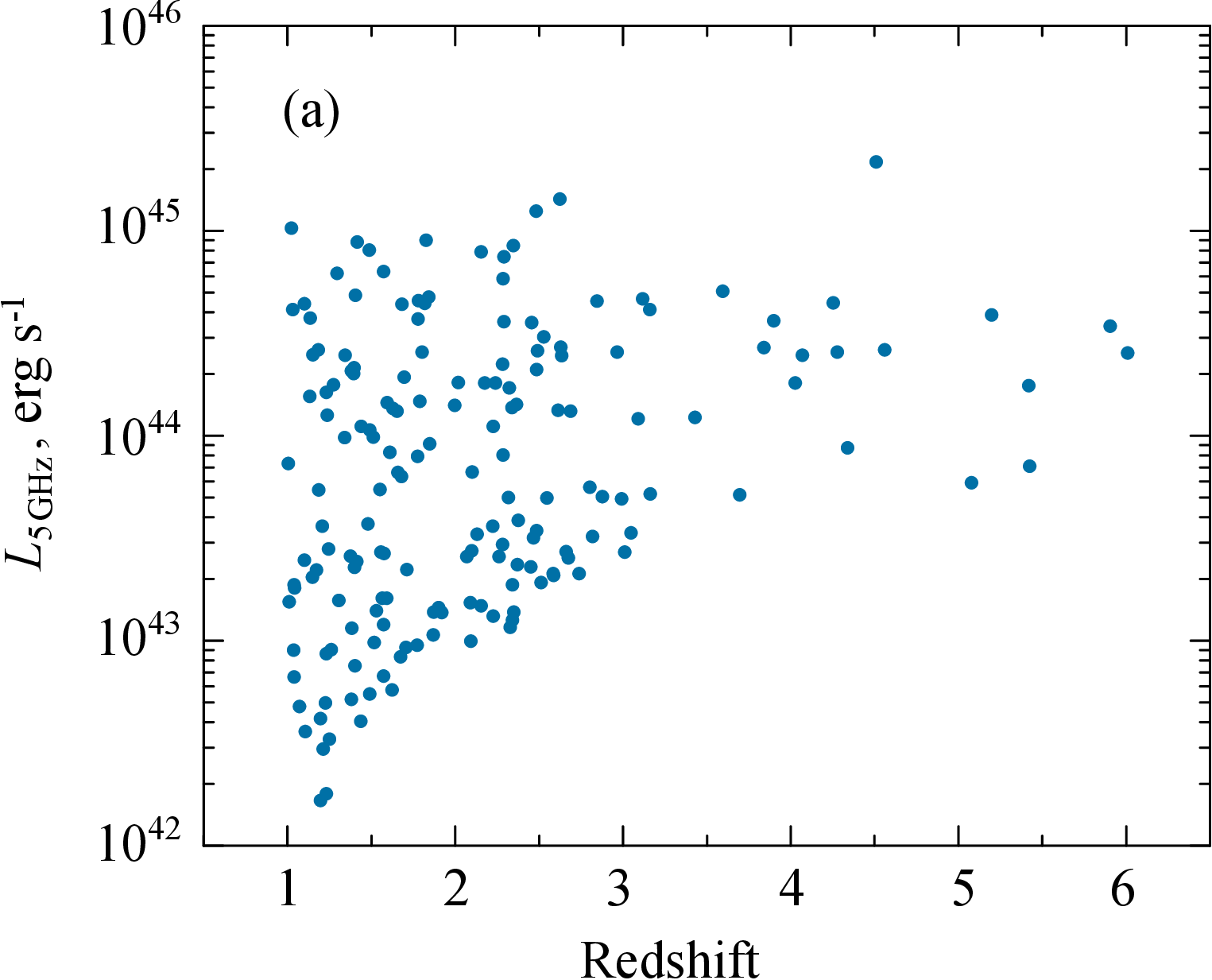}
\hspace{2mm}\includegraphics[width=0.47\textwidth]{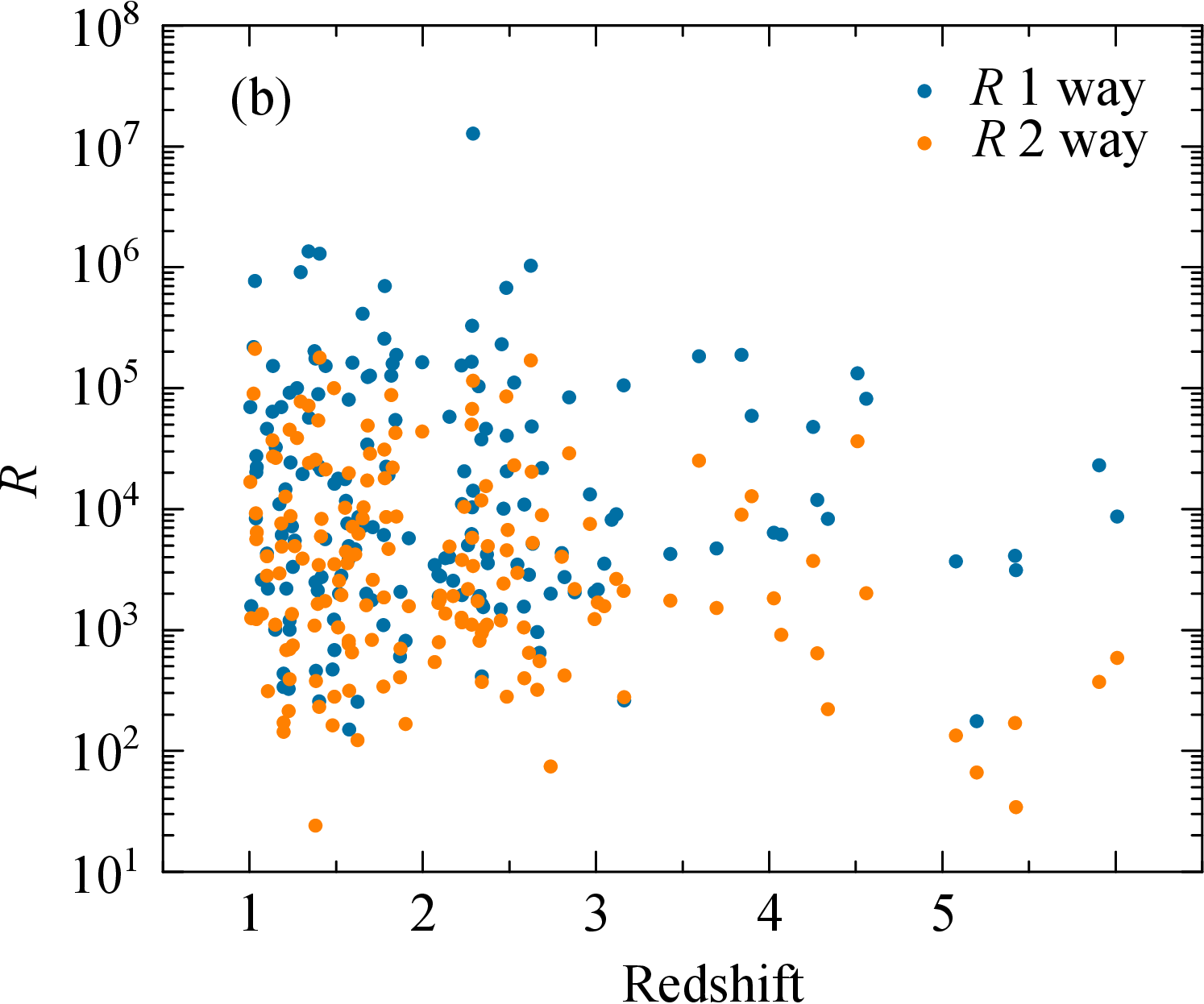}
\caption{\small Luminosity at 5~GHz versus redshift (panel~a) and radio loudness versus redshift (panel~b). 
The
blue and orange circles correspond to the radio loudness values calculated
in two different ways.}
\label{figLum_R_vs_z}
\end{figure*}

Figure~\ref{figLum_R_vs_z}a
presents a diagram of
the luminosity at 5~GHz 
versus redshift. The empty bottom region is explained by the selection effect (the minimum flux density level at 1.4~GHz is limited to 20 mJy).

  \begin{figure*}
\includegraphics[width=0.46\textwidth]{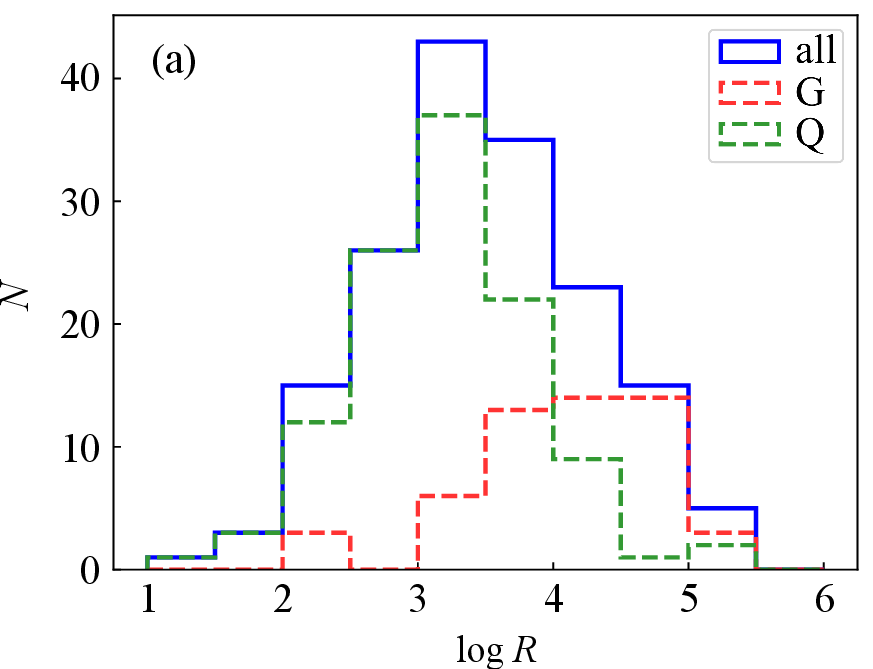}
\includegraphics[width=0.453\textwidth]{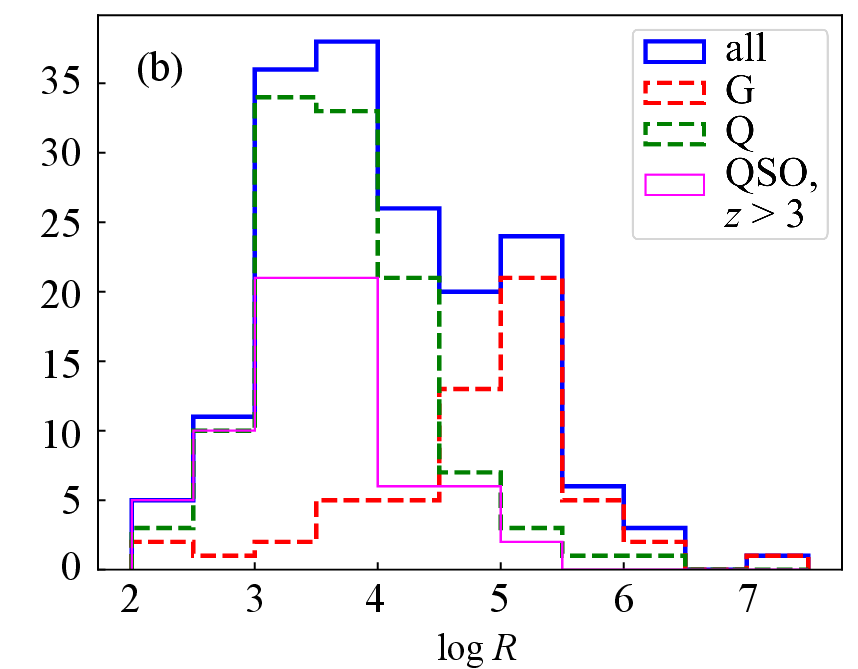}
\caption{\small Distributions of the radio loudness.  Panel~a shows results obtained by second approach, panel~b -- by the first
one. The radio loudness distribution of distant quasars at $z>3$ is marked in magenta \citep{2021MNRAS.508.2798S}.}
\label{fig:R_gist}
\end{figure*}

\subsection{Radio Loudness}

The more common
radio loudness is defined as \citep{1989AJ.....98.1195K}:
\begin{equation}
R = \frac{S_{5\text{GHz}}}{S_{4400\text{\AA}}}
\end{equation}
where $S_{5\text{GHz}}$ is the rest-frame radio flux density at 5 GHz and $S_{4400\text{\AA}}$ is the rest-frame optical flux density at $4400~\text{\AA}$ corresponding to 
the $B$ or $g$ filter.

Many authors 
suppose
the same spectral indices for objects of the whole sample 
in estimating the rest-frame flux densities 
based on 
the
observed 
flux densities
in both the radio and optical 
bands.
This assumption is often caused by a lack of knowledge about the spectral energy distribution. For example, \citet{2021MNRAS.508.2798S} estimated the radio loudness for 102 high-redshift quasars based on broadband radio measurements; however, 
the
authors took the 
same
optical spectral index for all 
the
sources. Here we calculated the radio loudness 
in
two ways: 
\begin{list}{}{
\setlength\leftmargin{5mm} \setlength\topsep{0mm}
\setlength\parsep{0mm} \setlength\itemsep{1mm} }
    \item [1)] similarly 
to
\citet{2021MNRAS.508.2798S}---in this way
we can compare the radio loudness for our sample and 
for the
102 high-redshift quasars;  \item [2)] the optical spectral index was calculated for each source based on 
the
photometric data at 9134~$\text{\AA}$ 
(the SDSS $z$ filter)
and 3.4~{$\mu$}m ($W1$ band of WISE)---as the sources
have redshifts 
greater
than 1, 
the light emitted at a wavelength of 4400~$\text{\AA}$
is received in the region between $z$ and $W1$, 
therefore
this 
approach
is more accurate.  \end{list}

The calculated values of radio loudness $R_1$ and $R_2$
versus redshift are shown in Figure~\ref{figLum_R_vs_z}b.
Figure~\ref{fig:R_gist}b presents the radio loudness distribution which was estimated 
using the second approach, $R_2$. 
The $\log~R$ value spans from 1.38 to 5.33, the median is 3.41. 
The
two subsamples (G and Q) have different radio loudness distributions with 
a
more radio loud G group (see also Table~\ref{lum_loud}).

\section{RADIO VARIABILITY}
\label{sec:var}

Flux density variability over a time period of up to 30 years was estimated at 5 and 11~GHz using the variability ($V_{S}$) and modulation ($M$) indices  \citep{1992ApJ...399...16A,2003A&A...401..161K}. The literature data were adopted within 10\% ranges of these two frequencies. For example, at $\nu = 5$~GHz
we used all data within $4.5$\,--\,$5.5$~GHz. The median values of 
the number of measurements 
at 5 and 11~GHz over the sample
are very small, $N_{\rm obs}=5$. 
The variability and modulation indices at 5~GHz and the number of observing epochs
($N_{\rm obs}$) are presented in Table~\ref{table2_appendix} in columns (8)--(10) (A short fragment is presented. The full version can be found in VizieR\protect\footnote{\url{https://cdsarc.cds.unistra.fr/viz-bin/cat/J/other/AstBu/78.443}}).

\begin{table*}\setlength{\tabcolsep}{2pt}
\centering
\caption{\label{table2_appendix} The sample parameters: source name at 2000.0 epoch, radio luminosity at 5~GHz $L_{5}$, logarithm of radio loudness $R$, calculated in two ways, spectral indices $\alpha_{\rm low}$ and $\alpha_{\rm high}$, radio spectrum type, modulation $M$ and variability $V$ indices at 5~GHz, number of observations $N_{\rm obs}$, and the observed peak frequency in the spectrum $\nu_{\rm peak,obs}$. A short fragment is presented.}
\begin{tabular}{c|c|c|c|c|c|c|c|c|c|c}
\hline
Name, &  $\log L_{5}$,  &  $\log R$, & $\log R$, & \multirow{2}{*}{$\alpha_{\rm low}$} & \multirow{2}{*}{$\alpha_{\rm high}$} & \multirow{2}{*}{Sp. type} & \multirow{2}{*}{$M_{5}$} & \multirow{2}{*}{$V_{5}$} & \multirow{2}{*}{$N_{\rm obs}$} & {$\nu_{\rm peak,obs}$,} \\
2000.0           &                    erg s$^{-1}$& 1 way  & 2 way &  & & &    & & &  GHz   \\ \hline
 1 & 2 & 3 & 4 & 5 & 6 & 7 & 8 & 9 & 10 & 11\\
\hline

J0030$+$2957  &  $44.59_{\rm +0.01}^{-0.01}$  &  $2.25_{\rm +0.01}^{-0.01}$  &  $1.82_{\rm +0.37}^{~~-}$  &  $-0.06\pm0.10$  &  $-0.06\pm0.10$  &  flat  &  --  &  --  & -- & -- \\ 
J0032$-$0414  &  $43.72_{\rm +0.05}^{-0.06}$  &  $2.42_{\rm +0.05}^{-0.06}$  &  $2.44_{\rm +0.17}^{-0.28}$  &  $-0.67\pm0.08$  &  $-0.67\pm0.08$  &  steep & --   &  --  & -- & -- \\ 
J0038$+$1227  &  $44.3_{\rm +0.04}^{-0.05}$  &  $3.33_{\rm +0.04}^{-0.05}$  &  $3.22_{\rm +0.08}^{-0.11}$  &  $-0.76\pm0.08$  &  $-0.50\pm0.01$  &  steep  & 0.09  & 0.05 & 9 &  -- \\ 
J0101$-$2831  &  $44.29_{\rm +0.05}^{-0.05}$  &  $5.1_{\rm +0.1}^{-0.12}$  &  $4.46_{\rm +0.09}^{-0.12}$  &  $+0.29\pm0.02$  &  $-0.59\pm0.05$  & peaked &  0.08  & 0.11 & 6 & 0.37 \\ 
J0113$+$1335  &  $43.43_{\rm +0.04}^{-0.05}$  &  $2.98_{\rm +0.06}^{-0.05}$  &  $2.51_{\rm +0.1}^{-0.13}$  &  $+0.23\pm0.13$  &  $+0.23\pm0.13$  & inverted &  0.15   & 0.07 & 3 &  -- \\ 
J0116$-$2052  &  $44.94_{\rm +0.02}^{-0.02}$  &  $3.44_{\rm +0.07}^{-0.09}$  &  $3.92_{\rm +0.04}^{-0.05}$  &  $+0.25\pm0.30$  &  $-1.08\pm0.10$  & peaked & 0.09 & 0.18 & 14 & 0.13 \\ 
J0117$+$0114  &  $43.71_{\rm +0.19}^{-0.35}$  &  $3.67_{\rm +0.21}^{-0.42}$  &  $3.18_{\rm +0.31}^{~~-}$  &  $+0.60\pm3.08$  &  $-0.77\pm0.04$  &  peaked  &    --   & -- & -- & 0.11 \\ 
J0122$+$1923  &  $44.16_{\rm +0.08}^{-0.1}$  &  $5.21_{\rm +0.2}^{-0.38}$  &  $3.85_{\rm +0.34}^{~~-}$  &  $+0.74\pm0.70$  &  $-1.56\pm1.00$  &  peaked  &  0.17 &  0.24 & 6 & 0.17 \\ 
J0125$+$0054  &  $43.35_{\rm +0.04}^{-0.04}$  &  $3.85_{\rm +0.06}^{-0.07}$  &  $3.41_{\rm +0.16}^{-0.24}$  &  $-0.88\pm0.02$  &  $-0.88\pm0.02$  &  steep  &  -- &  -- & -- & -- \\ 
J0130$-$2610  &  $44.93_{\rm +0.03}^{-0.04}$  &  --  &  --  &  $-0.84\pm0.02$  &  $-1.27\pm0.05$  &  ultra-steep  &    0.08   &  0.06  & 3 &  -- \\ 
\hline
\end{tabular}
\end{table*}

\begin{figure}
\includegraphics[width=0.47\textwidth]{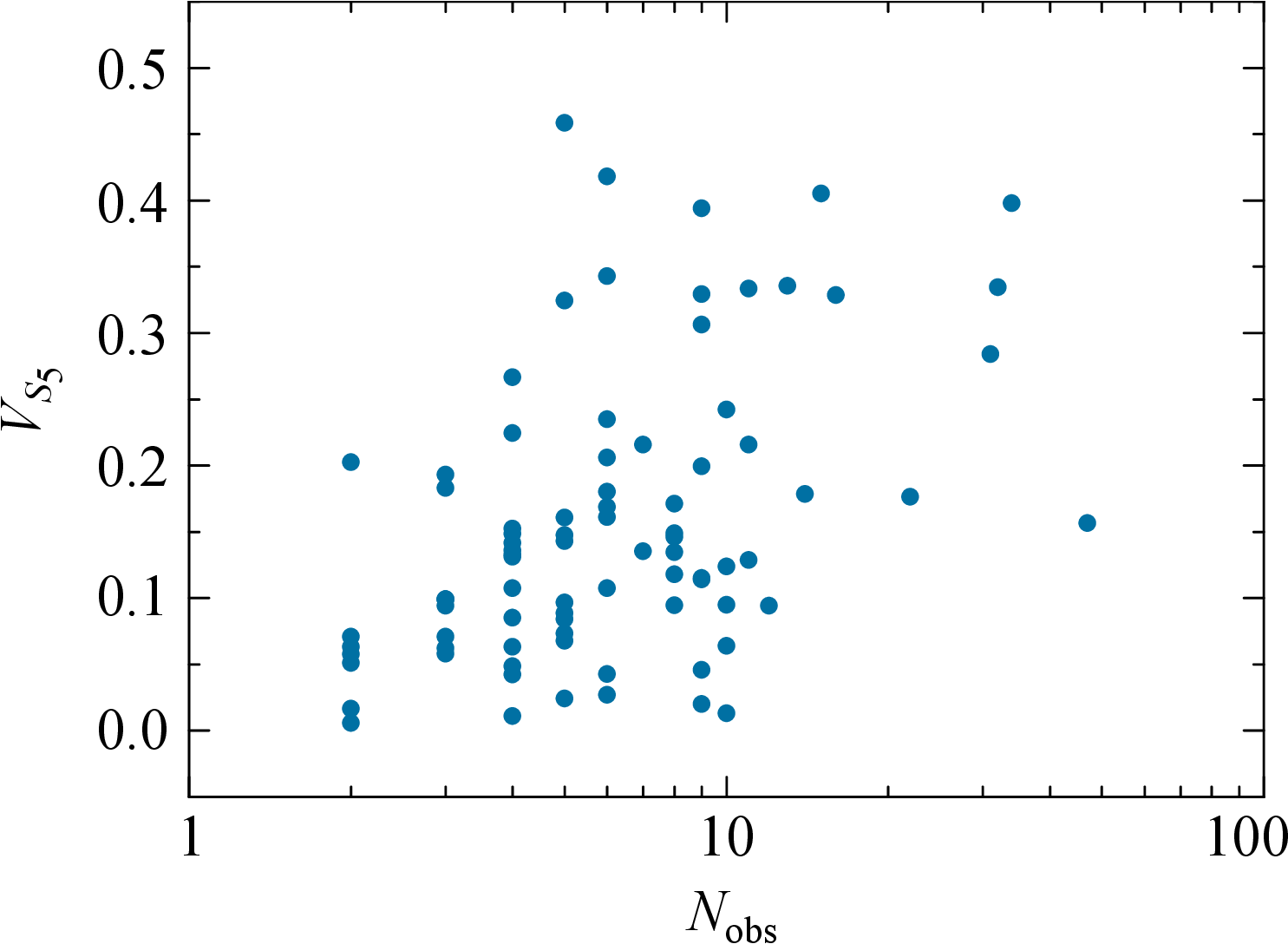}
\caption{The variability index $V_{S_{5}}$ versus the number of observations $N_{\rm obs}$.}
\label{fig:nobs}
\end{figure}

The number of observing epochs $N_{\rm obs}$ is an important parameter, which impacts the search for AGN variability. It is known that variability increases with the number of observations, which can be explained by the fact that peaks of variability are easily missed when sampling is very sparse \citep{2000AJ....120.2278T}. We found a significant correlation (Fig.~\ref{fig:nobs}) between $N_{\rm obs}$ and the variability index at 5~GHz (Spearman $\rho=0.43$, $p$-$value<0.005$). The observations are not homogeneous and differ also in timescales, affecting the revealed variability level.

\begin{figure}
\includegraphics[width=0.48\textwidth]{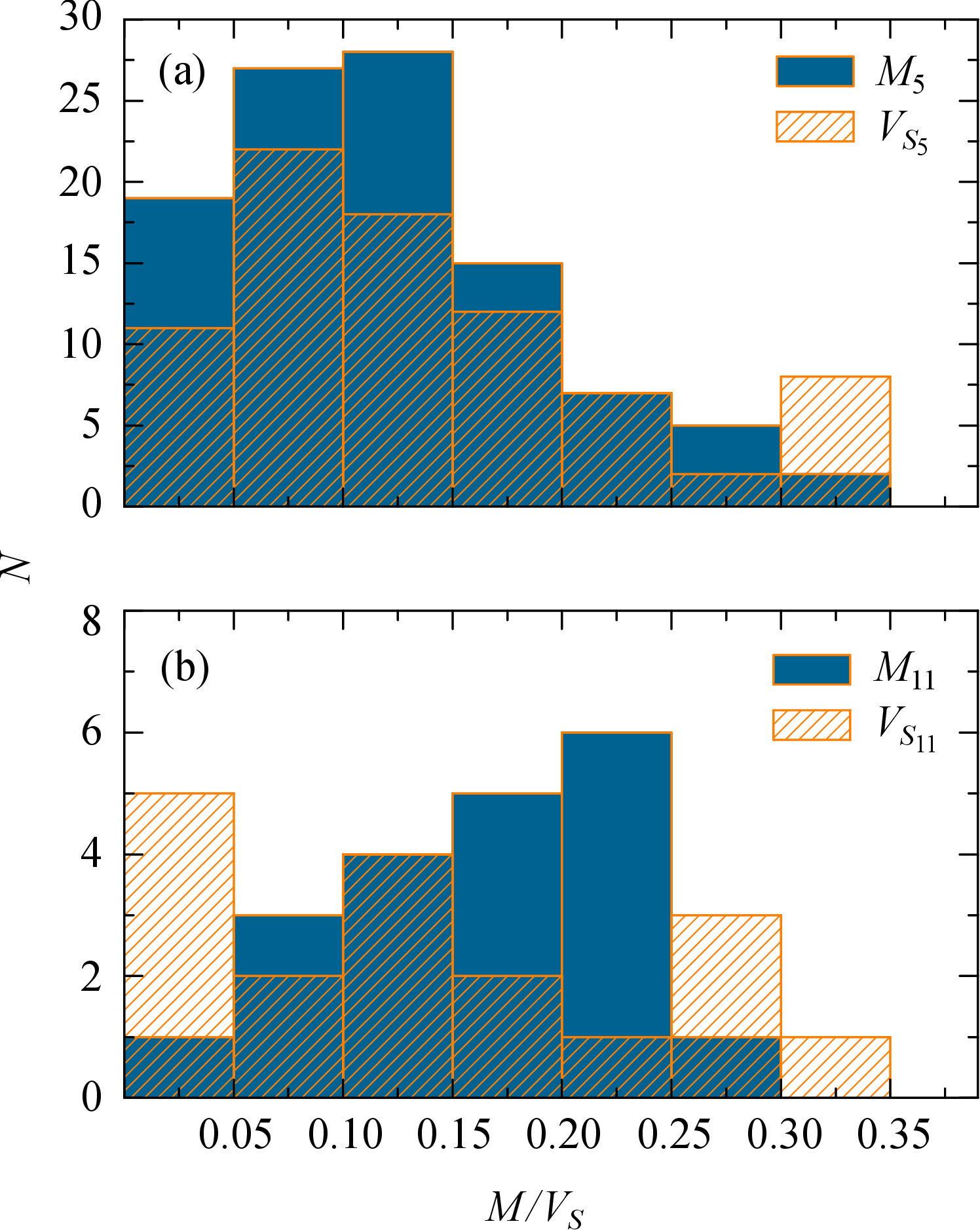}
\caption{\small The distributions of the modulation $M$ and variability $V_{S}$ indices at 5~GHz~(a) and 11~GHz~(b).}
\label{fig:MV}
\end{figure}

\begin{figure}
\includegraphics[width=0.47\textwidth]{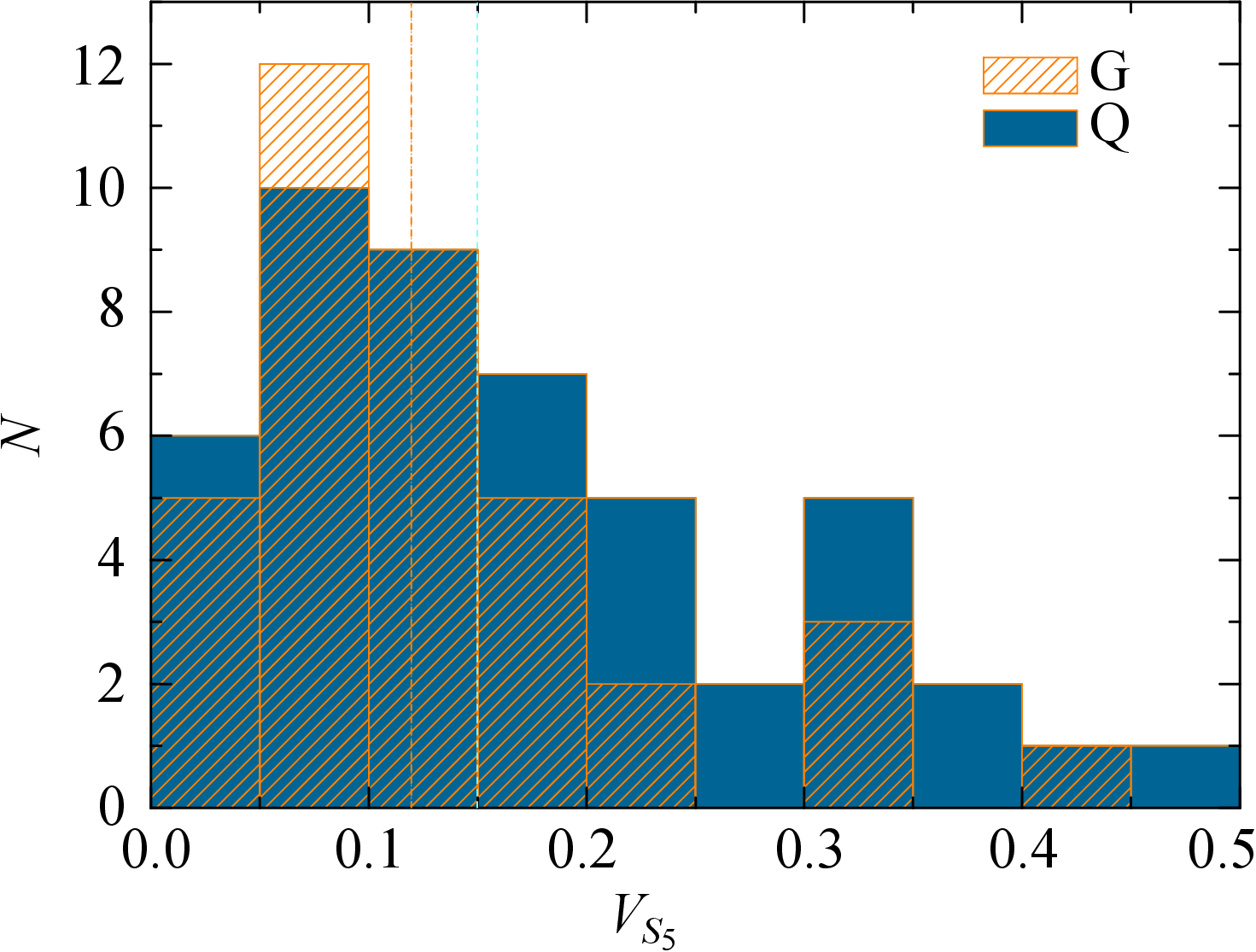}
\caption{\small 
The distributions of the variability index $V_{S_{5}}$ for galaxies and quasars; the median values are equal to 0.12 and 0.15, respectively (dotted orange and blue lines).}
\label{fig:V-g-qso}
\end{figure}

Table~\ref{var} lists the statistics of $V_{S}$ and $M$. The negative variability 
indices were excluded from the calculations. We obtained the average variability and modulation indices for the sample to be 0.11--0.16 at 5~GHz
and 0.14--0.18 at 11~GHz (Fig.~\ref{fig:MV}, Table~\ref{var}). The median $V_{S_{5}}$ values for galaxies and quasars are 0.12 and 0.15, respectively (Fig.~\ref{fig:V-g-qso}), but the distributions of both the variability and modulation indices for them are not significantly different (at the 0.05 level according to the Kolmogorov--Smirnov test). The most variable flux density is obtained for the galaxies with upturn, complex, inverted, and peaked spectra (Table~\ref{spec}).

\begin{figure}
\includegraphics[width=0.48\textwidth]{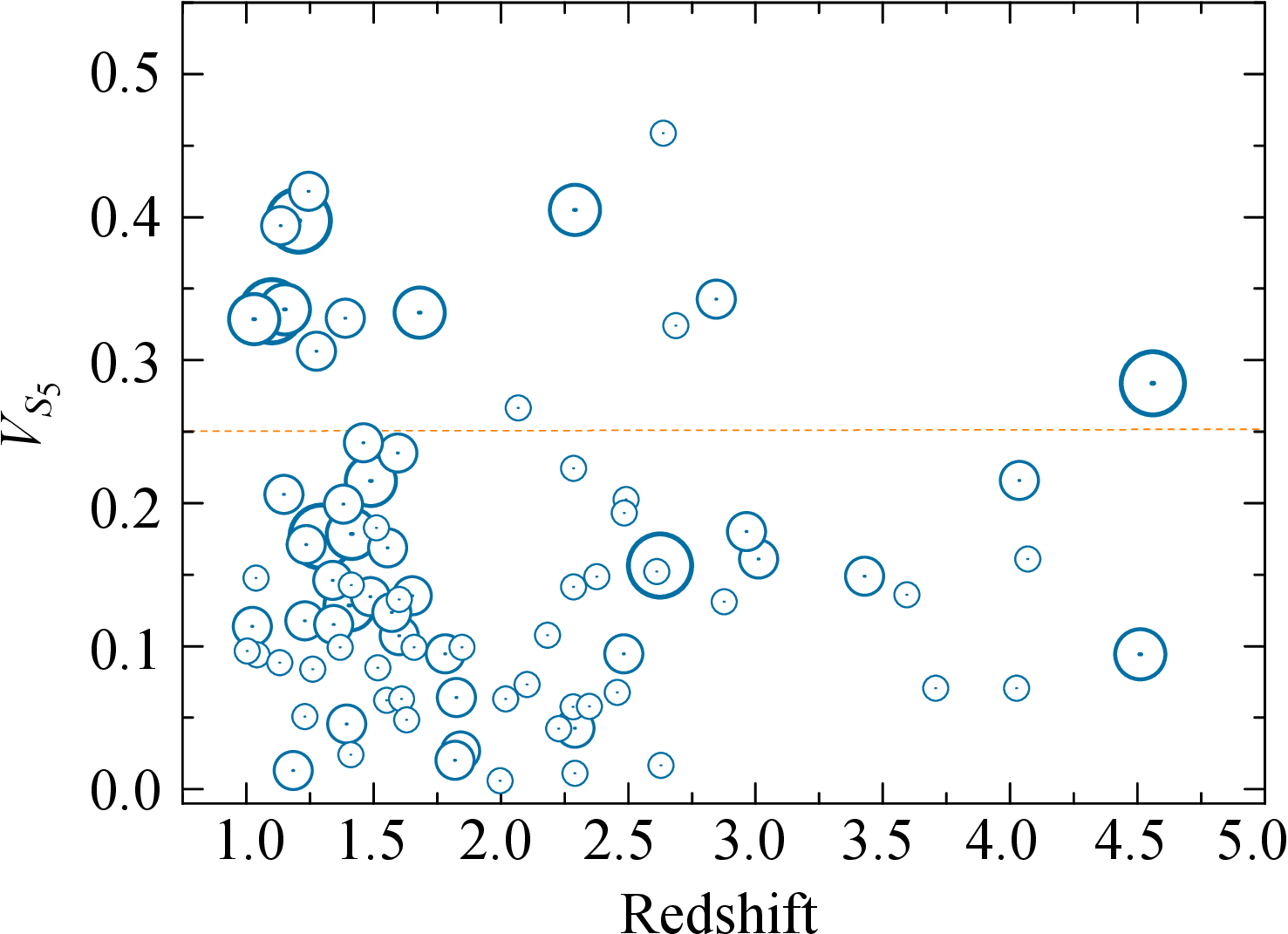}
\caption{\small The redshift versus the variability index $V_{S_{5}}$. Symbol sizes are proportional to $N_{\rm obs}$ and correspond to up to 5, 10, 20, and more number of observations.}
\label{fig:zM}
\end{figure}

Most radio sources have a quite low variability index, less than 0.25 (Fig.~\ref{fig:zM}). Poor statistics 
for redshifts $z > 3$ does not allow 
us to find any difference in variability at different cosmological epochs. 

\section{PEAKED-SPECTRUM GALAXIES}
\label{sec:PS}

 \setlength{\tabcolsep}{3.5pt}
\begin{table}
\caption{\label{var} \small The median, mean, maximum, and minimum values of the variability and modulation indices $V_{S}$ and $M$}
\begin{tabular}{c|c|c|c|l|l}
\hline
  Parameter   & $N$ & Median & Mean &  Min &Max \\
\hline
$V_{S_{5}}$ &  85 & 0.13 & 0.16$\pm$0.11 & 0.006 & 0.46 \\
$M_{5}$     & 103 & 0.11 & 0.12$\pm$0.07 & 0.002 & 0.33 \\
\hline
$V_{S_{11}}$ & 21 & 0.14 & 0.18$\pm$0.14 & 0.02 & 0.47 \\
$M_{11}$     & 24 & 0.17 & 0.16$\pm$0.10 & 0.001 & 0.40 \\
\hline
\end{tabular}
\end{table}

\setlength{\tabcolsep}{2.5pt}
\begin{table}
\caption{\label{spec} \small The median and mean values of $M_{5}$ for different spectral types (standard deviations are given in parentheses)}
\begin{tabular}{l|c|c|c}
\hline
 \multicolumn{1}{c|}{Spectral type}   & $N$ & Median & Mean  \\
\hline
flat    & 15 & 0.10 & 0.11 (0.08) \\
peaked  & 25 & 0.14 &  0.15 (0.07) \\
steep, ultra-steep  & 54 & 0.09 & 0.10 (0.06) \\
upturn,\,complex,\,inverted  & 9 & 0.18 & 0.17 (0.09) \\
\hline
\end{tabular}
\end{table}
\setlength{\tabcolsep}{2.5pt}

We have found 31 galaxies (18\%) having peaked radio spectra (PS). Among them 13~sources have already been known as GPS sources (Table~\ref{listPS}, column~(10)). All of them are bright radio sources with a flux density level about or more than 1~Jy in the~GHz band. To classify the genuine HFP/GPS/MPS sources, the following common criteria are usually used: the observed peak frequency $\nu_{\rm obs,peak}$ is less than 1~GHz for MPS and CSS sources, from 1 up to 5~GHz for GPS sources, and more than 5~GHz 
for HFPs; the spectral indices of the
optically thick $\alpha_{\rm thick}$ and thin $\alpha_{\rm thin}$ emission parts are close to $+0.5$ and $-0.7$, respectively (for the peaked spectra they match with $\alpha_{\rm low}$ and $\alpha_{\rm high}$); radio variability does not exceed 25\%  on a long timescale, as these objects are considered as the least variable compact extragalactic radio sources \citep{1990A&AS...82..261O,1991ApJ...380...66O,1997A&A...321..105D,1998PASP..110..493O,2004A&A...424...91E,2021A&ARv..29....3O}. MPS and CSS sources have different sizes, up to 1~kpc and about 20~kpc respectively. They are also 
located at different redshifts, MPSs at higher redshifts, $z>1$ \citep{2016MNRAS.459.2455C}, and CSS objects at closer distances.

\begin{table*}
\centering
\caption{\label{listPS} \small The list of PS galaxies. Columns: (1) source name, (2) the spectral types in accordance with the turnover frequency, (3-4) the spectral indices $\alpha_{\rm thick}$ and $\alpha_{\rm thin}$, (5) - the turnover frequency $\nu_{\rm peak}$ calculated in the observer’s frame of reference, (6) - the peak flux density $S_{\rm peak}$, (7) - the variability index $V_{S_{5}}$, (8) - RA and Dec projected linear sizes, (9) -  frequency on which projected linear sizes were calculated, (10) - the references to the galaxy size determination: [1]---Astrogeo database, [2]---\protect\cite{2022MNRAS.511.4572A}, [3]---\protect\cite{2017A&A...599A.123N}, [4]---\citet{2017ApJ...836..174C}, [5]---\citet{2011MNRAS.416.1135R}, [6]---\citet{1998PASP..110..493O}, [7]---\citet{1985A&A...152...38S}, [8]---\citet{1991ApJ...380...66O}, [9]---\citet{2002MNRAS.337..981S}, [10]---\citet{1995A&A...295..629S}}

\begin{tabular}{c|c|c|c|c|r|c|c|c|c}

\hline 
Name & sp. type & $\alpha_{\rm thin}$ & $\alpha_{\rm thick}$ & $\nu_{\rm peak}$, & $S_{\rm peak}$, & $V_{S_{5}}$ & LS, & Frequency, & {\multirow{2}{*}{Reference}}  \\
 &   &              &                &      GHz           & Jy         &           & kpc &   GHz           \\
1  &  2  &  3          &       4        &      5           & 6         &     7      & 8   &  9      &  10       \\
\hline 
J0101$-$2831 & MPS & $-0.77\pm0.05$ & $+0.35\pm0.04$ & $0.27\pm0.01$ &  $1.07\pm0.02$ & 0.11 & $0.022\times0.052$ & 8.7 & [1], [4] \\
J0116$-$2052 & MPS & $-1.04\pm0.04$ & $+0.10\pm0.04$ & $0.08\pm0.01$ & $13.36\pm0.09$ & 0.18 & $0.358\times0.152$ & 8.7 & [1], [4,5] \\
J0117$+$0114 & MPS & $-0.76\pm0.10$ & $+5.61\pm0.28$ & $0.11\pm0.01$ & $0.18\pm0.01$ & -- & -- &  -- &     -- \\
J0122$+$1923 & MPS & $-0.83\pm0.08$ & $+1.05\pm0.28$ & $0.17\pm0.01$ &  $1.38\pm0.03$ & 0.24 & $0.133\times0.800$ & 8.7  &  [1], [4] \\
J0148$+$1028 & MPS & $-1.20\pm0.11$ & $+0.41\pm0.06$ & $0.32\pm0.02$ & $1.05\pm0.03$  & 0.34 & $0.130\times0.152$ & 8.7 & [1], [4] \\
J0232$-$0742 & MPS & $-0.46\pm0.37$ & $+1.37\pm2.48$ & $0.17\pm0.06$ & $0.09\pm0.01$ & -- & -- & -- &       -- \\
J0432$+$4138 & MPS & $-0.72\pm0.02$ & $+0.25\pm0.06$ & $0.16\pm0.02$ & $17.62\pm0.24$ & 0.11 & $0.340\times0.059$ & 8.7  &  [1], [6] \\
J0745$+$1011 & GPS & $-0.99\pm0.03$ & $+0.44\pm0.02$ & $2.31\pm0.04$ & $4.17\pm0.02$ & 0.16 & $0.039\times0.075$ & 7.6   &  [7] \\
J0837$-$1951 & MPS & $-1.11\pm0.04$ & $+0.66\pm0.04$ & $0.34\pm0.01$ & $11.15\pm0.13$ & 0.33 & $0.160\times0.320$ & 7.6  &  [1], [4] \\
J0904$+$4727 & GPS & $-0.57\pm0.56$ & $+0.72\pm0.64$ & $1.68\pm0.54$ & $0.22\pm0.02$ & 0.22 & $0.073\times0.073$ &   4.8  &  [1] \\
J0909$+$4753 & MPS & $-0.53\pm0.16$ & $+0.56\pm0.37$ & $0.53\pm0.07$ & $0.31\pm0.02$ & 0.46 & $0.058\times0.054$ & 8.7   &  [1]  \\
J1002$+$0158 & MPS & $-2.60\pm- $ & $+0.33\pm-   $ & $1.42\pm-   $ &  $0.03\pm-$ & -- & & --  &      -- \\
J1109$+$3744 & MPS & $-0.53\pm0.07$ & $+0.71\pm0.45$ & $0.50\pm0.06$ & $1.45\pm0.06$ & 0.41 & $0.086\times0.138$ & 7.6  &  [1]  \\
J1126$+$3345 & MPS & $-1.30\pm0.08$ & $+0.26\pm0.05$ & $0.24\pm0.01$ & $3.57\pm0.06$ & 0.12 & $0.025\times0.071$ & 7.6  &  [1]  \\
J1129$+$5025 & MPS & $-0.89\pm0.08$ & $+0.52\pm0.43$ & $0.06\pm0.01$ & $7.23\pm0.18$ & 0.10 & $0.061\times0.168$ & 7.6  &  [1]  \\
J1133$+$2936 & MPS & $-0.75\pm0.04$ & $+5.49\pm0.08$ & $0.09\pm0.01$ & $1.30\pm0.03$ & --   & -- & -- &     -- \\
J1232$+$6644 & GPS & $-0.68\pm0.13$ & $+6.42\pm0.55$ & $0.51\pm0.01$ & $0.10\pm0.03$ & 0.05 & --  & -- &     -- \\
J1438$+$0150 & MPS & $-0.89\pm0.09$ & $+7.73\pm0.01$ & $0.09\pm0.01$ & $0.83\pm0.03$ & -- & --  &--  &      -- \\
J1459$+$4442 & HFP & $-0.77\pm0.21$ & $+0.35\pm0.06$ & $7.58\pm0.55$ & $0.21\pm0.01$ & 0.15 & $0.058\times0.042$ &  8.7 &   [1] \\
J1521$+$0430 & GPS & $-1.47\pm0.08$ & $+0.50\pm0.04$ & $0.84\pm0.03$ & $4.77\pm0.11$ & 0.18  & $0.047\times0.076$ &  8.7  &   [1], [8] \\
J1541$+$3840 & GPS & $-0.82\pm0.18$ & $+0.35\pm0.18$ & $0.74\pm0.09$ & $0.07\pm0.01$ & --   &-- & --  &      -- \\
J1545$+$4130 & GPS & $-0.17\pm0.03$ & $+3.11\pm1.66$ & $0.37\pm0.11$ & $0.09\pm0.01$ & 0.13 & $0.041\times0.100$ &  7.6&   [1] \\
J1550$+$4536 & GPS & $-1.97\pm0.39$ & $+0.35\pm0.06$ & $5.21\pm0.63$ & $0.07\pm0.01$ & 0.16 & $0.032\times0.062$ &  7.6   &  [1]  \\
J1602$+$3326 & GPS & $-0.64\pm0.05$ & $+0.20\pm0.05$ & $0.70\pm0.06$ & $2.79\pm0.06$ & 0.33 & $0.114\times0.033$ & 7.6  &   [1], [8] \\
J1606$+$3124 & GPS & $-0.84\pm0.09$ & $+0.33\pm0.10$ & $2.43\pm0.16$ & $0.73\pm0.02$ & 0.31 & 0.056 &  8.4 &  [2], [8] \\
J1640$+$1220 & MPS & $-0.69\pm0.07$ & $+0.01\pm0.09$ & $0.07\pm0.42$ & $4.05\pm2.28$ & 0.34 & $0.090\times0.073$ & 7.6  &   [1], [9] \\
J2037$-$0010 & MPS & $-0.98\pm0.22$ & $+0.93\pm0.24$ & $0.38\pm0.04$ & $0.92\pm0.09$ & 0.18 & --&  -- &     -- \\
J2058$+$0542 & MPS & $-0.97\pm0.02$ & $+0.73\pm0.05$ & $0.27\pm0.01$ & $2.89\pm0.04$ & 0.20 & $0.126\times0.337$ &  7.6 &   [1], [9] \\
J2227$-$2705 & MPS & $-0.96\pm0.07$ & $+1.09\pm0.17$ & $0.17\pm0.01$ & $1.42\pm0.01$ & --   & 3.3 &  4.8/8.4$^{a}$  &  [3], [4] \\
J2250$+$7129 & MPS & $-1.13\pm0.06$ & $+0.12\pm0.05$ & $0.05\pm0.01$ & $13.93\pm0.49$ & 0.03  & $0.100\times0.173$ &  7.6 &  [1], [10] \\
J2307$+$1450 & HFP & $-3.42\pm1.37$ & $+0.40\pm0.10$ & $7.24\pm1.46$ & $0.22\pm0.01$ & 0.42 & $0.025\times0.061$ & 7.6 &   [1] \\
\hline
\multicolumn{10}{l}{\footnotesize$^a$ VLA measurements}\\
\end{tabular}
\end{table*}

In our sample, among PS galaxies, there are two galaxies that are high frequency peaker (HFP) candidates, nine are gigahertz-peaked spectrum (GPS) candidates, and 20 are MPS candidates. The variability 
index $V_{S_{5}}$ was estimated for 24 PS galaxies, and it varies from 0.05 to 0.46. For eight of them the variability level higher than 0.25 at 5~GHz.

The projected linear sizes of the PS galaxies are shown in Table~\ref{listPS} (column~(8)). The data for two of them, J1606$+$3124 and J2227$-$2705, were taken from the literature. For 20 sources we estimated the linear sizes using the publicly available images in the Astrogeo database\footnote{\url{http://astrogeo.org}} at 7.6 and 8.7~GHz, because these are the frequencies available for most of our sources. For nine sources the linear sizes 
have not been estimated. The sizes span
from 0.022 to 0.8~kpc, the median is 0.073~kpc, excluding the size of 
J2227$-$2705 that was estimated based on 
a lower-resolution VLA image.

We found eleven new MPS, five new GPS, and two new HFP candidates. The objects
are  compact, but the radio spectra of most new PS galaxies are not well determined due to poor radio data at 
frequencies greater than several GHz, where larger radio variability of AGNs is expected than in the MHz domain.  Taking into account that we analyze the data from many telescopes with different angular resolutions and systematic
errors, multi-frequency and long-term radio measurements are needed for reliable classification of their radio spectra. 

It is clearly seen that there are almost no HFP galaxies in the sample. Even the sources with inverted or upturn radio spectra can be HFP candidates for which there are no high-frequency measurements, they also make up a small fraction, about 6\%, of galaxies in the sample (Table~\ref{tab:type}). One of the possible reasons for the absence of HFPs in the sample of bright radio galaxies can be the selection method that was used for sample construction. A relatively low frequency of the source selection (1.4~GHz) could lead to missing of the
inverted spectrum sources which had a flux density less than 20~mJy at MHz frequencies.

We fitted peaked spectra of PS galaxies with a function that describes synchrotron self absorption (SSA) radiation emitted by electrons with the power-law energy distribution in a homogeneous magnetic field
\citep{1970ranp.book.....P,1999A&A...349...45T}:
\begin{equation}
\begin{split}
S_{\nu} & = S_{\rm peak}\left(\cfrac{\nu}{\nu_{\rm peak}}\right)^{\alpha_{\rm thick}} \\
& \times \cfrac{1-\exp(-\tau_{m}(\nu/\nu_{\rm peak})^{\alpha_{\rm thin} -\alpha_{\rm thick}})} {1-\exp(-\tau_{m})},
\end{split}
\end{equation}
where $\tau_{m}\approx\cfrac{3}{2} \left(\sqrt{1-\cfrac{8\alpha_{\rm thin}}{3\alpha_{\rm thick}}}-1\right)$; $\nu_{\rm peak}$ is 
the turnover frequency; $S_{\rm peak}$ is the maximum flux density at the turnover frequency; $\alpha_{\rm thin}$ 
and $\alpha_{\rm thick}$ are the spectral indices of the optically thin and thick parts of the spectrum; $\tau_{m}$ is 
the optical depth at the turnover. 
The fitted parameters for the peaked spectra are presented in Table~\ref{listPS} (columns~(3)--(6)). Most of these spectra are well described with 
\mbox{$0.25 < \alpha_{\rm thick} < 0.75$} and \mbox{$-0.40 < \alpha_{\rm thin} < -1.20$}. Three sources (J0745$+$1011, J1521$+$0430, J1606$+$3124) have an upturn component in their spectra below 1~GHz. Optically thin part of the spectra has cut-off and requires an additional absorption for J0122$+$1923 above 9~GHz and for J2227$-$2705 above 2~GHz.

\section{DISCUSSION}

The obtained radio parameters for the sample are summarised in Table~\ref{table2_appendix}: radio luminosity at 5~GHz, logarithm of radio loudness, calculated using two approaches, spectral indices $\alpha_{\rm low}$ and $\alpha_{\rm high}$, radio spectrum type, and  the observed peak frequency in the spectrum. Below we discuss the dichotomy of the sources' types in the sample, mid-IR properties and classification, PS galaxies, and variability characteristics.

\subsection{Distant Galaxies vs Distant Quasars}

\citealt{2021MNRAS.508.2798S} studied the
radio properties of 102 distant quasars at redshifts $z\geq3$. Here we compare their characteristic radio features with similar properties of distant galaxies. We note that these samples have different spectral type distributions: among quasars at $z\geq3$ prevail objects with the peaked and flat (70\%  of the sample) radio spectra and only 15\% of the objects have steep spectra. At the same time, the galaxies mainly have steep and ultra steep radio spectra (60\% of the sample), and 33\% have peaked and flat radio spectra in total. These differences mean the dominating contribution of the compact bright radio core in the total spectrum for the quasars, while the radio core is less pronounced for the galaxies. The constructed average spectra well demonstrate the different spectral behavior for both samples. Firstly, the different values of the spectral index: the galaxies have steep ones, while the quasars have flat indices (Table~\ref{tab:aver_ind}). Secondly, the average quasar spectra have peaks varying from 0.5 GHz to 5.4 GHz (Table~\ref{tab:aver_ind}), while we do not reveal the peak in the average spectra of the galaxies, except for four cases.

Figure~\ref{fig:Lum_gist}
shows 
the
radio luminosity distributions  
for
distant galaxies and distant quasars. 
The
statistical properties of these distributions are presented in Table~\ref{lum_loud}. We note that distant quasars have 
a
similar radio luminosity distribution as the subsample G (and the high luminosity subgroup of 
the Q subsample). This fact can be explained by the selection effect: at high redshifts more bright quasars are detected easier.

We calculated the radio loudness 
using the first approach (the same optical spectral index for all the sources)
to compare
with quasars at $z\geq3$. 
Figure~\ref{fig:R_gist}a
demonstrates that 
the
radio loudness distributions of distant quasars and the 
Q~subsample
are similar (see also Table~\ref{lum_loud}). It is an expected result if 
the
distant quasars and 
the Q~subsample
belong to the 
same kind of the sources with a continuous distribution of radio properties.
At the same time, 
the
distant quasars have 
lower
radio loudness compared to the group~G. This fact is expected, because quasars 
are usually
brighter compared to radio galaxies in the 
optical
band.

\subsection{WISE Color Diagrams}
\label{sec:WISE}

We searched for the mid-IR counterparts in the Wide-field Infrared Survey Explorer (WISE, \citealt{2010AJ....140.1868W}) all-sky catalog for all the sources in our sample. We cross-matched the sources' positions with the WISE source catalog by adopting a search radius of 20$^{\prime\prime}$. For each source with several counterparts in the WISE catalog located within the aforementioned area, the object nearest to the coordinates was selected.

\begin{figure}
\includegraphics[width=.48\textwidth]{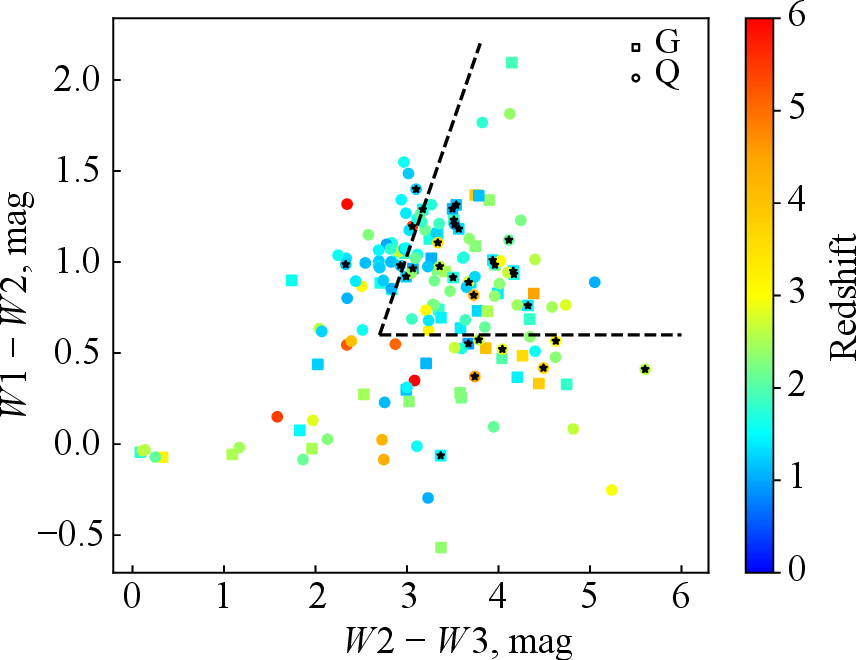} 
\caption{\small The WISE ``color\,--\,color'' diagram for our sample. The sources are colored according to their redshift. The dashed lines show the region for high-redshift objects \protect\cite{2005ApJ...631..163S,2018ApJS..235...10K}.
Black stars are the MPS objects.}
\label{fig:wise_color_space}
\end{figure}

Figure~\ref{fig:wise_color_space} shows the WISE ``color\,--\,color'' 
diagram for the sample ($W1-W2$ vs $W2-W3$, 
where $W1$, $W2$, and $W3$ are the WISE magnitudes at effective wavelengths of 3.4, 4.6, and 12~{$\mu$}m). 
The G and Q subsamples are marked by squares and circles respectively. The sources are colored according to their redshifts. The MPS objects are marked as black stars.

As expected, the Q population is faint 
\mbox{($W1 > 15$)} and red in the $W1-W2$ color. The G population has a medium $W1-W2$ 
color of 0.8.

In the paper of \cite{2018ApJS..235...10K}, Fig.~2 demonstrates 
the WISE ``color\,--\,color'' diagram of radio sources that have been identified as quasars in the SDSS spectroscopic catalog. To separate low-redshift quasars, the authors applied the following photometric cut:\linebreak \mbox{$W1 - W2 < 1.3\,(W2 - W3) - 3.04$} and \mbox{$W1 - W2 > 0.6$}.

We depicted the same region in Fig.~\ref{fig:wise_color_space} (where dashed black line shows the region of color-space where high-redshift objects can be found). As expected  \citep{2004ApJS..154..166L,2005ApJ...631..163S,2018ApJS..235...10K}, the Q subsample at $z>2$ dominates there (73\% of the sources in that region). In the same area there are relatively many objects with lower redshifts ($1<z<2$), which is most likely due to the incompleteness of the sample. The fact that almost all MPS objects are located in the designated area confirms that these objects are distant. 

\begin{figure}
\includegraphics[width=.48\textwidth]{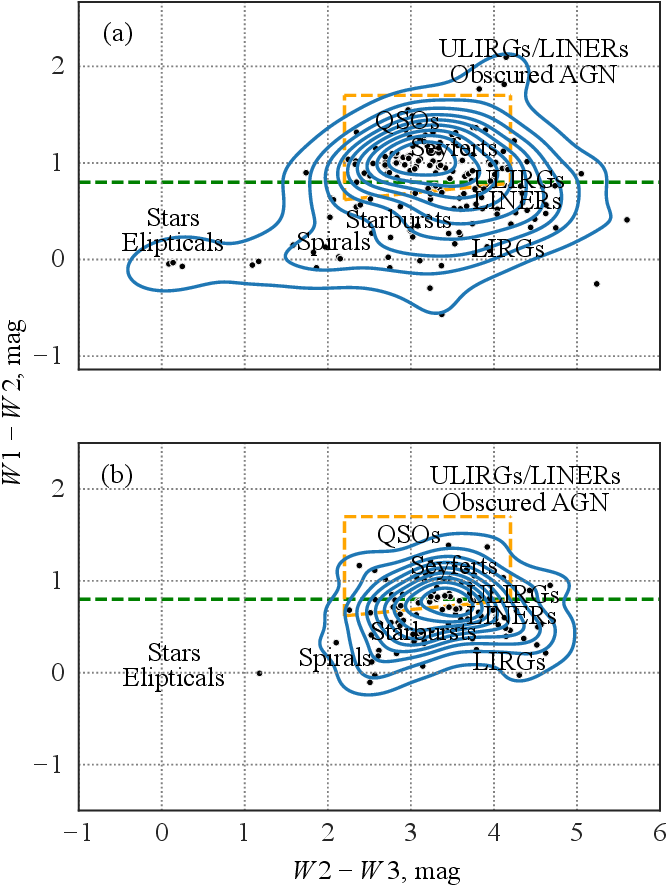}
\caption{\small
Panel (a): the WISE color diagram of our sample. Panel (b): the WISE color diagram for the bright quasars at $z\geq3$ from \protect\cite{2021MNRAS.508.2798S}. The green dashed line 
($W1 - W2 = 0.8$)
is the threshold above which QSOs are expected \protect\citep{2012ApJ...753...30S}. The orange dashed line is the ``AGN box'' \protect\cite{2011ApJ...735..112J}.
}
\label{fig:wise_G_QSO}
\end{figure}

Figure~\ref{fig:wise_G_QSO}
shows the WISE "сolor-color" diagrams for our sample and the sample of bright quasars at $z\geq3$ from \protect\cite{2021MNRAS.508.2798S} for comparison. The signatures indicate areas of different object types. By comparing the two panels of Fig.~\ref{fig:wise_G_QSO}, we reveal the wider distribution of IR properties for the galaxies from the sample. Despite the fact that the two samples are incomplete and different, it is clear that the collected list of the galaxies is heterogeneous and contains different types of AGNs.

It is unavoidable that a certain number of objects would drop out of their area. With the orange line, we marked the ``AGN'' region, which includes the WISE QSOs and Seyfert galaxies \citep{2011ApJ...735..112J}:\linebreak
$W2 - W3 > 2.2$ and $W2 - W3 < 4.2$,\linebreak
$W1 - W2 > (0.1 \times (W2 - W3) + 0.38)$
and $W1 - W2 < 1.7$.

Fifty-four percent of our objects, including most of the MPSs, lie within
this box. Most of the studied objects are located in the WISE QSO area above 
$W1 - W2 = 0.8$ \citep{2012ApJ...753...30S} (Fig.~15a).

\subsection{PS Population at High Redshifts}

We found 31 (18\%) galaxies with peaked radio spectra, which is significantly less than in the samples of bright distant quasars at $z\geq3$ in \cite{2021MNRAS.508.2798S}, 
where peaked spectrum sources accounted for 46\% of the sample. Still, it is larger than in some mixed galaxy and quasar samples: for example, \cite{1990MNRAS.245P..20O} found about 
a 10\% fraction of PS sources. A small fraction, not more than 10\%, 
of radio sources in the 1~Jy complete sample \citep{2011A&A...536A..14P}, showed peaked spectra, and they were mostly identified as blazars. In \cite{2013AstBu..68..262M}, 
the authors obtained only several per cents of sources with a constant peaked spectrum shape over a long period of time. Among 5890 radio sources from the complete sample of the AT20G survey
with a flux density more than 40~mJy, 
21\% of the sources were found to have peaked spectra \citep{2010MNRAS.402.2403M}.

Most of the PS galaxies in the sample under investigation are MPS candidates (20). Their redshifts and small linear sizes corresponding to the generally accepted MPS conception
allow us to consider all the 20 galaxies with \mbox{$\nu_{\rm peak,obs}<1$~GHz} as MPS candidates. The  median value of their redshift is 1.6. For reliable classification of all PS types in the sample, we need simultaneous multi-frequency radio data to define the turnover, preferably at several epochs, and long-term monitoring at several frequencies to determine whether there is variability at any or all of the frequencies.

The discovered peaked spectrum quasars are located at higher redshifts than quasars with other types of spectra. There are 18 PS quasars in the sample with the average $\langle z\rangle=2.74$ and the remaining 96 QSOs have $\langle z\rangle=1.83$. That confirms the plausibility of searching for young and high-redshift AGNs among the MPS sources (see, for example, \cite{2015MNRAS.450.1477C}). However, the same is not true for the galaxies.

\section{SUMMARY}
\label{sec:summary}

We have collected a sample of 173 bright HzRGs in the radio domain ($S_{1.4}\geqslant20$ mJy) at redshifts $z\geqslant1$ for the study. Using available literature measurements in a wide frequency range (74~MHz\,--\,22~GHz) on a time scale of 30 years, we have constructed radio spectra of the galaxies and determined their types, classifying the most of them as having a steep or ultra-steep radio spectrum (60\% of the sample). The spectral indices calculated at low and high frequencies vary on average from $-$0.6 to $-$1, the median values are $\alpha_{\rm low}=-0.77$ for the galaxies not identified with quasars and $\alpha_{\rm low}=-0.56$ for the galaxies with quasars (Section~\ref{sec:spectra}). The median  spectral index for the entire sample $\alpha_{\rm low}=-0.63$ agrees well with the previous studies (e.g., \citealt{2013MNRAS.435..650M,2021A&A...648A..14R}). In comparison with  the spectral indices of the radio sources in the revised 3C catalogue \citep{1969ApJ...157....1K}, where $\alpha_{0.75-5}=-0.85$ was obtained for galaxies and $\alpha_{0.75-5}=-0.79$ for quasars, we derived a similar slope for the galaxies and more flat spectra for the quasars at low frequencies $\alpha_{\rm low}$ (0.10--7.4~GHz).

We classified about 18\% (31) of spectra as PS  with predominant MPS type 
(20/31). Most of them fall into the high-redshift AGN box in the WISE color space diagram.

For the whole sample we have not found the evolution of radio spectra with redshift, the so called ``$\alpha$\,--\,$z$'' correlation, when more distant galaxies have steeper radio spectra; but suspecting the influence of quasars in our sample, we checked that relation for 59 galaxies not identified with quasars and blazars. In that case there exists, indeed, a significant anticorrelation between $\alpha_{\rm low}$ and $z$, and also between $\alpha_{5-10}$ calculated with k-correction at the rest-frame frequencies of 5--10 GHz and $z$, as it was reported in the previous studies (e.g., \citealt{2008AARv..15...67M,2011ApJ...743..122A}). 

The obtained spectral indices at frequencies of 1--10~GHz in the observer's frame of reference indicate optically thin radiation in the frequency range of 2\,--\,30~GHz in the source's reference frame for most of the objects. Sixteen per cent of the sample objects are characterized by a flat radio spectrum typical of the compact core emission in quasars. There is a systematic flattening of the spectrum at low frequencies ($ 5$~GHz) relative to the high frequencies ($ 11$~GHz). We assume that this is explained by the synchrotron self-absorption mechanism, a model of which explains fairly well the radio spectra shapes for the majority of the PS galaxies in our sample (Section~\ref{sec:PS}).

An analysis of the modulation and variability indices has shown that the median variability at 11~GHz is higher than at 5~GHz. Since the radio spectra of most objects (78\%) tend to steepen toward high frequencies (i.e., at higher frequencies the flux density decreases), we assume that this causes  a selection bias leading to the detection of sources with higher variability amplitude at higher frequencies. The median values of the flux density variability index are $V_{S_{11}}$=0.14 and $V_{S_{5}}$=0.13, which generally indicates a weak or moderate character of the long-term variability of the studied galaxies. Several blazars included in the sample demonstrate variability up to 50\%, but most of the quasars, AGNs, and blazars in the sample exhibit moderate variability of about 20\%, which is the typical value for galaxies with AGNs and quasars at lower redshifts \citep{2007A&A...462..547F}. Our work has shown that the variability of bright radio galaxies at large redshifts, up to $z=5$, does not differ significantly from the variability of galaxies at $z < 1$ (having characteristic values of 10\,--\,20\%), being caused by presumably similar physical phenomena in the galaxy centers, namely the processes around a massive central object, i.e., emission of the relativistic jet and the  repercussions of its propagation through the interstellar medium. For the AGN, such moderate variability on scales of tens of years presumably occurs in the radio core itself (e.g., \citealt{2008Natur.452..966M}). In order to clarify the processes and localization of the radio emission source in each specific case, an analysis of simultaneous measurements at several frequencies and the evolution of the radio spectrum is essential.

Selection of bright and distant radio galaxies using the NASA/IPAC Extragalactic Database (NED) revealed that the sample includes objects classified in a fairly wide range: from subclasses of blazars and AGNs to QSOs. We found that our sample has a very high rate of identification with quasars (114 of 173 objects). The scattering of the obtained characteristics (radio spectrum types, radio luminosity values, variability amplitudes, dispersion in the WISE colour diagram) confirms the diversity of radio emission sources in such galaxies.

\section*{ACKNOWLEDGEMENTS}
A part of the observed data was obtained with the RATAN-600 scientific facility. This research has made use of the NASA/IPAC Extragalactic Database (NED), which is operated by the Jet Propulsion Laboratory, California Institute of Technology, under contract with the National Aeronautics and Space Administration; the CATS data base, available on the Special Astrophysical Observatory website.

\section*{FUNDING}
The work was performed as part of the SAO RAS government contract approved by the Ministry of Science and Higher Education of the Russian Federation.\

\bibliographystyle{mnras}
\bibliography{Khabibullina_HzRG}

\begin{thebibliography}{}
\makeatletter
\relax
\def\mn@urlcharsother{\let\do\@makeother \do\$\do\&\do\#\do\^\do\_\do\%\do\~}
\def\mn@doi{\begingroup\mn@urlcharsother \@ifnextchar [ {\mn@doi@} {\mn@doi@[]}}
\def\mn@doi@[#1]#2{\def\@tempa{#1}\ifx\@tempa\@empty \href {http://dx.doi.org/#2} {doi:#2}\else \href {http://dx.doi.org/#2} {#1}\fi \endgroup}
\def\mn@eprint#1#2{\mn@eprint@#1:#2::\@nil}
\def\mn@eprint@arXiv#1{\href {http://arxiv.org/abs/#1} {{\tt arXiv:#1}}}
\def\mn@eprint@dblp#1{\href {http://dblp.uni-trier.de/rec/bibtex/#1.xml} {dblp:#1}}
\def\mn@eprint@#1:#2:#3:#4\@nil{\def\@tempa {#1}\def\@tempb {#2}\def\@tempc {#3}\ifx \@tempc \@empty \let \@tempc \@tempb \let \@tempb \@tempa \fi \ifx \@tempb \@empty \def\@tempb {arXiv}\fi \@ifundefined {mn@eprint@\@tempb}{\@tempb:\@tempc}{\expandafter \expandafter \csname mn@eprint@\@tempb\endcsname \expandafter{\@tempc}}}

\bibitem[\protect\citeauthoryear{{Afonso} et~al.,}{{Afonso} et~al.}{2011}]{2011ApJ...743..122A}
{Afonso} J.,  et~al., 2011, \mn@doi [\apj] {10.1088/0004-637X/743/2/122}, \href {https://ui.adsabs.harvard.edu/abs/2011ApJ...743..122A} {743, 122}

\bibitem[\protect\citeauthoryear{{Aller}, {Aller}  \& {Hughes}}{{Aller} et~al.}{1992}]{1992ApJ...399...16A}
{Aller} M.~F.,  {Aller} H.~D.,   {Hughes} P.~A.,  1992, \mn@doi [\apj] {10.1086/171898}, \href {http://adsabs.harvard.edu/abs/1992ApJ...399...16A} {399, 16}

\bibitem[\protect\citeauthoryear{{An}, {Wang}, {Zhang}, {Aditya}, {Hong}  \& {Cui}}{{An} et~al.}{2022}]{2022MNRAS.511.4572A}
{An} T.,  {Wang} A.,  {Zhang} Y.,  {Aditya} J.~N.~H.~S.,  {Hong} X.,   {Cui} L.,  2022, \mn@doi [\mnras] {10.1093/mnras/stac205}, \href {https://ui.adsabs.harvard.edu/abs/2022MNRAS.511.4572A} {511, 4572}

\bibitem[\protect\citeauthoryear{{Balayan} \& {Verkhodanov}}{{Balayan} \& {Verkhodanov}}{2004}]{2004Ap.....47..505B}
{Balayan} S.~K.,  {Verkhodanov} O.~V.,  2004, \mn@doi [Astrophysics] {10.1023/B:ASYS.0000049789.18676.81}, \href {https://ui.adsabs.harvard.edu/abs/2004Ap.....47..505B} {47, 505}

\bibitem[\protect\citeauthoryear{{Becker}, {White}  \& {Helfand}}{{Becker} et~al.}{1994}]{1994ASPC...61..165B}
{Becker} R.~H.,  {White} R.~L.,   {Helfand} D.~J.,  1994, in {Crabtree} D.~R.,  {Hanisch} R.~J.,   {Barnes} J.,  eds,  Astronomical Society of the Pacific Conference Series Vol. 61, Astronomical Data Analysis Software and Systems III. p.~165

\bibitem[\protect\citeauthoryear{{Bryant}, {Johnston}, {Broderick}, {Hunstead}, {De Breuck}  \& {Gaensler}}{{Bryant} et~al.}{2009}]{2009MNRAS.395.1099B}
{Bryant} J.~J.,  {Johnston} H.~M.,  {Broderick} J.~W.,  {Hunstead} R.~W.,  {De Breuck} C.,   {Gaensler} B.~M.,  2009, \mn@doi [\aapr] {10.1111/j.1365-2966.2009.14607.x}, \href {https://ui.adsabs.harvard.edu/abs/2009MNRAS.395.1099B} {395, 1099}

\bibitem[\protect\citeauthoryear{{Bursov}, {Lipovka}, {Soboleva}, {Temirova}, {Gol'Neva}, {Parijskaya}  \& {Savastenya}}{{Bursov} et~al.}{1996}]{1996BSAO...42....5B}
{Bursov} N.~N.,  {Lipovka} N.~M.,  {Soboleva} N.~S.,  {Temirova} A.~V.,  {Gol'Neva} N.~E.,  {Parijskaya} E.~Y.,   {Savastenya} A.~V.,  1996, Bulletin of the Special Astrophysics Observatory, \href {https://ui.adsabs.harvard.edu/abs/1996BSAO...42....5B} {42, 5}

\bibitem[\protect\citeauthoryear{{Callingham} et~al.,}{{Callingham} et~al.}{2017}]{2017ApJ...836..174C}
{Callingham} J.~R.,  et~al., 2017, \mn@doi [\apj] {10.3847/1538-4357/836/2/174}, \href {https://ui.adsabs.harvard.edu/abs/2017ApJ...836..174C} {836, 174}

\bibitem[\protect\citeauthoryear{{Cohen}, {Lane}, {Cotton}, {Kassim}, {Lazio}, {Perley}, {Condon}  \& {Erickson}}{{Cohen} et~al.}{2007}]{2007AJ....134.1245C}
{Cohen} A.~S.,  {Lane} W.~M.,  {Cotton} W.~D.,  {Kassim} N.~E.,  {Lazio} T.~J.~W.,  {Perley} R.~A.,  {Condon} J.~J.,   {Erickson} W.~C.,  2007, \mn@doi [\aj] {10.1086/520719}, \href {https://ui.adsabs.harvard.edu/abs/2007AJ....134.1245C} {134, 1245}

\bibitem[\protect\citeauthoryear{{Condon}, {Cotton}, {Greisen}, {Yin}, {Perley}, {Taylor}  \& {Broderick}}{{Condon} et~al.}{1998}]{1998AJ....115.1693C}
{Condon} J.~J.,  {Cotton} W.~D.,  {Greisen} E.~W.,  {Yin} Q.~F.,  {Perley} R.~A.,  {Taylor} G.~B.,   {Broderick} J.~J.,  1998, \mn@doi [\aj] {10.1086/300337}, \href {https://ui.adsabs.harvard.edu/abs/1998AJ....115.1693C} {115, 1693}

\bibitem[\protect\citeauthoryear{{Coppejans}, {Cseh}, {Williams}, {van Velzen}  \& {Falcke}}{{Coppejans} et~al.}{2015}]{2015MNRAS.450.1477C}
{Coppejans} R.,  {Cseh} D.,  {Williams} W.~L.,  {van Velzen} S.,   {Falcke} H.,  2015, \mn@doi [\mnras] {10.1093/mnras/stv681}, \href {https://ui.adsabs.harvard.edu/abs/2015MNRAS.450.1477C} {450, 1477}

\bibitem[\protect\citeauthoryear{{Coppejans} et~al.,}{{Coppejans} et~al.}{2016a}]{2016MNRAS.459.2455C}
{Coppejans} R.,  et~al., 2016a, \mn@doi [\mnras] {10.1093/mnras/stw799}, \href {http://adsabs.harvard.edu/abs/2016MNRAS.459.2455C} {459, 2455}

\bibitem[\protect\citeauthoryear{{Coppejans} et~al.,}{{Coppejans} et~al.}{2016b}]{2016MNRAS.463.3260C}
{Coppejans} R.,  et~al., 2016b, \mn@doi [\mnras] {10.1093/mnras/stw2236}, \href {http://adsabs.harvard.edu/abs/2016MNRAS.463.3260C} {463, 3260}

\bibitem[\protect\citeauthoryear{{Coppejans} et~al.,}{{Coppejans} et~al.}{2017}]{2017MNRAS.467.2039C}
{Coppejans} R.,  et~al., 2017, \mn@doi [\mnras] {10.1093/mnras/stx215}, \href {https://ui.adsabs.harvard.edu/abs/2017MNRAS.467.2039C} {467, 2039}

\bibitem[\protect\citeauthoryear{{D'Abrusco} et~al.,}{{D'Abrusco} et~al.}{2019}]{2019ApJS..242....4D}
{D'Abrusco} R.,  et~al., 2019, \mn@doi [\apjs] {10.3847/1538-4365/ab16f4}, \href {https://ui.adsabs.harvard.edu/abs/2019ApJS..242....4D/abstract} {242, 15}

\bibitem[\protect\citeauthoryear{{Dallacasa}, {Stanghellini}, {Centonza}  \& {Fanti}}{{Dallacasa} et~al.}{2000}]{2000A&A...363..887D}
{Dallacasa} D.,  {Stanghellini} C.,  {Centonza} M.,   {Fanti} R.,  2000, \aap, \href {https://ui.adsabs.harvard.edu/abs/2000A&A...363..887D} {363, 887}

\bibitem[\protect\citeauthoryear{{De Breuck}, {van Breugel}, {R{\"o}ttgering}  \& {Miley}}{{De Breuck} et~al.}{2000}]{2000A&AS..143..303D}
{De Breuck} C.,  {van Breugel} W.,  {R{\"o}ttgering} H.~J.~A.,   {Miley} G.,  2000, \mn@doi [\aaps] {10.1051/aas:2000181}, \href {https://ui.adsabs.harvard.edu/abs/2000A&AS..143..303D} {143, 303}

\bibitem[\protect\citeauthoryear{{De Breuck} et~al.,}{{De Breuck} et~al.}{2010}]{2010ApJ...725...36D}
{De Breuck} C.,  et~al., 2010, \mn@doi [\apj] {10.1088/0004-637X/725/1/36}, \href {https://ui.adsabs.harvard.edu/abs/2008A&26ARv..15...67M} {725, 36}

\bibitem[\protect\citeauthoryear{{Douglas}, {Bash}, {Bozyan}, {Torrence}  \& {Wolfe}}{{Douglas} et~al.}{1996}]{1996AJ....111.1945D}
{Douglas} J.~N.,  {Bash} F.~N.,  {Bozyan} F.~A.,  {Torrence} G.~V.,   {Wolfe} C.,  1996, \mn@doi [\aj] {10.1086/117932}, \href {http://adsabs.harvard.edu/abs/1996AJ....111.1945D} {111, 1945}

\bibitem[\protect\citeauthoryear{{Edwards} \& {Tingay}}{{Edwards} \& {Tingay}}{2004}]{2004A&A...424...91E}
{Edwards} P.~G.,  {Tingay} S.~J.,  2004, \mn@doi [\aap] {10.1051/0004-6361:20035749}, \href {http://adsabs.harvard.edu/abs/2004A&A...424...91E} {424, 91}

\bibitem[\protect\citeauthoryear{{Falcke}, {K{\"o}rding}  \& {Nagar}}{{Falcke} et~al.}{2004}]{2004NewAR..48.1157F}
{Falcke} H.,  {K{\"o}rding} E.,   {Nagar} N.~M.,  2004, \mn@doi [\nar] {10.1016/j.newar.2004.09.029}, \href {https://ui.adsabs.harvard.edu/abs/2004NewAR..48.1157F} {48, 1157}

\bibitem[\protect\citeauthoryear{{Fan} et~al.,}{{Fan} et~al.}{2007}]{2007A&A...462..547F}
{Fan} J.~H.,  et~al., 2007, \mn@doi [\aap] {10.1051/0004-6361:20054775}, \href {https://ui.adsabs.harvard.edu/abs/2007A&A...462..547F} {462, 547}

\bibitem[\protect\citeauthoryear{{Fanti}, {Fanti}, {Schilizzi}, {Spencer}, {Nan Rendong}, {Parma}, {van Breugel}  \& {Venturi}}{{Fanti} et~al.}{1990}]{1990A&A...231..333F}
{Fanti} R.,  {Fanti} C.,  {Schilizzi} R.~T.,  {Spencer} R.~E.,  {Nan Rendong} {Parma} P.,  {van Breugel} W.~J.~M.,   {Venturi} T.,  1990, \aap, \href {https://ui.adsabs.harvard.edu/abs/1990A&A...231..333F} {231, 333}

\bibitem[\protect\citeauthoryear{{Frey}, {Paragi}, {Gurvits}, {Cseh}  \& {Gab{\'a}nyi}}{{Frey} et~al.}{2010}]{2010A&A...524A..83F}
{Frey} S.,  {Paragi} Z.,  {Gurvits} L.~I.,  {Cseh} D.,   {Gab{\'a}nyi} K.~{\'E}.,  2010, \mn@doi [\aap] {10.1051/0004-6361/201015554}, \href {https://ui.adsabs.harvard.edu/abs/2010A&A...524A..83F} {524, A83}

\bibitem[\protect\citeauthoryear{{Gab{\'a}nyi} et~al.,}{{Gab{\'a}nyi} et~al.}{2021}]{2021AN....342.1092G}
{Gab{\'a}nyi} K.~{\'E}.,  et~al., 2021, \mn@doi [Astronomische Nachrichten] {10.1002/asna.20210057}, \href {https://ui.adsabs.harvard.edu/abs/2021AN....342.1092G} {342, 1092}

\bibitem[\protect\citeauthoryear{{Gendre} \& {Wall}}{{Gendre} \& {Wall}}{2008}]{2008MNRAS.390..819G}
{Gendre} M.~A.,  {Wall} J.~V.,  2008, \mn@doi [\mnras] {10.1111/j.1365-2966.2008.13792.x}, \href {https://ui.adsabs.harvard.edu/abs/2008MNRAS.390..819G} {390, 819}

\bibitem[\protect\citeauthoryear{{Ghisellini}, {Celotti}, {Tavecchio}, {Haardt}  \& {Sbarrato}}{{Ghisellini} et~al.}{2014}]{2014MNRAS.438.2694G}
{Ghisellini} G.,  {Celotti} A.,  {Tavecchio} F.,  {Haardt} F.,   {Sbarrato} T.,  2014, \mn@doi [\mnras] {10.1093/mnras/stt2394}, \href {https://ui.adsabs.harvard.edu/abs/2014MNRAS.438.2694G} {438, 2694}

\bibitem[\protect\citeauthoryear{{Ginsburg} et~al.,}{{Ginsburg} et~al.}{2019}]{2019AJ....157...98G}
{Ginsburg} A.,  et~al., 2019, \mn@doi [\aj] {10.3847/1538-3881/aafc33}, \href {https://ui.adsabs.harvard.edu/abs/2019AJ....157...98G} {157, 98}

\bibitem[\protect\citeauthoryear{{Gorokhov} \& {Verkhodanov}}{{Gorokhov} \& {Verkhodanov}}{1994}]{1994AstL...20..671G}
{Gorokhov} V.~L.,  {Verkhodanov} O.~V.,  1994, Astronomy Letters, \href {https://ui.adsabs.harvard.edu/abs/1994AstL...20..671G} {20, 671}

\bibitem[\protect\citeauthoryear{{Gower}, {Scott}  \& {Wills}}{{Gower} et~al.}{1967}]{1967MmRAS..71...49G}
{Gower} J.~F.~R.,  {Scott} P.~F.,   {Wills} D.,  1967, \memras, \href {http://adsabs.harvard.edu/abs/1967MmRAS..71...49G} {71, 49}

\bibitem[\protect\citeauthoryear{{Gregory}, {Scott}, {Douglas}  \& {Condon}}{{Gregory} et~al.}{1996}]{1996ApJS..103..427G}
{Gregory} P.~C.,  {Scott} W.~K.,  {Douglas} K.,   {Condon} J.~J.,  1996, \mn@doi [\apjs] {10.1086/192282}, \href {http://adsabs.harvard.edu/abs/1996ApJS..103..427G} {103, 427}

\bibitem[\protect\citeauthoryear{{Harwood}, {Hardcastle}, {Croston}  \& {Goodger}}{{Harwood} et~al.}{2013}]{2013MNRAS.435.3353H}
{Harwood} J.~J.,  {Hardcastle} M.~J.,  {Croston} J.~H.,   {Goodger} J.~L.,  2013, \mn@doi [\mnras] {10.1093/mnras/stt1526}, \href {https://ui.adsabs.harvard.edu/abs/2013MNRAS.435.3353H} {435, 3353}

\bibitem[\protect\citeauthoryear{{Harwood}, {Hardcastle}  \& {Croston}}{{Harwood} et~al.}{2015}]{2015MNRAS.454.3403H}
{Harwood} J.~J.,  {Hardcastle} M.~J.,   {Croston} J.~H.,  2015, \mn@doi [\mnras] {10.1093/mnras/stv2194}, \href {https://ui.adsabs.harvard.edu/abs/2015MNRAS.454.3403H} {454, 3403}

\bibitem[\protect\citeauthoryear{{Healey}, {Romani}, {Taylor}, {Sadler}, {Ricci}, {Murphy}, {Ulvestad}  \& {Winn}}{{Healey} et~al.}{2007}]{2007ApJS..171...61H}
{Healey} S.~E.,  {Romani} R.~W.,  {Taylor} G.~B.,  {Sadler} E.~M.,  {Ricci} R.,  {Murphy} T.,  {Ulvestad} J.~S.,   {Winn} J.~N.,  2007, \mn@doi [\apjs] {10.1086/513742}, \href {https://ui.adsabs.harvard.edu/abs/2007ApJS..171...61H} {171, 61}

\bibitem[\protect\citeauthoryear{{Healey} et~al.,}{{Healey} et~al.}{2008}]{2008ApJS..175...97H}
{Healey} S.~E.,  et~al., 2008, \mn@doi [\apjs] {10.1086/523302}, \href {https://ui.adsabs.harvard.edu/abs/2008ApJS..175...97H} {175, 97}

\bibitem[\protect\citeauthoryear{{Hurley-Walker} et~al.,}{{Hurley-Walker} et~al.}{2017}]{2017MNRAS.464.1146H}
{Hurley-Walker} N.,  et~al., 2017, \mn@doi [\mnras] {10.1093/mnras/stw2337}, \href {https://ui.adsabs.harvard.edu/abs/2017MNRAS.464.1146H} {464, 1146}

\bibitem[\protect\citeauthoryear{{Intema}, {Jagannathan}, {Mooley}  \& {Frail}}{{Intema} et~al.}{2017}]{2017A&A...598A..78I}
{Intema} H.~T.,  {Jagannathan} P.,  {Mooley} K.~P.,   {Frail} D.~A.,  2017, \mn@doi [\aap] {10.1051/0004-6361/201628536}, \href {https://ui.adsabs.harvard.edu/abs/2017A&A...598A..78I} {598, A78}

\bibitem[\protect\citeauthoryear{{Jarrett} et~al.,}{{Jarrett} et~al.}{2011}]{2011ApJ...735..112J}
{Jarrett} T.~H.,  et~al., 2011, \mn@doi [\aj] {10.1088/0004-637X/735/2/112}, \href {https://ui.adsabs.harvard.edu/abs/2011ApJ...735..112J/abstract} {735, 33}

\bibitem[\protect\citeauthoryear{{Kellermann}}{{Kellermann}}{1964}]{1964ApJ...140..969K}
{Kellermann} K.~I.,  1964, \mn@doi [\apj] {10.1086/147998}, \href {https://ui.adsabs.harvard.edu/abs/1964ApJ...140..969K/abstract} {140, 969}

\bibitem[\protect\citeauthoryear{{Kellermann}, {Pauliny-Toth}  \& {Williams}}{{Kellermann} et~al.}{1969}]{1969ApJ...157....1K}
{Kellermann} K.~I.,  {Pauliny-Toth} I.~I.~K.,   {Williams} P.~J.~S.,  1969, \mn@doi [\apj] {10.1086/150046}, \href {https://ui.adsabs.harvard.edu/abs/1969ApJ...157....1K} {157, 1}

\bibitem[\protect\citeauthoryear{{Kellermann}, {Sramek}, {Schmidt}, {Shaffer}  \& {Green}}{{Kellermann} et~al.}{1989}]{1989AJ.....98.1195K}
{Kellermann} K.~I.,  {Sramek} R.,  {Schmidt} M.,  {Shaffer} D.~B.,   {Green} R.,  1989, \mn@doi [\aj] {10.1086/115207}, \href {https://ui.adsabs.harvard.edu/abs/1989AJ.....98.1195K} {98, 1195}

\bibitem[\protect\citeauthoryear{{Ker}, {Best}, {Rigby}, {R{\"o}ttgering}  \& {Gendre}}{{Ker} et~al.}{2012}]{2012MNRAS.420.2644K}
{Ker} L.~M.,  {Best} P.~N.,  {Rigby} E.~E.,  {R{\"o}ttgering} H.~J.~A.,   {Gendre} M.~A.,  2012, \mn@doi [\mnras] {10.1111/j.1365-2966.2011.20235.x}, \href {https://ui.adsabs.harvard.edu/abs/2012MNRAS.420.2644K} {420, 2644}

\bibitem[\protect\citeauthoryear{{Khabibullina} \& {Verkhodanov}}{{Khabibullina} \& {Verkhodanov}}{2009a}]{2009AstBu..64..123K}
{Khabibullina} M.~L.,  {Verkhodanov} O.~V.,  2009a, \mn@doi [\bsao] {10.1134/S1990341309020023}, \href {https://ui.adsabs.harvard.edu/abs/2009AstBu..64..123K} {64, 123}

\bibitem[\protect\citeauthoryear{{Khabibullina} \& {Verkhodanov}}{{Khabibullina} \& {Verkhodanov}}{2009b}]{2009AstBu..64..276K}
{Khabibullina} M.~L.,  {Verkhodanov} O.~V.,  2009b, \mn@doi [\bsao] {10.1134/S1990341309030079}, \href {https://ui.adsabs.harvard.edu/abs/2009AstBu..64..276K} {64, 276}

\bibitem[\protect\citeauthoryear{{Khabibullina} \& {Verkhodanov}}{{Khabibullina} \& {Verkhodanov}}{2009c}]{2009AstBu..64..340K}
{Khabibullina} M.~L.,  {Verkhodanov} O.~V.,  2009c, \mn@doi [\bsao] {10.1134/S199034130904004X}, \href {https://ui.adsabs.harvard.edu/abs/2009AstBu..64..340K} {64, 340}

\bibitem[\protect\citeauthoryear{{Khabibullina}, {Verkhodanov}  \& {Parijskij}}{{Khabibullina} et~al.}{2008}]{2008AstBu..63...95K}
{Khabibullina} M.~L.,  {Verkhodanov} O.~V.,   {Parijskij} Y.~N.,  2008, \mn@doi [\bsao] {10.1134/S1990341308020016}, \href {https://ui.adsabs.harvard.edu/abs/2008AstBu..63...95K} {63, 95}

\bibitem[\protect\citeauthoryear{{Kiikov}, {Mingaliev}, {Stolyarov}  \& {Stupalov}}{{Kiikov} et~al.}{2002}]{2002BSAO...54....5K}
{Kiikov} S.~O.,  {Mingaliev} M.~G.,  {Stolyarov} V.~A.,   {Stupalov} M.~S.,  2002, Bulletin of the Special Astrophysics Observatory, \href {https://ui.adsabs.harvard.edu/abs/2002BSAO...54....5K} {54, 5}

\bibitem[\protect\citeauthoryear{{Klamer}, {Ekers}, {Bryant}, {Hunstead}, {Sadler}  \& {De Breuck}}{{Klamer} et~al.}{2006}]{2006MNRAS.371..852K}
{Klamer} I.~J.,  {Ekers} R.~D.,  {Bryant} J.~J.,  {Hunstead} R.~W.,  {Sadler} E.~M.,   {De Breuck} C.,  2006, \mn@doi [\mnras] {10.1111/j.1365-2966.2006.10714.x}, \href {https://ui.adsabs.harvard.edu/abs/2006MNRAS.371..852K} {371, 852}

\bibitem[\protect\citeauthoryear{{Kopylov}, {Goss}, {Pariĭskiĭ}, {Soboleva}, {Verkhodanov}, {Temirova}  \& {Zhelenkova}}{{Kopylov} et~al.}{2006}]{2006AstL...32..433K}
{Kopylov} A.~I.,  {Goss} W.~M.,  {Pariĭskiĭ} Y.~N.,  {Soboleva} N.~S.,  {Verkhodanov} O.~V.,  {Temirova} A.~V.,   {Zhelenkova} O.~P.,  2006, \mn@doi [Astronomy Letters] {10.1134/S1063773706070012}, \href {https://ui.adsabs.harvard.edu/abs/2006AstL...32..433K} {32, 433}

\bibitem[\protect\citeauthoryear{{Kovalev}, {Nizhelsky}, {Kovalev}, {Berlin}, {Zhekanis}, {Mingaliev}  \& {Bogdantsov}}{{Kovalev} et~al.}{1999}]{1999A&AS..139..545K}
{Kovalev} Y.~Y.,  {Nizhelsky} N.~A.,  {Kovalev} Y.~A.,  {Berlin} A.~B.,  {Zhekanis} G.~V.,  {Mingaliev} M.~G.,   {Bogdantsov} A.~V.,  1999, \mn@doi [\aaps] {10.1051/aas:1999406}, \href {https://ui.adsabs.harvard.edu/abs/1999A&AS..139..545K} {139, 545}

\bibitem[\protect\citeauthoryear{{Kraus} et~al.,}{{Kraus} et~al.}{2003}]{2003A&A...401..161K}
{Kraus} A.,  et~al., 2003, \mn@doi [\aap] {10.1051/0004-6361:20030118}, \href {http://adsabs.harvard.edu/abs/2003A&A...401..161K} {401, 161}

\bibitem[\protect\citeauthoryear{{Krogager} et~al.,}{{Krogager} et~al.}{2018}]{2018ApJS..235...10K}
{Krogager} J.~K.,  et~al., 2018, \mn@doi [\apjs] {10.3847/1538-4365/aaab51}, \href {https://ui.adsabs.harvard.edu/abs/2018ApJS..235...10K/abstract} {235, 10}

\bibitem[\protect\citeauthoryear{{Labiano}, {Barthel}, {O'Dea}, {de Vries}, {P{\'e}rez}  \& {Baum}}{{Labiano} et~al.}{2007}]{2007A&A...463...97L}
{Labiano} A.,  {Barthel} P.~D.,  {O'Dea} C.~P.,  {de Vries} W.~H.,  {P{\'e}rez} I.,   {Baum} S.~A.,  2007, \mn@doi [\aap] {10.1051/0004-6361:20066183}, \href {https://ui.adsabs.harvard.edu/abs/2007A&A...463...97L} {463, 97}

\bibitem[\protect\citeauthoryear{{Lacy} et~al.,}{{Lacy} et~al.}{2004}]{2004ApJS..154..166L}
{Lacy} M.,  et~al., 2004, \mn@doi [\apjs] {10.1086/422816}, \href {https://ui.adsabs.harvard.edu/abs/2004ApJS..154..166L} {154, 166}

\bibitem[\protect\citeauthoryear{{Lacy} et~al.,}{{Lacy} et~al.}{2020}]{2020PASP..132c5001L}
{Lacy} M.,  et~al., 2020, \mn@doi [\pasp] {10.1088/1538-3873/ab63eb}, \href {https://ui.adsabs.harvard.edu/abs/2020PASP..132c5001L} {132, 035001}

\bibitem[\protect\citeauthoryear{{Lazio}, {Waltman}, {Ghigo}, {Fiedler}, {Foster}  \& {Johnston}}{{Lazio} et~al.}{2001}]{2001ApJS..136..265L}
{Lazio} T. J.~W.,  {Waltman} E.~B.,  {Ghigo} F.~D.,  {Fiedler} R.~L.,  {Foster} R.~S.,   {Johnston} K.~J.,  2001, \mn@doi [\apjs] {10.1086/322531}, \href {https://ui.adsabs.harvard.edu/abs/2001ApJS..136..265L} {136, 265}

\bibitem[\protect\citeauthoryear{{Liu}, {Jiang}, {Gu}  \& {Gurvits}}{{Liu} et~al.}{2016}]{2016AN....337..101L}
{Liu} Y.,  {Jiang} D.~R.,  {Gu} M.,   {Gurvits} L.~I.,  2016, \mn@doi [Astronomische Nachrichten] {10.1002/asna.201512273}, \href {https://ui.adsabs.harvard.edu/abs/2016AN....337..101L} {337, 101}

\bibitem[\protect\citeauthoryear{{Lyke} et~al.,}{{Lyke} et~al.}{2020}]{2020ApJS..250....8L}
{Lyke} B.~W.,  et~al., 2020, \mn@doi [\apjsupp] {10.3847/1538-4365/aba623}, \href {https://ui.adsabs.harvard.edu/abs/2020ApJS..250....8L} {250, 24}

\bibitem[\protect\citeauthoryear{{Majorova}}{{Majorova}}{2008}]{2008AstBu..63...56M}
{Majorova} E.~K.,  2008, \mn@doi [\bsao] {10.1007/s11755-008-1006-6}, \href {https://ui.adsabs.harvard.edu/abs/2008AstBu..63...56M} {63, 56}

\bibitem[\protect\citeauthoryear{{Marscher} et~al.,}{{Marscher} et~al.}{2008}]{2008Natur.452..966M}
{Marscher} A.~P.,  et~al., 2008, \mn@doi [\nat] {10.1038/nature06895}, \href {https://ui.adsabs.harvard.edu/abs/2008Natur.452..966M} {452, 966}

\bibitem[\protect\citeauthoryear{{Matthews}, {Morgan}  \& {Schmidt}}{{Matthews} et~al.}{1964}]{1964ApJ...140...35M}
{Matthews} T.~A.,  {Morgan} W.~W.,   {Schmidt} M.,  1964, \mn@doi [\apj] {10.1086/147890}, \href {https://ui.adsabs.harvard.edu/abs/1964ApJ...140...35M} {140, 35}

\bibitem[\protect\citeauthoryear{{Mauch}, {Kl{\"o}ckner}, {Rawlings}, {Jarvis}, {Hardcastle}, {Obreschkow}, {Saikia}  \& {Thompson}}{{Mauch} et~al.}{2013}]{2013MNRAS.435..650M}
{Mauch} T.,  {Kl{\"o}ckner} H.-R.,  {Rawlings} S.,  {Jarvis} M.,  {Hardcastle} M.~J.,  {Obreschkow} D.,  {Saikia} D.~J.,   {Thompson} M.~A.,  2013, \mn@doi [\mnras] {10.1093/mnras/stt1323}, \href {https://ui.adsabs.harvard.edu/abs/2013MNRAS.435..650M} {435, 650}

\bibitem[\protect\citeauthoryear{{Miley} \& {De Breuck}}{{Miley} \& {De Breuck}}{2008}]{2008AARv..15...67M}
{Miley} G.,  {De Breuck} C.,  2008, \mn@doi [\aapr] {10.1007/s00159-007-0008-z}, \href {https://ui.adsabs.harvard.edu/abs/2008A&ARv..15...67M/abstract} {15, 67}

\bibitem[\protect\citeauthoryear{{Mingaliev}, {Botashev}  \& {Stolyarov}}{{Mingaliev} et~al.}{1998}]{1998BSAO...46...28M}
{Mingaliev} M.~G.,  {Botashev} A.~M.,   {Stolyarov} V.~A.,  1998, Bulletin of the Special Astrophysics Observatory, \href {https://ui.adsabs.harvard.edu/abs/1998BSAO...46...28M} {46, 28}

\bibitem[\protect\citeauthoryear{{Mingaliev}, {Stolyarov}, {Davies}, {Melhuish}, {Bursov}  \& {Zhekanis}}{{Mingaliev} et~al.}{2001}]{2001A&A...370...78M}
{Mingaliev} M.~G.,  {Stolyarov} V.~A.,  {Davies} R.~D.,  {Melhuish} S.~J.,  {Bursov} N.~A.,   {Zhekanis} G.~V.,  2001, \mn@doi [\aap] {10.1051/0004-6361:20010215}, \href {http://adsabs.harvard.edu/abs/2001A&A...370...78M} {370, 78}

\bibitem[\protect\citeauthoryear{{Mingaliev}, {Sotnikova}, {Torniainen}, {Tornikoski}  \& {Udovitskiy}}{{Mingaliev} et~al.}{2012}]{2012A&A...544A..25M}
{Mingaliev} M.~G.,  {Sotnikova} Y.~V.,  {Torniainen} I.,  {Tornikoski} M.,   {Udovitskiy} R.~Y.,  2012, \mn@doi [\aap] {10.1051/0004-6361/201118506}, \href {http://adsabs.harvard.edu/abs/2012A&A...544A..25M} {544, A25}

\bibitem[\protect\citeauthoryear{{Mingaliev}, {Sotnikova}, {Mufakharov}, {Erkenov}  \& {Udovitskiy}}{{Mingaliev} et~al.}{2013}]{2013AstBu..68..262M}
{Mingaliev} M.~G.,  {Sotnikova} Y.~V.,  {Mufakharov} T.~V.,  {Erkenov} A.~K.,   {Udovitskiy} R.~Y.,  2013, \mn@doi [Astrophysical Bulletin] {10.1134/S1990341313030036}, \href {http://adsabs.harvard.edu/abs/2013AstBu..68..262M} {68, 262}

\bibitem[\protect\citeauthoryear{{Mingaliev} et~al.,}{{Mingaliev} et~al.}{2017}]{2017AN....338..700M}
{Mingaliev} M.,  et~al., 2017, \mn@doi [Astronomische Nachrichten] {10.1002/asna.201713361}, \href {https://ui.adsabs.harvard.edu/abs/2017AN....338..700M} {338, 700}

\bibitem[\protect\citeauthoryear{{Morabito} \& {Harwood}}{{Morabito} \& {Harwood}}{2018}]{2018MNRAS.480.2726M}
{Morabito} L.~K.,  {Harwood} J.~J.,  2018, \mn@doi [\mnras] {10.1093/mnras/sty2019}, \href {https://ui.adsabs.harvard.edu/abs/2018MNRAS.480.2726M} {480, 2726}

\bibitem[\protect\citeauthoryear{{Mufakharov}, {Mikhailov}, {Sotnikova}, {Mingaliev}, {Stolyarov}, {Erkenov}, {Nizhelskij}  \& {Tsybulev}}{{Mufakharov} et~al.}{2021}]{2021MNRAS.503.4662M}
{Mufakharov} T.,  {Mikhailov} A.,  {Sotnikova} Y.,  {Mingaliev} M.,  {Stolyarov} V.,  {Erkenov} A.,  {Nizhelskij} N.,   {Tsybulev} P.,  2021, \mn@doi [\mnras] {10.1093/mnras/staa3688}, \href {https://ui.adsabs.harvard.edu/abs/2021MNRAS.503.4662M} {503, 4662}

\bibitem[\protect\citeauthoryear{{Murphy} et~al.,}{{Murphy} et~al.}{2010}]{2010MNRAS.402.2403M}
{Murphy} T.,  et~al., 2010, \mn@doi [\mnras] {10.1111/j.1365-2966.2009.15961.x}, \href {https://ui.adsabs.harvard.edu/abs/2010MNRAS.402.2403M} {402, 2403}

\bibitem[\protect\citeauthoryear{{Nesvadba}, {De Breuck}, {Lehnert}, {Best}  \& {Collet}}{{Nesvadba} et~al.}{2017}]{2017A&A...599A.123N}
{Nesvadba} N.~P.~H.,  {De Breuck} C.,  {Lehnert} M.~D.,  {Best} P.~N.,   {Collet} C.,  2017, \mn@doi [\aap] {10.1051/0004-6361/201528040}, \href {https://ui.adsabs.harvard.edu/abs/2017A&A...599A.123N/abstract} {599, 44}

\bibitem[\protect\citeauthoryear{{O'Dea}}{{O'Dea}}{1990}]{1990MNRAS.245P..20O}
{O'Dea} C.~P.,  1990, \mnras, \href {https://ui.adsabs.harvard.edu/abs/1990MNRAS.245P..20O} {245, 20P}

\bibitem[\protect\citeauthoryear{{O'Dea}}{{O'Dea}}{1998}]{1998PASP..110..493O}
{O'Dea} C.~P.,  1998, \mn@doi [\pasp] {10.1086/316162}, \href {http://adsabs.harvard.edu/abs/1998PASP..110..493O} {110, 493}

\bibitem[\protect\citeauthoryear{{O'Dea} \& {Baum}}{{O'Dea} \& {Baum}}{1997}]{1997AJ....113..148O}
{O'Dea} C.~P.,  {Baum} S.~A.,  1997, \mn@doi [\aj] {10.1086/118241}, \href {http://adsabs.harvard.edu/abs/1997AJ....113..148O} {113, 148}

\bibitem[\protect\citeauthoryear{{O'Dea} \& {Saikia}}{{O'Dea} \& {Saikia}}{2021}]{2021A&ARv..29....3O}
{O'Dea} C.~P.,  {Saikia} D.~J.,  2021, \mn@doi [\aapr] {10.1007/s00159-021-00131-w}, \href {https://ui.adsabs.harvard.edu/abs/2021A&ARv..29....3O} {29, 3}

\bibitem[\protect\citeauthoryear{{O'Dea}, {Baum}  \& {Morris}}{{O'Dea} et~al.}{1990}]{1990A&AS...82..261O}
{O'Dea} C.~P.,  {Baum} S.~A.,   {Morris} G.~B.,  1990, \aaps, \href {http://adsabs.harvard.edu/abs/1990A&AS...82..261O} {82, 261}

\bibitem[\protect\citeauthoryear{{O'Dea}, {Baum}  \& {Stanghellini}}{{O'Dea} et~al.}{1991}]{1991ApJ...380...66O}
{O'Dea} C.~P.,  {Baum} S.~A.,   {Stanghellini} C.,  1991, \mn@doi [\apj] {10.1086/170562}, \href {http://adsabs.harvard.edu/abs/1991ApJ...380...66O} {380, 66}

\bibitem[\protect\citeauthoryear{{Ochsenbein}, {Bauer}  \& {Marcout}}{{Ochsenbein} et~al.}{2000}]{2000A&AS..143...23O}
{Ochsenbein} F.,  {Bauer} P.,   {Marcout} J.,  2000, \mn@doi [\aaps] {10.1051/aas:2000169}, \href {https://ui.adsabs.harvard.edu/abs/2000A&AS..143...23O} {143, 23}

\bibitem[\protect\citeauthoryear{{Pacholczyk}}{{Pacholczyk}}{1970}]{1970ranp.book.....P}
{Pacholczyk} A.~G.,  1970, {Radio astrophysics. Nonthermal processes in galactic and extragalactic sources}

\bibitem[\protect\citeauthoryear{{Parijskij}, {Goss}, {Verkhodanov}, {Kopylov}, {Soboleva}  \& {Temirova}}{{Parijskij} et~al.}{1999}]{1999BSAO...48....5P}
{Parijskij} Y.~N.,  {Goss} W.~M.,  {Verkhodanov} O.~V.,  {Kopylov} A.~I.,  {Soboleva} N.~S.,   {Temirova} A.~V.,  1999, \bsao, \href {https://ui.adsabs.harvard.edu/abs/1999BSAO...48....5P} {48, 5}

\bibitem[\protect\citeauthoryear{{Parijskij} et~al.,}{{Parijskij} et~al.}{2014}]{2014MNRAS.439.2314P}
{Parijskij} Y.~N.,  et~al., 2014, \mn@doi [\mnras] {10.1093/mnras/stu047}, \href {https://ui.adsabs.harvard.edu/abs/2014MNRAS.439.2314P/abstract} {439, 2314}

\bibitem[\protect\citeauthoryear{{P{\^a}ris} et~al.,}{{P{\^a}ris} et~al.}{2018}]{2018A&A...613A..51P}
{P{\^a}ris} I.,  et~al., 2018, \mn@doi [\aap] {10.1051/0004-6361/201732445}, \href {https://ui.adsabs.harvard.edu/abs/2018A&A...613A..51P/abstract} {613, 17}

\bibitem[\protect\citeauthoryear{{Planck Collaboration}}{{Planck Collaboration}}{2016}]{2016A&A...594A..13P}
{Planck Collaboration} 2016, \mn@doi [\aap] {10.1051/0004-6361/201525830}, \href {https://ui.adsabs.harvard.edu/abs/2016A&A...594A..13P} {594, A13}

\bibitem[\protect\citeauthoryear{{Planck Collaboration} et~al.,}{{Planck Collaboration} et~al.}{2011}]{2011A&A...536A..14P}
{Planck Collaboration} et~al., 2011, \mn@doi [\aap] {10.1051/0004-6361/201116475}, \href {https://ui.adsabs.harvard.edu/abs/2011A&A...536A..14P} {536, A14}

\bibitem[\protect\citeauthoryear{{Ramasawmy} et~al.,}{{Ramasawmy} et~al.}{2021}]{2021A&A...648A..14R}
{Ramasawmy} J.,  et~al., 2021, \mn@doi [\aap] {10.1051/0004-6361/202039858}, \href {https://ui.adsabs.harvard.edu/abs/2021A&A...648A..14R} {648, A14}

\bibitem[\protect\citeauthoryear{{Randall}, {Hopkins}, {Norris}  \& {Edwards}}{{Randall} et~al.}{2011}]{2011MNRAS.416.1135R}
{Randall} K.~E.,  {Hopkins} A.~M.,  {Norris} R.~P.,   {Edwards} P.~G.,  2011, \mn@doi [\mnras] {10.1111/j.1365-2966.2011.19116.x}, \href {https://ui.adsabs.harvard.edu/abs/2011MNRAS.416.1135R} {416, 1135}

\bibitem[\protect\citeauthoryear{{Rengelink}, {Tang}, {de Bruyn}, {Miley}, {Bremer}, {Roettgering}  \& {Bremer}}{{Rengelink} et~al.}{1997}]{1997A&AS..124..259R}
{Rengelink} R.~B.,  {Tang} Y.,  {de Bruyn} A.~G.,  {Miley} G.~K.,  {Bremer} M.~N.,  {Roettgering} H.~J.~A.,   {Bremer} M.~A.~R.,  1997, \mn@doi [\aaps] {10.1051/aas:1997358}, \href {https://ui.adsabs.harvard.edu/abs/1997A&AS..124..259R} {124, 259}

\bibitem[\protect\citeauthoryear{{Richards} et~al.,}{{Richards} et~al.}{2011}]{2011ApJS..194...29R}
{Richards} J.~L.,  et~al., 2011, \mn@doi [\apjs] {10.1088/0067-0049/194/2/29}, \href {http://adsabs.harvard.edu/abs/2011ApJS..194...29R} {194, 29}

\bibitem[\protect\citeauthoryear{{Romani}, {Sowards-Emmerd}, {Greenhill}  \& {Michelson}}{{Romani} et~al.}{2004}]{2004ApJ...610L...9R}
{Romani} R.~W.,  {Sowards-Emmerd} D.,  {Greenhill} L.,   {Michelson} P.,  2004, \mn@doi [\apjl] {10.1086/423201}, \href {https://ui.adsabs.harvard.edu/abs/2004ApJ...610L...9R} {610, L9}

\bibitem[\protect\citeauthoryear{{Sadler} et~al.,}{{Sadler} et~al.}{2006}]{2006MNRAS.371..898S}
{Sadler} E.~M.,  et~al., 2006, \mn@doi [\mnras] {10.1111/j.1365-2966.2006.10729.x}, \href {http://cdsads.u-strasbg.fr/abs/2006MNRAS.371..898S} {371, 898}

\bibitem[\protect\citeauthoryear{{Sadler}, {Chhetri}, {Morgan}, {Mahony}, {Jarrett}  \& {Tingay}}{{Sadler} et~al.}{2019}]{2019MNRAS.483.1354S}
{Sadler} E.~M.,  {Chhetri} R.,  {Morgan} J.,  {Mahony} E.~K.,  {Jarrett} T.~H.,   {Tingay} S.,  2019, \mn@doi [\mnras] {10.1093/mnras/sty3033}, \href {https://ui.adsabs.harvard.edu/abs/2019MNRAS.483.1354S/abstract} {483, 1354}

\bibitem[\protect\citeauthoryear{{Sanghera}, {Saikia}, {Luedke}, {Spencer}, {Foulsham}, {Akujor}  \& {Tzioumis}}{{Sanghera} et~al.}{1995}]{1995A&A...295..629S}
{Sanghera} H.~S.,  {Saikia} D.~J.,  {Luedke} E.,  {Spencer} R.~E.,  {Foulsham} P.~A.,  {Akujor} C.~E.,   {Tzioumis} A.~K.,  1995, \aap, \href {https://ui.adsabs.harvard.edu/abs/1995A&A...295..629S} {295, 629}

\bibitem[\protect\citeauthoryear{{Sbarrato}, {Ghisellini}, {Tagliaferri}, {Foschini}, {Nardini}, {Tavecchio}  \& {Gehrels}}{{Sbarrato} et~al.}{2015}]{2015MNRAS.446.2483S}
{Sbarrato} T.,  {Ghisellini} G.,  {Tagliaferri} G.,  {Foschini} L.,  {Nardini} M.,  {Tavecchio} F.,   {Gehrels} N.,  2015, \mn@doi [\mnras] {10.1093/mnras/stu2269}, \href {https://ui.adsabs.harvard.edu/abs/2015MNRAS.446.2483S} {446, 2483}

\bibitem[\protect\citeauthoryear{{Seymour} et~al.,}{{Seymour} et~al.}{2007}]{2007ApJS..171..353S}
{Seymour} N.,  et~al., 2007, \mn@doi [\apjs] {10.1086/517887}, \href {https://ui.adsabs.harvard.edu/abs/2007ApJS..171..353S} {171, 353}

\bibitem[\protect\citeauthoryear{{Singh} et~al.,}{{Singh} et~al.}{2014}]{2014A&A...569A..52S}
{Singh} V.,  et~al., 2014, \mn@doi [\aap] {10.1051/0004-6361/201423644}, \href {https://ui.adsabs.harvard.edu/abs/2014A&A...569A..52S} {569, A52}

\bibitem[\protect\citeauthoryear{{Smol{\v{c}}i{\'c}} et~al.,}{{Smol{\v{c}}i{\'c}} et~al.}{2014}]{2014MNRAS.443.2590S}
{Smol{\v{c}}i{\'c}} V.,  et~al., 2014, \mn@doi [\mnras] {10.1093/mnras/stu1331}, \href {https://ui.adsabs.harvard.edu/abs/2014MNRAS.443.2590S} {443, 2590}

\bibitem[\protect\citeauthoryear{{Snellen}, {Schilizzi}, {de Bruyn}, {Miley}, {Rengelink}, {Roettgering}  \& {Bremer}}{{Snellen} et~al.}{1998}]{1998A&AS..131..435S}
{Snellen} I.~A.~G.,  {Schilizzi} R.~T.,  {de Bruyn} A.~G.,  {Miley} G.~K.,  {Rengelink} R.~B.,  {Roettgering} H.~J.,   {Bremer} M.~N.,  1998, \mn@doi [\aaps] {10.1051/aas:1998281}, \href {http://adsabs.harvard.edu/abs/1998A&AS..131..435S} {131, 435}

\bibitem[\protect\citeauthoryear{{Snellen}, {Lehnert}, {Bremer}  \& {Schilizzi}}{{Snellen} et~al.}{2002}]{2002MNRAS.337..981S}
{Snellen} I.~A.~G.,  {Lehnert} M.~D.,  {Bremer} M.~N.,   {Schilizzi} R.~T.,  2002, \mn@doi [\mnras] {10.1046/j.1365-8711.2002.05978.x}, \href {http://adsabs.harvard.edu/abs/2002MNRAS.337..981S} {337, 981}

\bibitem[\protect\citeauthoryear{{Solovyov} \& {Verkhodanov}}{{Solovyov} \& {Verkhodanov}}{2014a}]{2014AstL...40..606S}
{Solovyov} D.~I.,  {Verkhodanov} O.~V.,  2014a, Astronomy Letters, \href {https://ui.adsabs.harvard.edu/abs/2014AstL...40..606S} {40, 606}

\bibitem[\protect\citeauthoryear{{Solovyov} \& {Verkhodanov}}{{Solovyov} \& {Verkhodanov}}{2014b}]{2014ARep...58..506S}
{Solovyov} D.~I.,  {Verkhodanov} O.~V.,  2014b, \mn@doi [Astronomy Reports] {10.1134/S106377291408006X}, \href {https://ui.adsabs.harvard.edu/abs/2014ARep...58..506S} {58, 506}

\bibitem[\protect\citeauthoryear{{Sotnikova}, {Mufakharov}, {Majorova}, {Mingaliev}, {Udovitskii}, {Bursov}  \& {Semenova}}{{Sotnikova} et~al.}{2019}]{2019AstBu..74..348S}
{Sotnikova} Y.~V.,  {Mufakharov} T.~V.,  {Majorova} E.~K.,  {Mingaliev} M.~G.,  {Udovitskii} R.~Y.,  {Bursov} N.~N.,   {Semenova} T.~A.,  2019, \mn@doi [Astrophysical Bulletin] {10.1134/S1990341319040023}, \href {https://ui.adsabs.harvard.edu/abs/2019AstBu..74..348S} {74, 348}

\bibitem[\protect\citeauthoryear{{Sotnikova} et~al.,}{{Sotnikova} et~al.}{2021}]{2021MNRAS.508.2798S}
{Sotnikova} Y.,  et~al., 2021, \mn@doi [\mnras] {10.1093/mnras/stab2114}, \href {https://ui.adsabs.harvard.edu/abs/2021MNRAS.508.2798S} {508, 2798}

\bibitem[\protect\citeauthoryear{{Souchay} et~al.,}{{Souchay} et~al.}{2015}]{2015A&A...583A..75S}
{Souchay} J.,  et~al., 2015, \mn@doi [\aap] {10.1051/0004-6361/201526092}, \href {https://ui.adsabs.harvard.edu/abs/2015A&A...583A..75S/abstract} {583, 9}

\bibitem[\protect\citeauthoryear{{Spingola}, {Dallacasa}, {Belladitta}, {Caccianiga}, {Giroletti}, {Moretti}  \& {Orienti}}{{Spingola} et~al.}{2020}]{2020A&A...643L..12S}
{Spingola} C.,  {Dallacasa} D.,  {Belladitta} S.,  {Caccianiga} A.,  {Giroletti} M.,  {Moretti} A.,   {Orienti} M.,  2020, \mn@doi [\aap] {10.1051/0004-6361/202039458}, \href {https://ui.adsabs.harvard.edu/abs/2020A&A...643L..12S} {643, L12}

\bibitem[\protect\citeauthoryear{{Spoelstra}, {Patnaik}  \& {Gopal-Krishna}}{{Spoelstra} et~al.}{1985}]{1985A&A...152...38S}
{Spoelstra} T.~A.~T.,  {Patnaik} A.~R.,   {Gopal-Krishna} 1985, \aap, \href {http://adsabs.harvard.edu/abs/1985A&A...152...38S} {152, 38}

\bibitem[\protect\citeauthoryear{{Stern}, {Dey}, {Spinrad}, {Maxfield}, {Dickinson}, {Schlegel}  \& {Gonz{\'a}lez}}{{Stern} et~al.}{1999}]{1999AJ....117.1122S}
{Stern} D.,  {Dey} A.,  {Spinrad} H.,  {Maxfield} L.,  {Dickinson} M.,  {Schlegel} D.,   {Gonz{\'a}lez} R.~A.,  1999, \mn@doi [\aj] {10.1086/300770}, \href {https://ui.adsabs.harvard.edu/abs/1999AJ....117.1122S/abstract} {171, 1122}

\bibitem[\protect\citeauthoryear{{Stern} et~al.,}{{Stern} et~al.}{2005}]{2005ApJ...631..163S}
{Stern} D.,  et~al., 2005, \mn@doi [\apj] {10.1086/432523}, \href {https://ui.adsabs.harvard.edu/abs/2005ApJ...631..163S} {631, 163}

\bibitem[\protect\citeauthoryear{{Stern} et~al.,}{{Stern} et~al.}{2012}]{2012ApJ...753...30S}
{Stern} D.,  et~al., 2012, \mn@doi [\apj] {10.1088/0004-637X/753/1/30}, \href {https://ui.adsabs.harvard.edu/abs/2012ApJ...753...30S/abstract} {753, 18}

\bibitem[\protect\citeauthoryear{{Tielens}, {Miley}  \& {Willis}}{{Tielens} et~al.}{1979}]{1979A&AS...35..153T}
{Tielens} A.~G.~G.~M.,  {Miley} G.~K.,   {Willis} A.~G.,  1979, \aaps, \href {https://ui.adsabs.harvard.edu/abs/1979A&AS...35..153T} {35, 153}

\bibitem[\protect\citeauthoryear{{Tornikoski}, {Lainela}  \& {Valtaoja}}{{Tornikoski} et~al.}{2000}]{2000AJ....120.2278T}
{Tornikoski} M.,  {Lainela} M.,   {Valtaoja} E.,  2000, \mn@doi [\aj] {10.1086/316809}, \href {http://adsabs.harvard.edu/abs/2000AJ....120.2278T} {120, 2278}

\bibitem[\protect\citeauthoryear{{Trushkin}}{{Trushkin}}{2003}]{2003BSAO...55...90T}
{Trushkin} S.~A.,  2003, \mn@doi [\bsao] {10.1134/S1063773706070012}, \href {https://ui.adsabs.harvard.edu/abs/2003BSAO...55...90T} {55, 90 }

\bibitem[\protect\citeauthoryear{{Tucci} et~al.,}{{Tucci} et~al.}{2008}]{2008MNRAS.386.1729T}
{Tucci} M.,  et~al., 2008, \mn@doi [\mnras] {10.1111/j.1365-2966.2008.13161.x}, \href {http://adsabs.harvard.edu/abs/2008MNRAS.386.1729T} {386, 1729}

\bibitem[\protect\citeauthoryear{{T{\"u}rler}, {Courvoisier}  \& {Paltani}}{{T{\"u}rler} et~al.}{1999}]{1999A&A...349...45T}
{T{\"u}rler} M.,  {Courvoisier} T.~J.~L.,   {Paltani} S.,  1999, \mn@doi [\aap] {10.48550/arXiv.astro-ph/9906274}, \href {https://ui.adsabs.harvard.edu/abs/1999A&A...349...45T} {349, 45}

\bibitem[\protect\citeauthoryear{{Verkhodanov}}{{Verkhodanov}}{1994}]{1994ARep...38..307V}
{Verkhodanov} O.~V.,  1994, Astronomy Reports, \href {https://ui.adsabs.harvard.edu/abs/1994ARep...38..307V} {38, 307}

\bibitem[\protect\citeauthoryear{Verkhodanov}{Verkhodanov}{1997}]{1997..conf..1V}
Verkhodanov O.~V.,  1997, in "Problems of Modern Radio Astronomy". p.~322

\bibitem[\protect\citeauthoryear{{Verkhodanov} \& {Parijskij}}{{Verkhodanov} \& {Parijskij}}{2008}]{Verkhodanov_and_Parijskij}
{Verkhodanov} O.~V.,  {Parijskij} Y.,  2008, Fiz.Mat.Lit.

\bibitem[\protect\citeauthoryear{{Verkhodanov} \& {Trushkin}}{{Verkhodanov} \& {Trushkin}}{2000}]{2000BSAO...50..115V}
{Verkhodanov} O.~V.,  {Trushkin} S.~A.,  2000, \bsao, \href {https://ui.adsabs.harvard.edu/abs/2000BSAO...50..115V} {50, 115}

\bibitem[\protect\citeauthoryear{{Verkhodanov} \& {Verkhodanova}}{{Verkhodanov} \& {Verkhodanova}}{1999}]{1999ARep...43..417V}
{Verkhodanov} O.~V.,  {Verkhodanova} N.~V.,  1999, Astronomy Reports, \href {https://ui.adsabs.harvard.edu/abs/1999ARep...43..417V} {483, 417}

\bibitem[\protect\citeauthoryear{{Verkhodanov}, {Trushkin}  \& {Chernenkov}}{{Verkhodanov} et~al.}{1997}]{1997BaltA...6..275V}
{Verkhodanov} O.~V.,  {Trushkin} S.~A.,   {Chernenkov} V.~N.,  1997, Baltic Astronomy, \href {http://adsabs.harvard.edu/abs/1997BaltA...6..275V} {6, 275}

\bibitem[\protect\citeauthoryear{{Verkhodanov}, {Kopylov}, {Parijskij}, {Soboleva}  \& {Temirova}}{{Verkhodanov} et~al.}{1999}]{1999BSAO...48...41V}
{Verkhodanov} O.~V.,  {Kopylov} A.~I.,  {Parijskij} Y.~N.,  {Soboleva} N.~S.,   {Temirova} A.~V.,  1999, \bsao, \href {https://ui.adsabs.harvard.edu/abs/1999BSAO...48...41V} {48, 41}

\bibitem[\protect\citeauthoryear{{Verkhodanov}, {Opylov}, {Zhelenkova}, {Verkhodanova}, {Chernenkov}, {Parijskij}, {Soboleva}  \& {Temirova}}{{Verkhodanov} et~al.}{2000a}]{2000A&AT...19..663V}
{Verkhodanov} O.~V.,  {Opylov} A.~I.,  {Zhelenkova} O.~P.,  {Verkhodanova} N.~V.,  {Chernenkov} V.~N.,  {Parijskij} Y.,  {Soboleva} N.~S.,   {Temirova} A.~V.,  2000a, \mn@doi [\aat] {10.1080/10556790008238611}, \href {https://ui.adsabs.harvard.edu/abs/2000A&26AT...19..663V} {19, 663}

\bibitem[\protect\citeauthoryear{{Verkhodanov}, {Andernach}  \& {Verkhodanova}}{{Verkhodanov} et~al.}{2000b}]{2000BSAO...49...53V}
{Verkhodanov} O.~V.,  {Andernach} H.,   {Verkhodanova} N.~V.,  2000b, \bsao, \href {https://ui.adsabs.harvard.edu/abs/2000BSAO...49...53V} {29, 53}

\bibitem[\protect\citeauthoryear{{Verkhodanov}, {Parijskij}, {Soboleva}, {Kopylov}, {Temirova}, {Zhelenkova}  \& {Goss}}{{Verkhodanov} et~al.}{2001}]{2001BSAO...52....5V}
{Verkhodanov} O.~V.,  {Parijskij} Y.~N.,  {Soboleva} N.~S.,  {Kopylov} A.~I.,  {Temirova} A.~V.,  {Zhelenkova} O.~P.,   {Goss} W.~V.,  2001, \bsao, \href {https://ui.adsabs.harvard.edu/abs/2001BSAO...52....5V} {52, 3 }

\bibitem[\protect\citeauthoryear{{Verkhodanov}, {Chavushyan}, {M{\'u}jica}, {Trushkin}  \& {Vald{\'e}s}}{{Verkhodanov} et~al.}{2003}]{2003ARep...47..119V}
{Verkhodanov} O.~V.,  {Chavushyan} V.~H.,  {M{\'u}jica} R.,  {Trushkin} S.~A.,   {Vald{\'e}s} J.~R.,  2003, \mn@doi [Astronomy Reports] {10.1134/1.1545575}, \href {https://ui.adsabs.harvard.edu/abs/2003ARep...47..119V} {47, 119}

\bibitem[\protect\citeauthoryear{{Verkhodanov}, {Trushkin}, {Andernach}  \& {Chernenkov}}{{Verkhodanov} et~al.}{2005}]{2005BSAO...58..118V}
{Verkhodanov} O.~V.,  {Trushkin} S.~A.,  {Andernach} H.,   {Chernenkov} V.~N.,  2005, Bulletin of the Special Astrophysics Observatory, \href {https://ui.adsabs.harvard.edu/abs/2005BSAO...58..118V} {58, 118}

\bibitem[\protect\citeauthoryear{{Verkhodanov}, {Khabibullina}, {Singh}, {Pirya}, {Verkhodanova}  \& {Nandi}}{{Verkhodanov} et~al.}{2008}]{2008pc2..conf..247V}
{Verkhodanov} O.~V.,  {Khabibullina} M.~L.,  {Singh} M.,  {Pirya} A.,  {Verkhodanova} N.~V.,   {Nandi} S.,  2008, in Problems of Practical Cosmology, Volume 2. pp 247--450

\bibitem[\protect\citeauthoryear{{Verkhodanov}, {Verkhodanova}  \& {Andernach}}{{Verkhodanov} et~al.}{2009}]{2009AstBu..64...72V}
{Verkhodanov} O.~V.,  {Verkhodanova} N.~V.,   {Andernach} H.,  2009, \mn@doi [\bsao] {10.1134/S1990341309010052}, \href {https://ui.adsabs.harvard.edu/abs/2009AstBu..64...72V} {64, 72}

\bibitem[\protect\citeauthoryear{{Verkhodanov}, {Kozlova}  \& {Sotnikova}}{{Verkhodanov} et~al.}{2018}]{2018AstBu..73..393V}
{Verkhodanov} O.~V.,  {Kozlova} D.~D.,   {Sotnikova} Y.~V.,  2018, \mn@doi [Astrophysical Bulletin] {10.1134/S1990341318040016}, \href {https://ui.adsabs.harvard.edu/abs/2018AstBu..73..393V} {73, 393}

\bibitem[\protect\citeauthoryear{{White} et~al.,}{{White} et~al.}{2012}]{2012MNRAS.427.1830W}
{White} G.~J.,  et~al., 2012, \mn@doi [\mnras] {10.1111/j.1365-2966.2012.21684.x}, \href {https://ui.adsabs.harvard.edu/abs/2012MNRAS.427.1830W} {427, 1830}

\bibitem[\protect\citeauthoryear{{Wright}, {Wark}, {Troup}, {Otrupcek}, {Jennings}, {Hunt}  \& {Cooke}}{{Wright} et~al.}{1991}]{1991MNRAS.251..330W}
{Wright} A.~E.,  {Wark} R.~M.,  {Troup} E.,  {Otrupcek} R.,  {Jennings} D.,  {Hunt} A.,   {Cooke} D.~J.,  1991, \mn@doi [\mnras] {10.1093/mnras/251.2.330}, \href {https://ui.adsabs.harvard.edu/abs/1991MNRAS.251..330W} {251, 330}

\bibitem[\protect\citeauthoryear{{Wright} et~al.,}{{Wright} et~al.}{2010}]{2010AJ....140.1868W}
{Wright} E.~L.,  et~al., 2010, \mn@doi [\aj] {10.1088/0004-6256/140/6/1868}, \href {https://ui.adsabs.harvard.edu/abs/2010AJ....140.1868W/abstract} {140, 1868}

\bibitem[\protect\citeauthoryear{{Wrobel}, {Patnaik}, {Browne}  \& {Wilkinson}}{{Wrobel} et~al.}{1998}]{1998AAS...193.4004W}
{Wrobel} J.~M.,  {Patnaik} A.~R.,  {Browne} I.~W.~A.,   {Wilkinson} P.~N.,  1998, in American Astronomical Society Meeting Abstracts. p. 40.04

\bibitem[\protect\citeauthoryear{{de Vries}, {Barthel}  \& {O'Dea}}{{de Vries} et~al.}{1997}]{1997A&A...321..105D}
{de Vries} W.~H.,  {Barthel} P.~D.,   {O'Dea} C.~P.,  1997, \aap, \href {http://adsabs.harvard.edu/abs/1997A&A...321..105D} {321, 105}

\bibitem[\protect\citeauthoryear{{van Breugel}, {De Breuck}, {Stanford}, {Stern}, {R{\"o}ttgering}  \& {Miley}}{{van Breugel} et~al.}{1999}]{1999ApJ...518L..61V}
{van Breugel} W.,  {De Breuck} C.,  {Stanford} S.~A.,  {Stern} D.,  {R{\"o}ttgering} H.,   {Miley} G.,  1999, \mn@doi [\apj] {10.1086/312080}, \href {https://ui.adsabs.harvard.edu/abs/1999ApJ...518L..61V} {518, L61}

\makeatother
\end{thebibliography}

\appendix
\onecolumn

\section{The radio continuum spectra of the galaxies.}

\begin{figure}
\includegraphics[width=180mm,bb=12 130 550 760,clip]{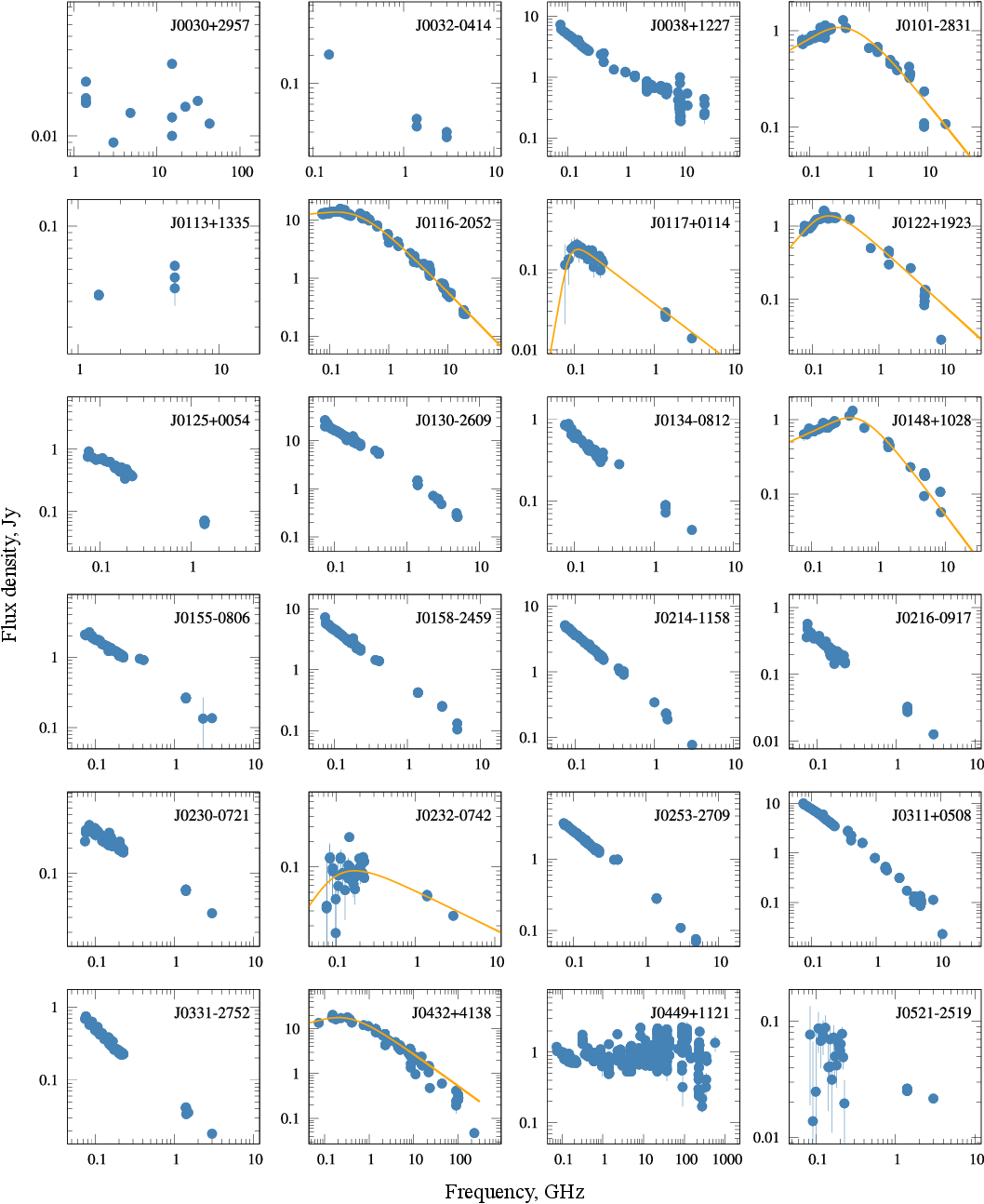}
\caption{The radio continuum spectra of the galaxies.}
\label{fig:A1}
\end{figure}

\newpage

\begin{figure} \addtocounter{figure}{-1}
\includegraphics[width=180mm,bb=9 130 550 760,clip]{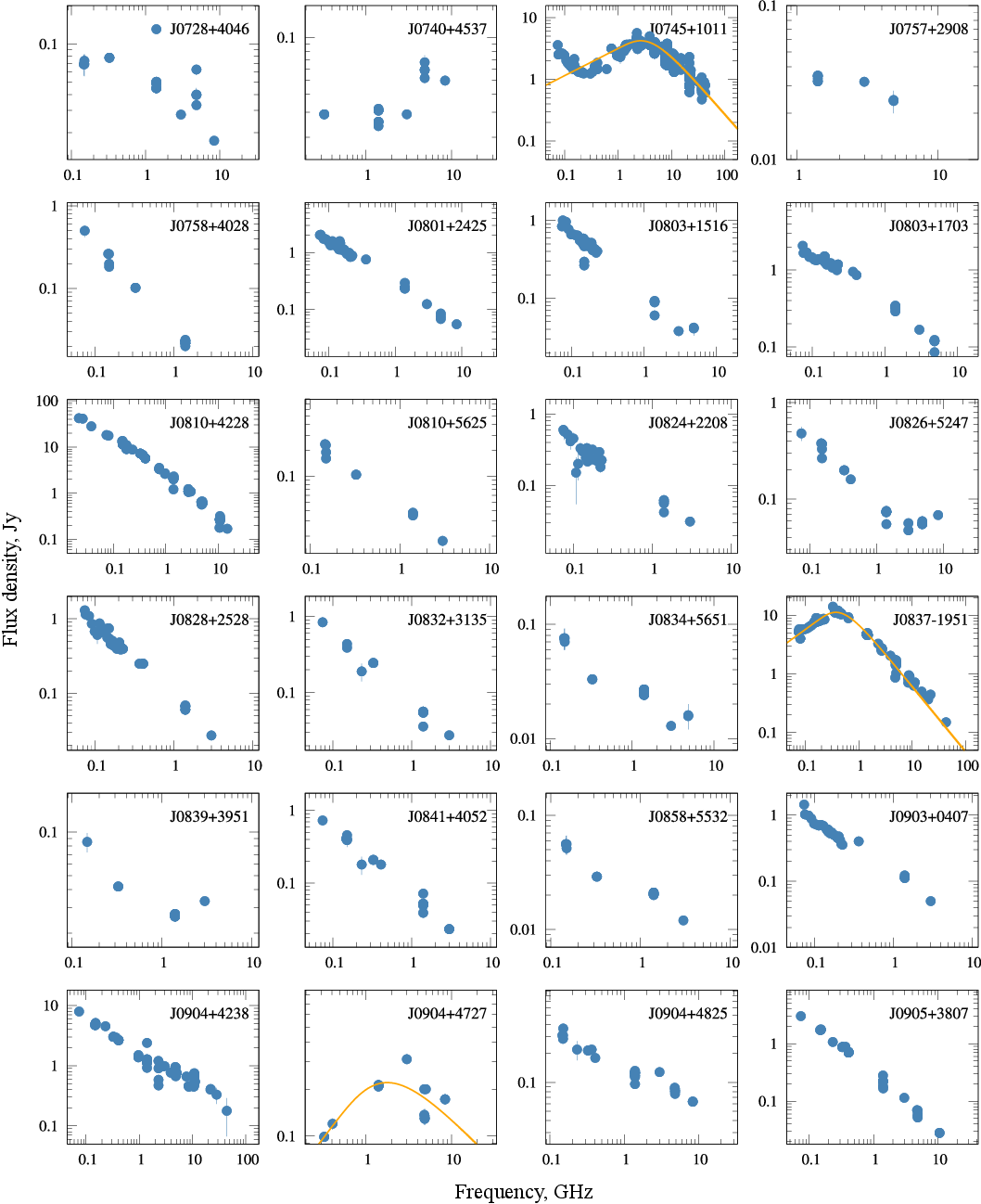}
\caption{(Contd.)}
\label{fig:A2}
\end{figure}

\newpage

\begin{figure} \addtocounter{figure}{-1}
\includegraphics[width=180mm,bb=9 130 550 760,clip]{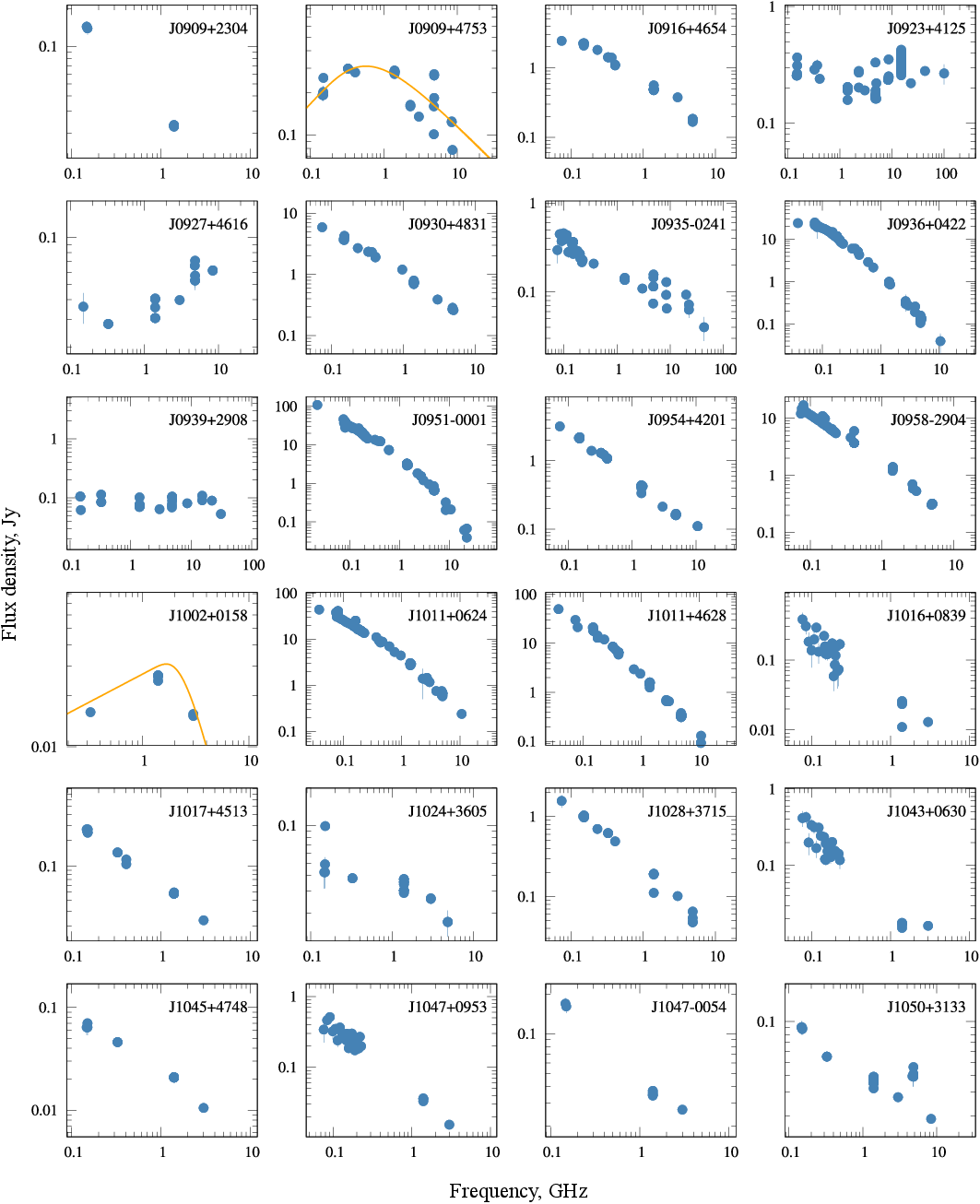}
\caption{(Contd.)}
\label{fig:A3}
\end{figure}

\newpage

\begin{figure} \addtocounter{figure}{-1}
\includegraphics[width=180mm,bb=9 130 550 760,clip]{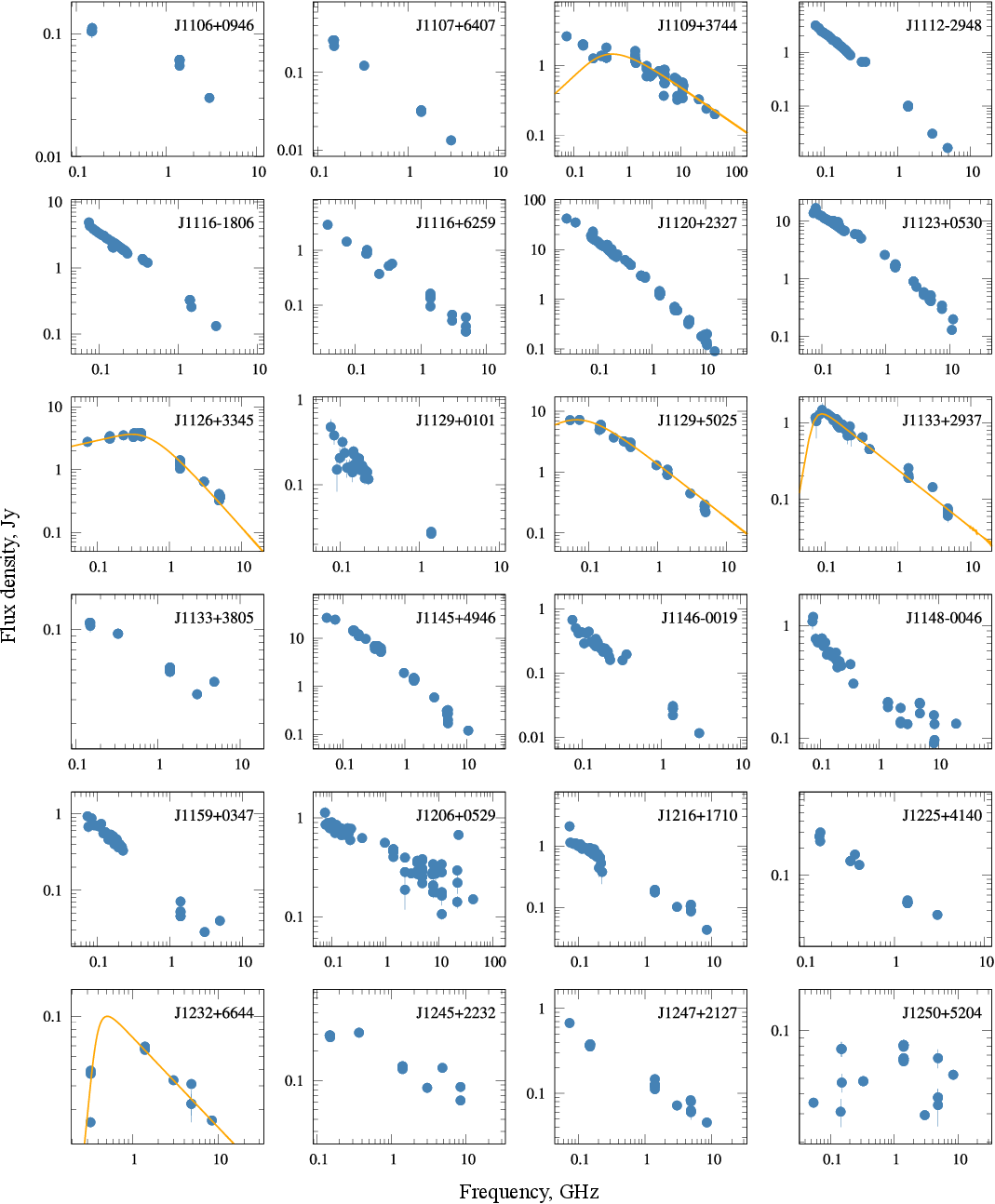}
\caption{(Contd.)}
\label{fig:A4}
\end{figure}

\newpage

\begin{figure} \addtocounter{figure}{-1}
\includegraphics[width=180mm,bb=9 130 550 760,clip]{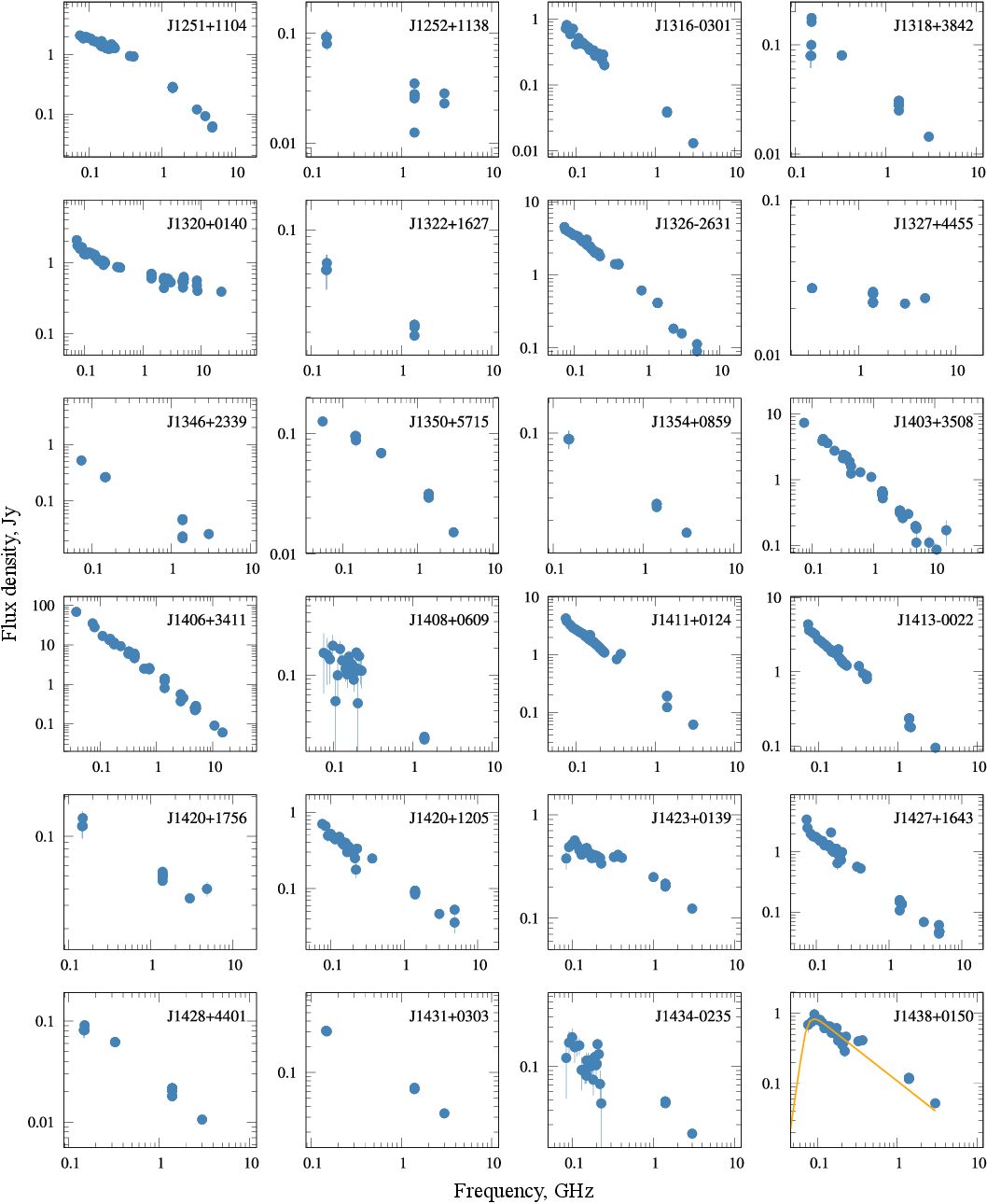}
\caption{(Contd.)}
\label{fig:A5}
\end{figure}

\newpage

\begin{figure} \addtocounter{figure}{-1}
\includegraphics[width=180mm,bb=9 130 550 760,clip]{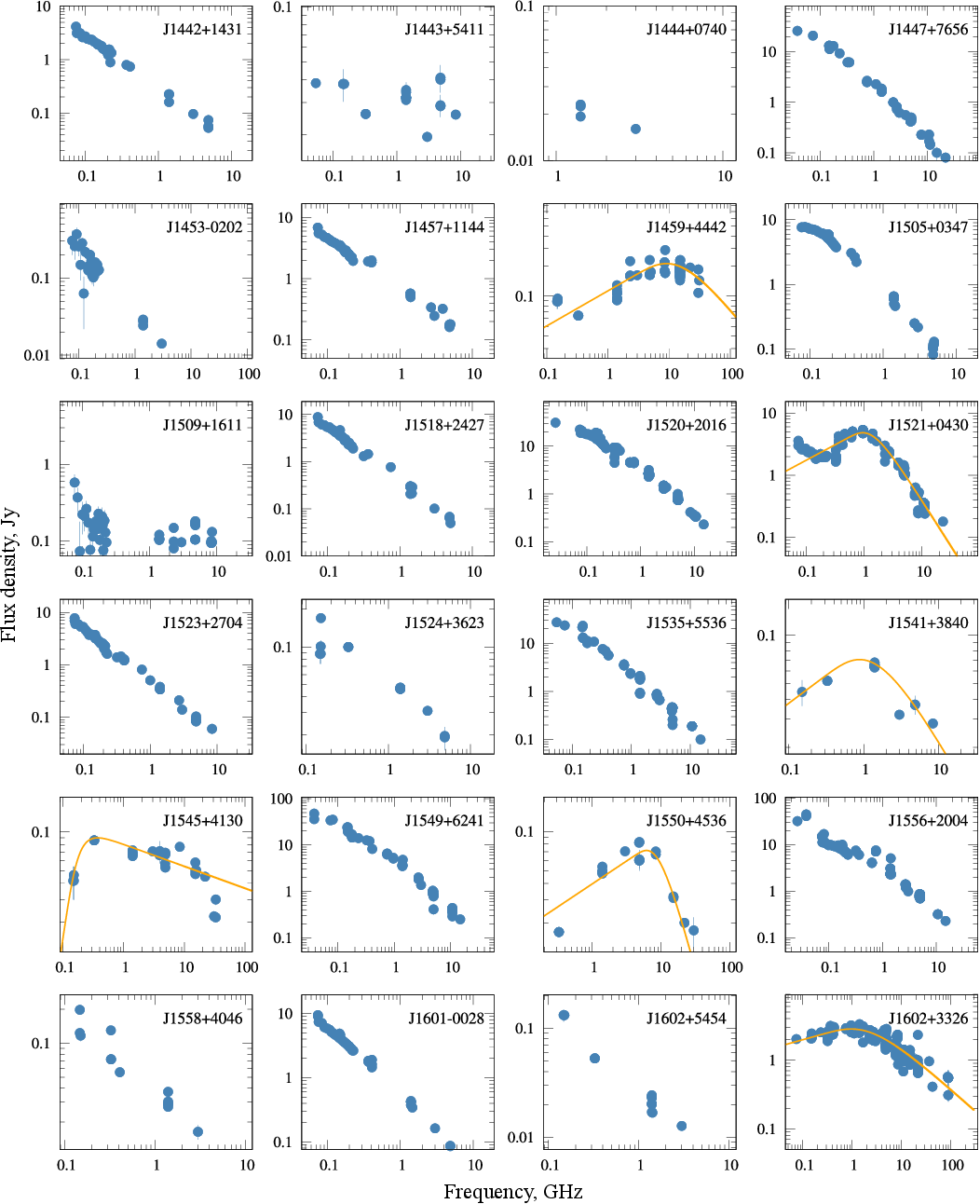}
\caption{(Contd.)}
\label{fig:A6}
\end{figure}

\newpage

\begin{figure} \addtocounter{figure}{-1}
\includegraphics[width=180mm,bb=9 130 550 760,clip]{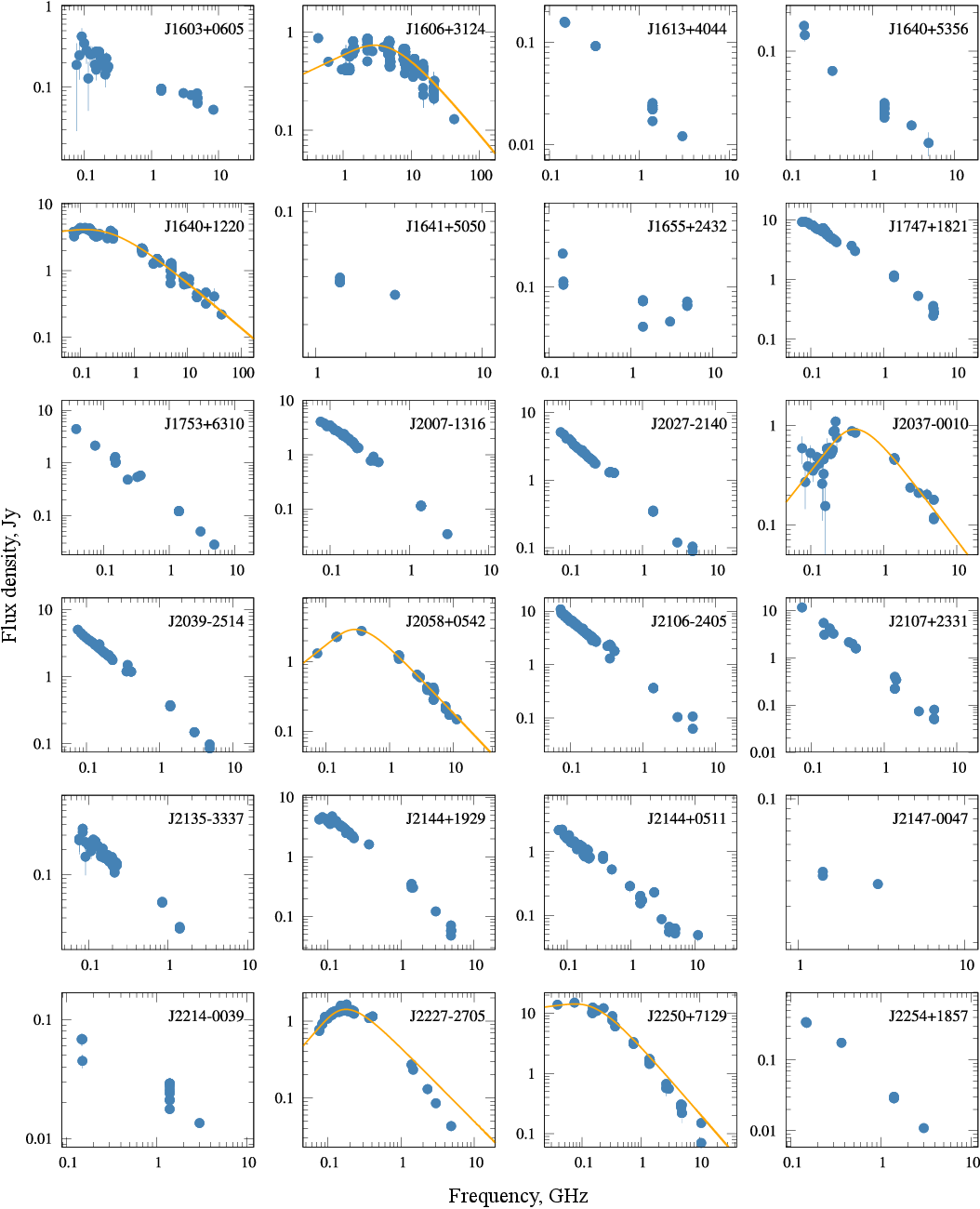}
\caption{(Contd.)}
\label{fig:A7}
\end{figure}

\newpage

\begin{figure} \addtocounter{figure}{-1}
\includegraphics[width=169mm]{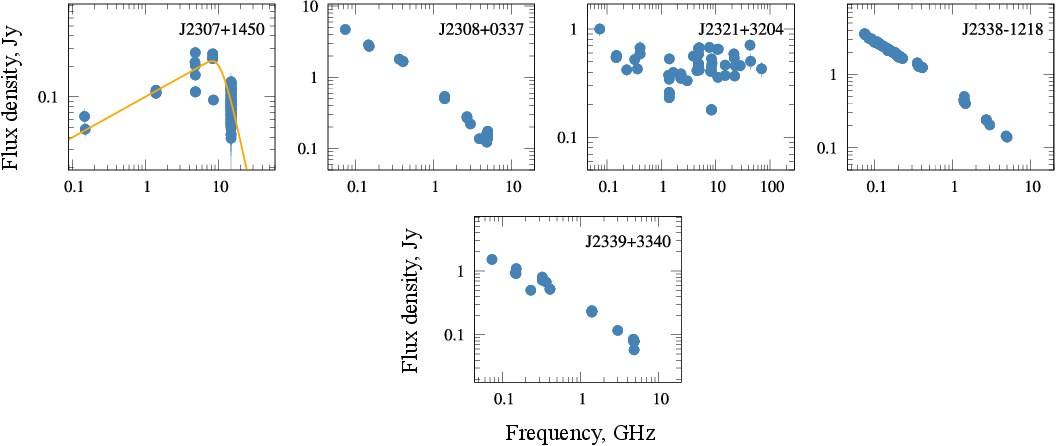}
\caption{(Contd.)}
\label{fig:A8}
\end{figure}

\end{document}